\documentclass[iop]{emulateapj}

\slugcomment{Submitted to the Astrophysical Journal}
\shorttitle{}
\shortauthors{}

\begin{document}  


\title{COMPARISON OF PRESTELLAR CORE ELONGATIONS
AND LARGE-SCALE MOLECULAR CLOUD STRUCTURES IN THE LUPUS I REGION}
\shorttitle{}
\shortauthors{}


\author{POIDEVIN Fr\'ed\'erick}
\affil{UCL, KLB, Department of Physics \& Astronomy, Gower Place,
  London WC1E 6BT, United Kingdom; Instituto de Astrofísica de Canarias, E-38200 La Laguna, Tenerife, Spain; 
Universidad de La Laguna, Dept. Astrofísica, E-38206 La Laguna, Tenerife, Spain.}
\email{fpoidevin@iac.es}

\author{ADE Peter A.R.}
\affil{Cardiff University, School of Physics and Astronomy, Queens
Buildings, The Parade, Cardiff, CF24 3AA, U.K.}

\author{ANGILE Francesco E.}
\affil{Department of Physics and Astronomy, University of Pennsylvania,
209 South 33rd Street, Philadelphia, PA, 19104, U.S.A.}

\author{BENTON Steven J.}
\affil{Department of Physics, University of Toronto, 60 St. George St, Toronto, ON, M5S 1A7, Canada.}

\author{CHAPIN Edward L.}
\affil{
XMM SOC, ESAC, Apartado 78, 28691 Villanueva de la Can\~{a}da, Madrid, Spain.}

\author{DEVLIN Mark J.}
\affil{Department of Physics and Astronomy, University of Pennsylvania,
209 South 33rd Street, Philadelphia, PA, 19104, U.S.A.}

\author{FISSEL Laura M.}
\affil{Department of Astronomy and Astrophysics, University of Toronto,
50 St. George Street, Toronto, ON M5S 3H4, Canada and Department of Physics and Astronomy,
Northwestern University, 2145 Sheridan Road, Evanston, IL 60208.}

\author{FUKUI Yasuo}
\affil{Department of Physics, Nagoya University, Chikusa-ku, Nagoya, Aichi, 464-8601, Japan.}

\author{GANDILO Natalie N.}
\affil{Department of Astronomy and Astrophysics, University of Toronto,
50 St. George Street, Toronto, ON M5S 3H4, Canada.}

\author{GUNDERSEN Joshua O.}
\affil{Department of Physics, University of Miami, 1320 Campo Sano Drive,
Coral Gables, FL, 33146, U.S.A.}

\author{HARGRAVE Peter C.}
\affil{Cardiff University, School of Physics and Astronomy, Queens
Buildings, The Parade, Cardiff, CF24 3AA, U.K.}

\author{KLEIN Jeffrey}
\affil{Department of Physics and Astronomy, University of Pennsylvania,
209 South 33rd Street, Philadelphia, PA, 19104, U.S.A.}

\author{KOROTKOV Andrei L.}
\affil{Department pof Physics, Brown University, 182 Hope Street,
Providence, RI, 02912, U.S.A.}

\author{MATTHEWS Tristan G.}
\affil{Center for Interdisciplinary Exploration and Research in
Astrophysics (CIERA) and Department of Physics and Astronomy,
Northwestern University, 2145 Sheridan Road, Evanston, IL 60208.}

\author{MONCELSI Lorenzo}
\affil{California Institute of Technology, 1200 E. California Blvd.,
Pasadena, CA, 91125, U.S.A.}

\author{MROCZKOWSKI Tony K.}
\affil{California Institute of Technology, 1200 E. California Blvd.,
Pasadena, CA, 91125, U.S.A.}

\author{NETTERFIELD Calvin B.}
\affil{Department of Astronomy and Astrophysics, University of Toronto,
50 St. George Street, Toronto, ON M5S 3H4, Canada; 
Department of Physics, University of Toronto, 60 St. George St, Toronto, ON, M5S 1A7, Canada.}

\author{NOVAK Giles}
\affil{Center for Interdisciplinary Exploration and Research in
Astrophysics (CIERA) and Department of Physics and Astronomy,
Northwestern University, 2145 Sheridan Road, Evanston, IL 60208.}

\author{NUTTER David}
\affil{Cardiff University, School of Physics and Astronomy, Queens
Buildings, The Parade, Cardiff, CF24 3AA, U.K.}

\author{OLMI Luca}
\affil{University of Puerto Rico, Rio Piedras Campus, Physics
Department, Box 23343, UPR station, San Juan, Puerto Rico AND Osservatorio Astrofisico de Arcetri, INAF, Largo E. Fermi 5,
I-50125, Firenze, Italy. }

\author{PASCALE Enzo}
\affil{Cardiff University, School of Physics and Astronomy, Queens
Buildings, The Parade, Cardiff, CF24 3AA, U.K.}

\author{SAVINI Giorgio}
\affil{UCL, KLB, Department of Physics \& Astronomy, Gower Place, London WC1E 6BT, U.K.}

\author{SCOTT Douglas}
\affil{Department of Physics and Astronomy, University of British
Colombia, 6224 Agricultural Road, Vancouver, BC V6T 1Z1, Canada.}

\author{SHARIFF Jamil A.}
\affil{Department of Astronomy and Astrophysics, University of Toronto,
50 St. George Street, Toronto, ON M5S 3H4, Canada.}

\author{SOLER Juan Diego}
\affil{Department of Astronomy and Astrophysics, University of Toronto,
50 St. George Street, Toronto, ON M5S 3H4, Canada; CNRS - Institut d'Astrophysique Spatiale, Université Paris-XI, Orsay, France.}

\author{TACHIHARA Kengo}
\affil{Department of Physics, Nagoya University, Chikusa-ku, Nagoya, Aichi, 464-8601, Japan.}

\author{THOMAS Nicholas E.}
\affil{Department of Physics, University of Miami, 1320 Campo Sano Drive,
Coral Gables, FL, 33146, U.S.A.}

\author{TRUCH Matthew D.P.}
\affil{Department of Physics and Astronomy, University of Pennsylvania,
209 South 33rd Street, Philadelphia, PA, 19104, U.S.A.}

\author{TUCKER Carole E.}
\affil{Cardiff University, School of Physics and Astronomy, Queens
Buildings, The Parade, Cardiff, CF24 3AA, U.K.}

\author{TUCKER Gregory S.}
\affil{Department pof Physics, Brown University, 182 Hope Street,
Providence, RI, 02912, U.S.A.}

\author{WARD-THOMPSON Derek}
\affil{Jeremiah Horrocks Institute, University of Central Lancashire,
PR1 2HE, U.K.}

\begin{abstract}
Turbulence and magnetic fields are expected to be important for regulating molecular
cloud formation and evolution. However, their effects on subparsec
to 100 parsec scales, leading to the formation of
starless cores, is not well understood. 
We investigate the prestellar core structure morphologies obtained from
analysis of the {\it Herschel}-SPIRE 350 $\mu$m maps of the Lupus I cloud.
This distribution is first compared on a statistical basis to the 
large scale shape of the main filament.
We find the distribution of the elongation position angle of the cores 
to be consistent with a random distribution, which means no specific orientation of the morphology 
of the cores is observed with respect to a large-scale filament shape model 
for Lupus I, or relative to a large-scale 
bent filament model. This distribution is also compared to the 
mean orientation of the large-scale magnetic fields probed 
at 350 $\mu$m with the Balloon-borne Large 
Aperture Telescope for Polarimetry (BLASTPol) during its 2010
campaign. Here again we do not find any correlation 
between the core morphology distribution and the average 
orientation of the magnetic fields on parsec scales.
Our main conclusion is that the local filament dynamics -- including
secondary filaments that often run orthogonally to the primary
filament -- and possibly small-scale variations in the local magnetic 
field direction, could be the dominant factors for explaining the final orientation of each core.

\end{abstract}

\newpage








\section{INTRODUCTION}
 
Understanding the processes leading to the formation of stars in our Galaxy
is one of the great challenges which, despite much progress \citep[e.g.,][]{mol14}, still
remains open. At sub-parsec scales, the detailed mechanisms remain
ellusive through 
which gravitational collapse occurs, leading to the
formation of a prestellar core which eventually will give birth to one
or more stars. 
Recent advances have shown that turbulence is a
key ingredient and plays a dual role, both creating overdensities to
initiate core formation and counteracting the effects of gravity into 
the denser regions of these objects \citep[e.g.,][]{mck07}.
In addition to gravity and turbulence, other physical processes are likely to play a
significant role. Specifically, magnetic field and dynamical chemistry
networks are expected to be relevant for understanding the general
phenomenology of star-formation \citep[see for example][and references
therein]{lea13, gir13, tas12a, tas12b, tas12c}. 
However, all in all, it is
currently unclear which of all these mechanisms dominates
over the other ones, 
and over which spatial and temporal scales.

On larger spatial scales, the formation and evolution of molecular clouds is not well
understood and there is abundant literature on the subject. 
In particular, several simulation approaches addressing 
these questions have been developed over the last two decades
\citep[e.g.,][]{ost01,gam03,fal08,hei09,nak11,bon13}.
While these analyses sometimes use very different lines of reasoning, 
almost all of them include and/or show that 
the combined effects of magnetic fields and turbulence are 
key ingredients to understanding core mass 
function (CMF) estimates for our Galaxy. However, the impact 
of magnetic fields on different spatial and density scales has
not yet been established with regards to explaining the observed 
star formation rate (SFR).
Recent work suggests that magnetic fields are 
regulating cloud formation \citep[e.g.,][]{hei09,nak11},
with different scenarii depending on the 
magnetic field strength and orientation with respect to 
outflow-driven turbulence.

In practice, characterization of cloud structure properties 
and star-formation efficiency
based on map analysis show strong variations
from one cloud to another \citep[e.g.,][]{sch13, ryg13}. 
Recent studies may provide clues pointing toward the general mechanisms 
dominating the processes in different regions. In one such study, \citet{bal11} propose
that bound molecular clouds could be in a state of hierachical and chaotic 
gravitationnal collapse. 
On the other hand, \citet{poi13} show 
that simple ideal, isothermal and non-self graviting MHD 
simulations are sufficient to describe the large-scale 
observed physical properties of the envelopes of four different 
molecular clouds. This result is consistent with some of the molecular
clouds not being necessarily gravitationally bound as
discussed by \citet{war14}.  
These findings raise important questions regarding the range of spatial
scales and density regimes which are involved with non-self
gravitating MHD and effective local collapse leading to core formation.

The Lupus I molecular cloud complex has already been well studied \citep[e.g.,][]{har99}. 
We do not know if these clouds are bound or unbound. However, 
related to the questions above, one can study the relation between
the large-scale structures of the filamentary molecular clouds and 
the average distribution of the prestellar core
structure morphologies associated with that region. This is the central question addressed in
this work. We focus our study on the structural morphologies of the prestellar cores 
obtained from the analysis of the 350 $\mu$m SPIRE images associated
with these clouds \citep[see][]{ryg13}. The morphology
of the large-scale filamentary structure of Lupus I is discussed  in a separate paper by
\citet{mat13}. 
In addition to photometric imaging, 
we use the submm polarimetry data set on Lupus I obtained from the
2010 BLASTPol flight \citep[see][]{mat13}.
Ideally, it would be important to compare each identified core morphology with
the local magnetic field. In practice, the resolution of the 2010
BLASTPol data precludes this approach, and
only the average core morphology (taken to be the average orientation
of the core long axes) can be compared to average direction of the
cloud magnetic field.
Therefore, we limit our analysis to these statistical properties.

In the following, Section \ref{lupus} gives a brief description of the
Lupus I region.
The  data used in this analysis are described
in Section \ref{data_sets}.
We use the list of cores detected by \citet{ryg13} in the SPIRE data,
and Section \ref{data_analysis} discusses the methodology adopted to
determine the average elongation of these sources.
A comparison of the average filamentary molecular cloud structures and the sample of
prestellar cores for which it is possible
to define accurate average elongation position angle (EPA) is given in Section
\ref{stat}. A comparison of the mean magnetic field orientation and
the prestellar core orientations is also given.
Our results and their implication are discussed in Section
\ref{discussion} with conclusions provided in section \ref{conclusion}.

\section{THE LUPUS I REGION} \label{lupus}

The Lupus I molecular cloud region is a well studied site of
star formation.
It is one of the closest star-forming regions at 
(155 $\pm$ 8) pc \citep [see][]{lom08}, lying close to the
position centered at RA (J2000)$=15^{\rm h} 42^{\rm m} 00^{\rm s}$,
Dec (J2000)$=-34^{\circ} 12{\arcmin}  00{\arcsec}$.
Using the existing {\it Spitzer} catalog and SPIRE imaging at 250, 350 
and 500 $\mu$m, \citet{ryg13} detected
cores located in this region and discuss their
evolutionary classification.
These authors show that Lupus I is undergoing a large star
formation event, as estimated by the increased 
SFR and by the large number of prestellar objects when compared to more
evolved structures. However, the mechanism 
behind this surge in star formation is not well understood.

\section{Data Sets} \label{data_sets}

\subsection{Prestellar Cores Sample} \label{core_sample}

In our analysis we use the sample of prestellar cores
identified by \citet{ryg13} with their combined analysis of the 
{\it Herschel} 70 - 500 $\mu$m maps. 
The paper does not display an explicit list of coordinates of 
the cores they discuss. Therefore,
for identifying the coordinates of the prestellar cores, 
we used the central position of the maps displayed in their figure
A.2. Most of the time the peak location of the core was 
obvious, being no further than a few arcseconds from the 
center of their map. In such cases we report these coordinates, and if not 
we report the center coordinates of their map. 
The intensity maps shown in our Fig. \ref{stamps} (discussed further
below) compare well with the intensity maps shown by \citet{ryg13}
and we believe our core position
coordinate estimates are accurate to $\approx 10{\arcsec}$. 
These core position coordinate estimates
are provided in columns 2 and 3 of Tables \ref{tabcore1} and
\ref{tabcore2}, respectively. 

\subsection{BLASTPol Polarimetry Data Set} \label{polarimetry_data_set}

One of the best ways for probing magnetic fields in molecular clouds
is through submillimeter
(submm) polarimetry, where the radiation from aspherical dust grains,
aligned by the local magnetic field \citep[see][for a review]{laz07},
is detected in polarization. Therefore, submm polarimetry data
provide information about the
mean projected component of the magnetic field on the plane-of-the-sky
(POS). We use BLASTPol \citep[see][]{pas12, ang14} submm
polarimetry to
infer the POS magnetic field orientations.

Details about the BLAST and BLASTpol experiment, instruments 
and flights are given by \citet{pas12}, \citet{mon14} and \citet{ang14}.
\citet{mat13} provides a detailed discussion of the 
polarimetry analysis of the 2010 BLASTPol data at 250, 
350 and 500 $\mu$m.
Most of the modified blackbody fits of the prestellar cores discussed by \citet{ryg13} 
peak (in S$_{\nu}$ units) at a wavelength close to 350  $\mu$m. For this reason, we
focus our analysis on the polarimetry data set obtained at this wavelength.
This data set is similar to the one used by \citet{mat13} in
their analysis.

\section{DATA ANALYSIS} \label{data_analysis}

In this section the position angles (PA), whether they refer to
core elongation position angle (EPA) as seen on the POS, or to mean magnetic field
orientations, are counted positively from north in an anticlockwise direction.

The polarization PA is 
periodic and is defined to wrap around in a $[0^{\circ},180^{\circ}]$
period. The median values retained in our analysis correspond
to the mean and median estimates obtained such that the dispersions 
of the distributions are found to be the smallest.

\begin{figure*}
\centering
\includegraphics[width=4cm]{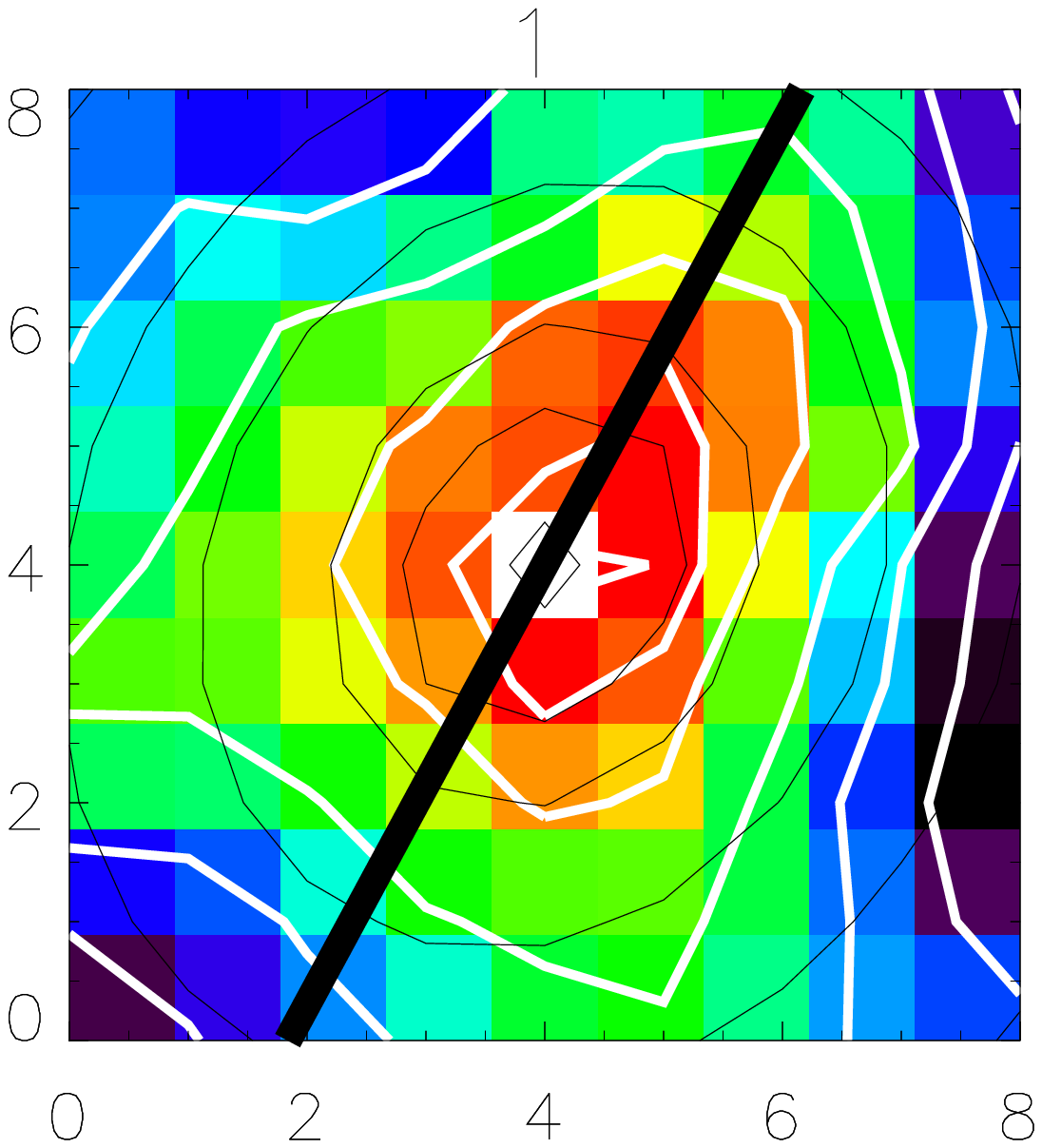}
\includegraphics[width=4cm]{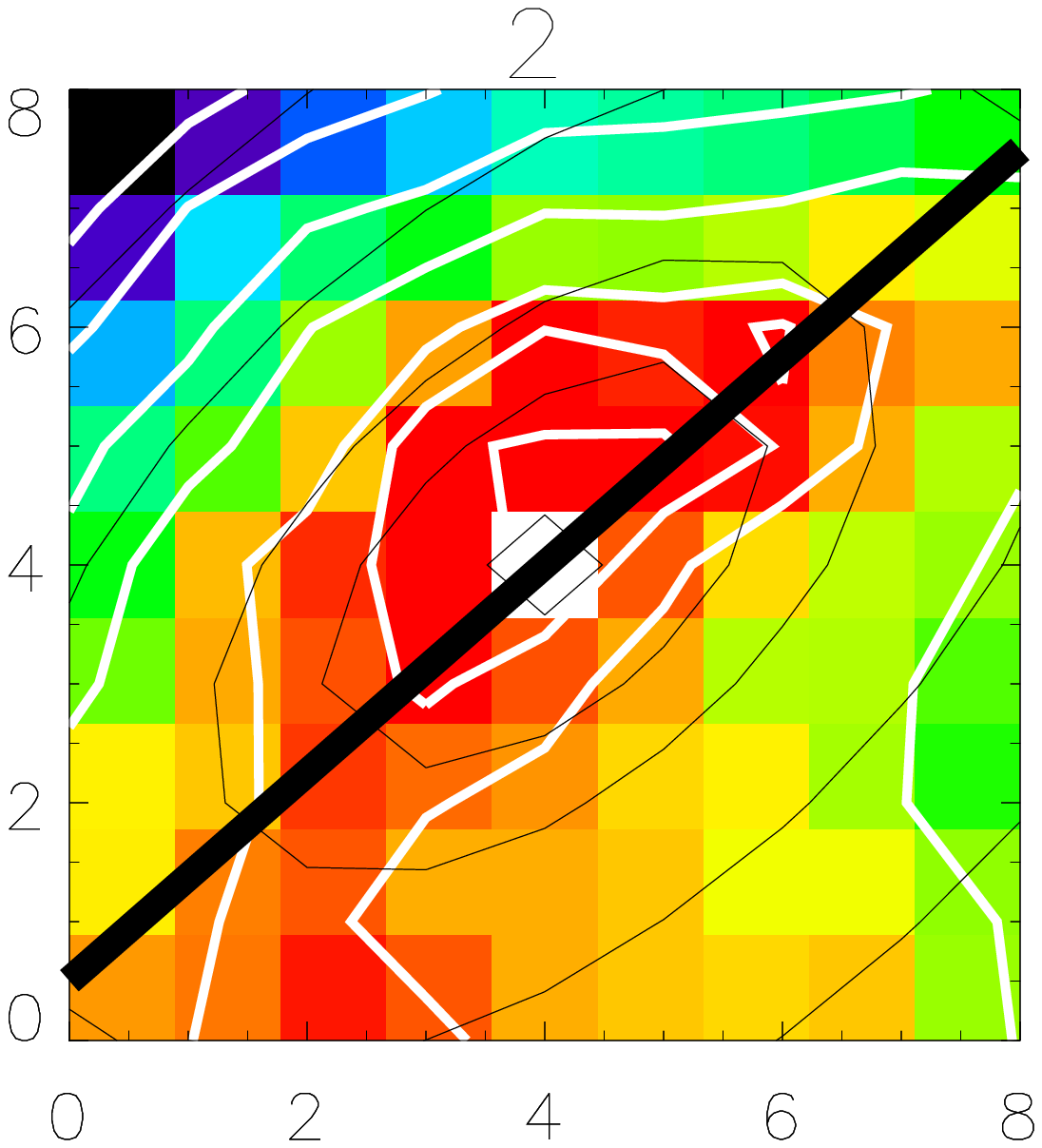}
\includegraphics[width=4cm]{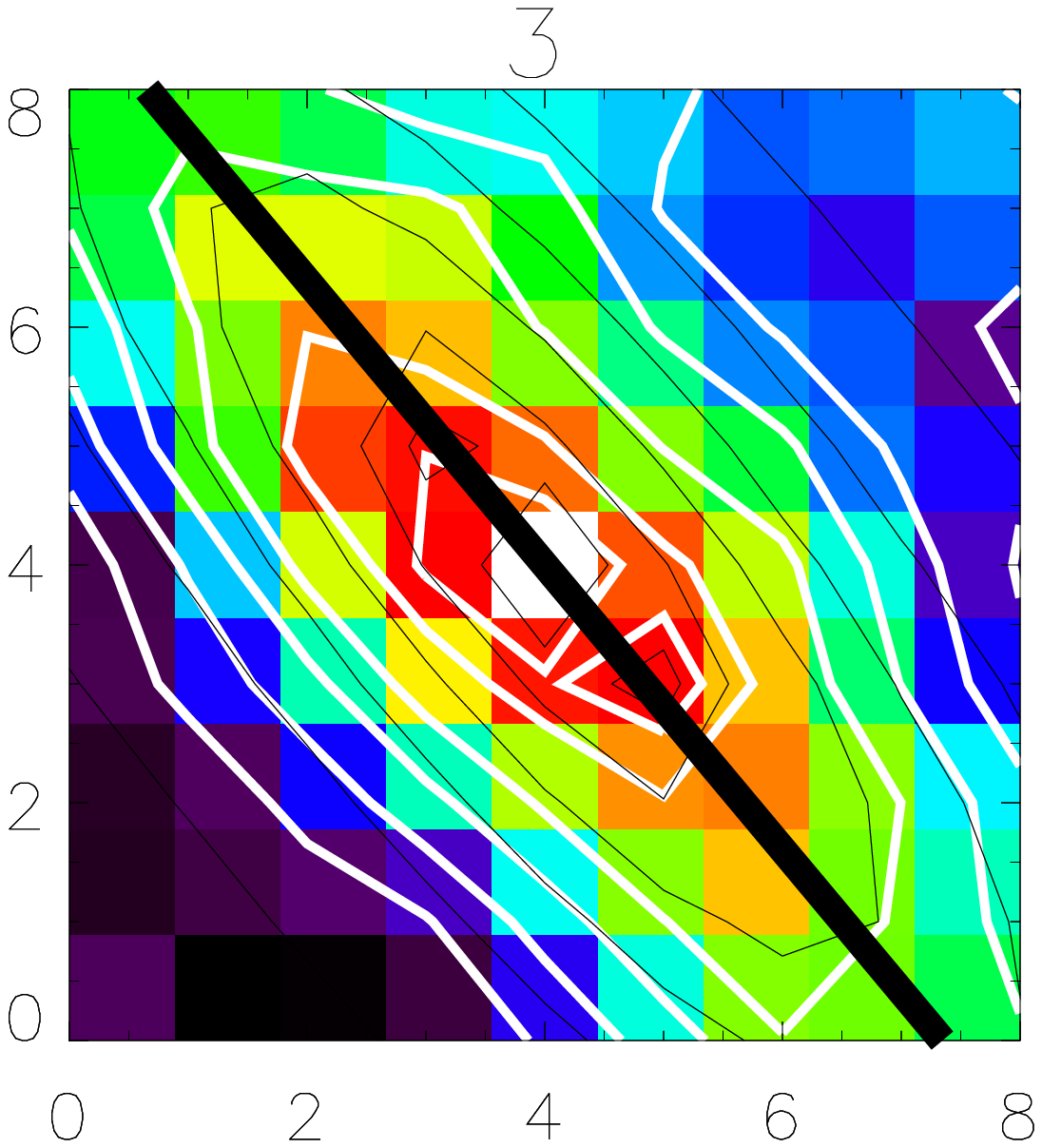}
\includegraphics[width=4cm]{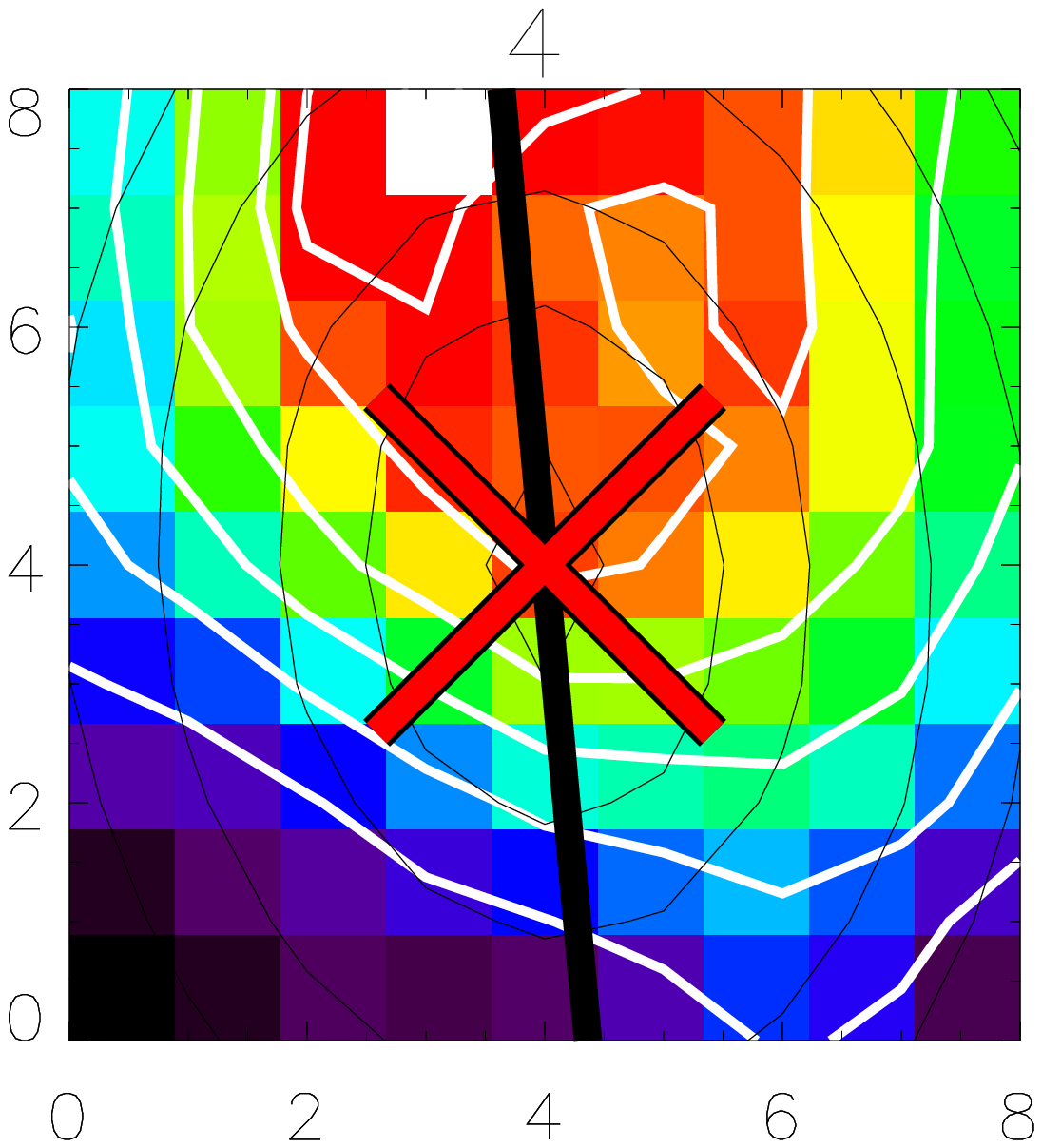}
\includegraphics[width=4cm]{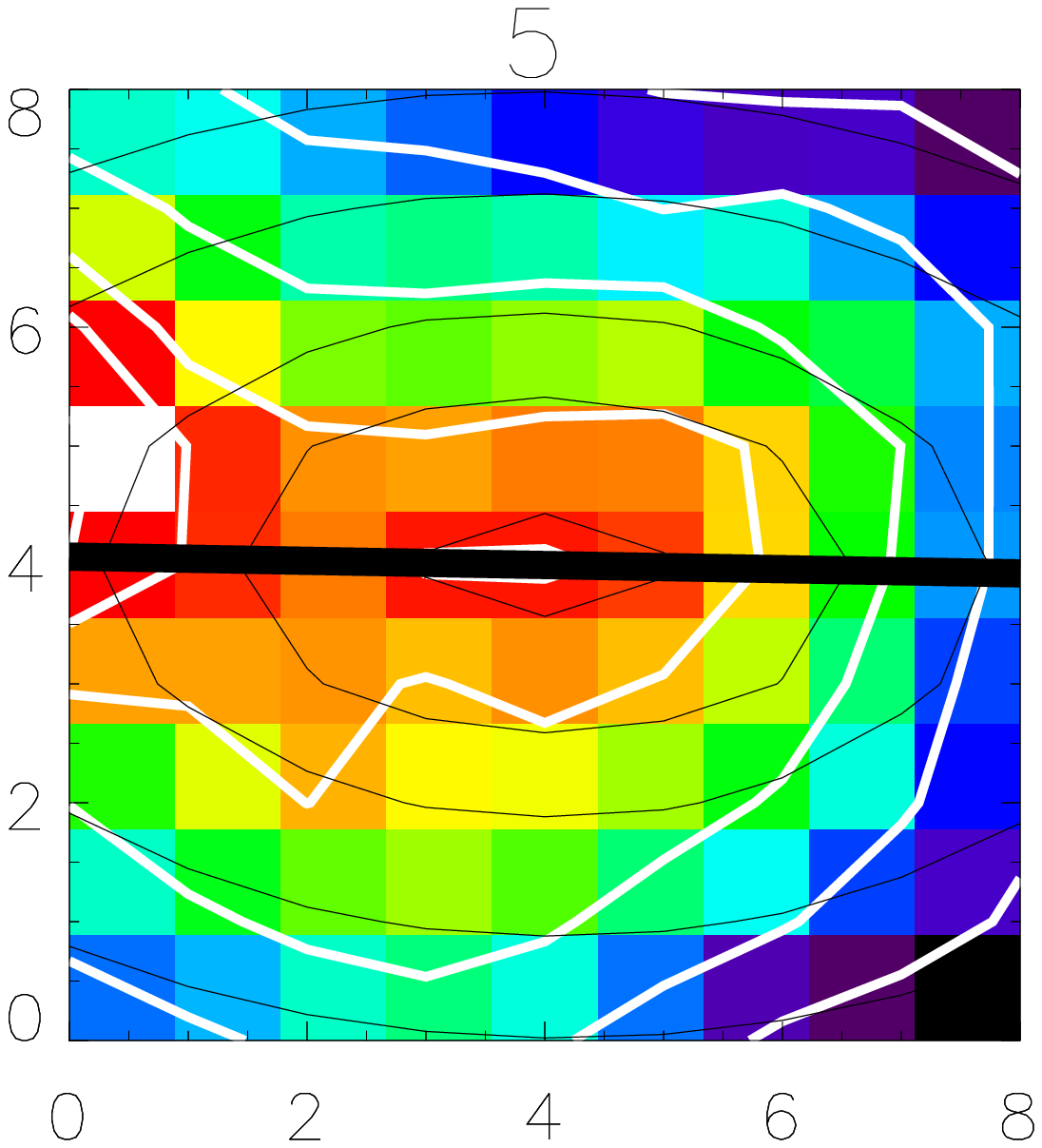}
\includegraphics[width=4cm]{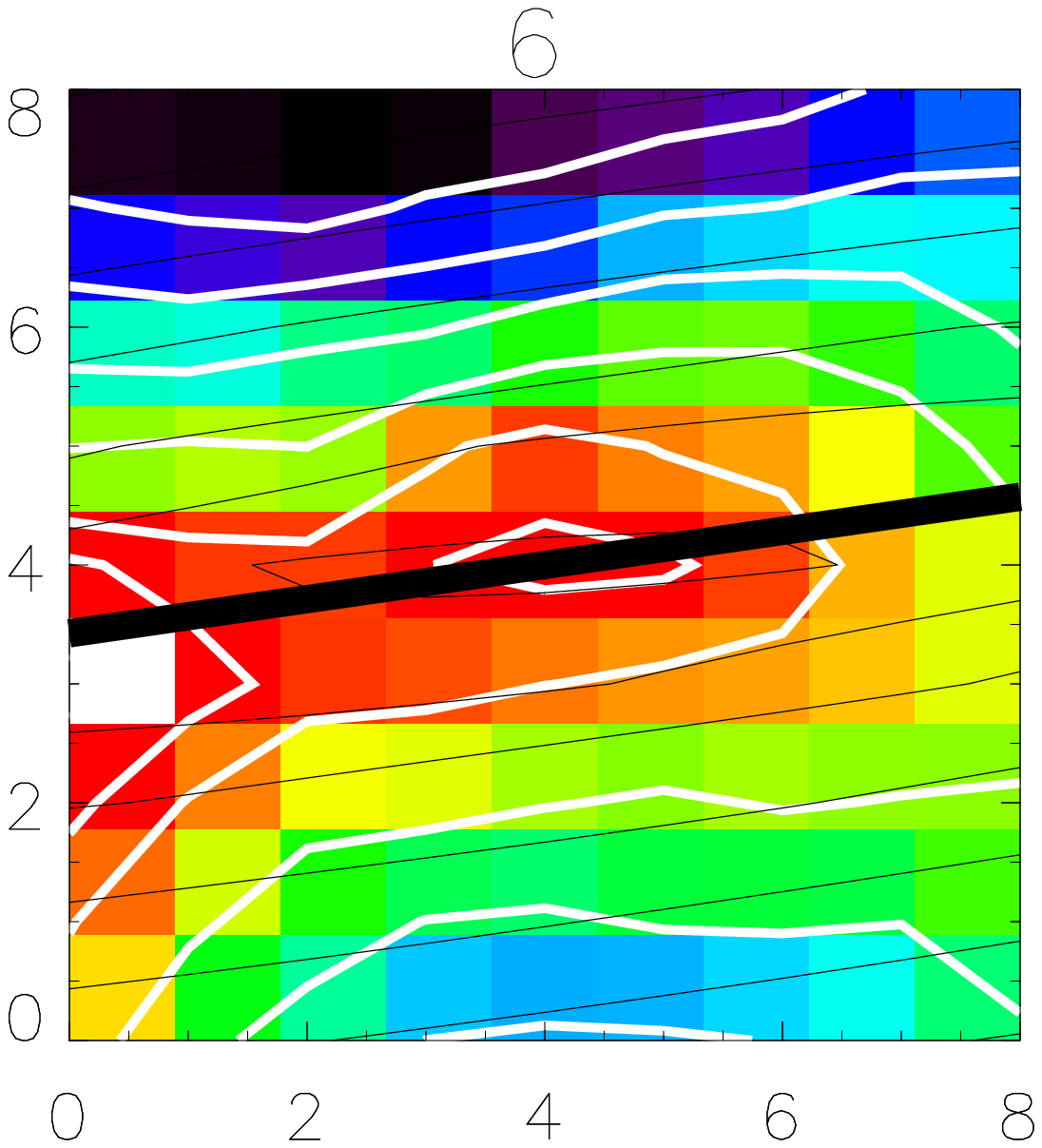}
\includegraphics[width=4cm]{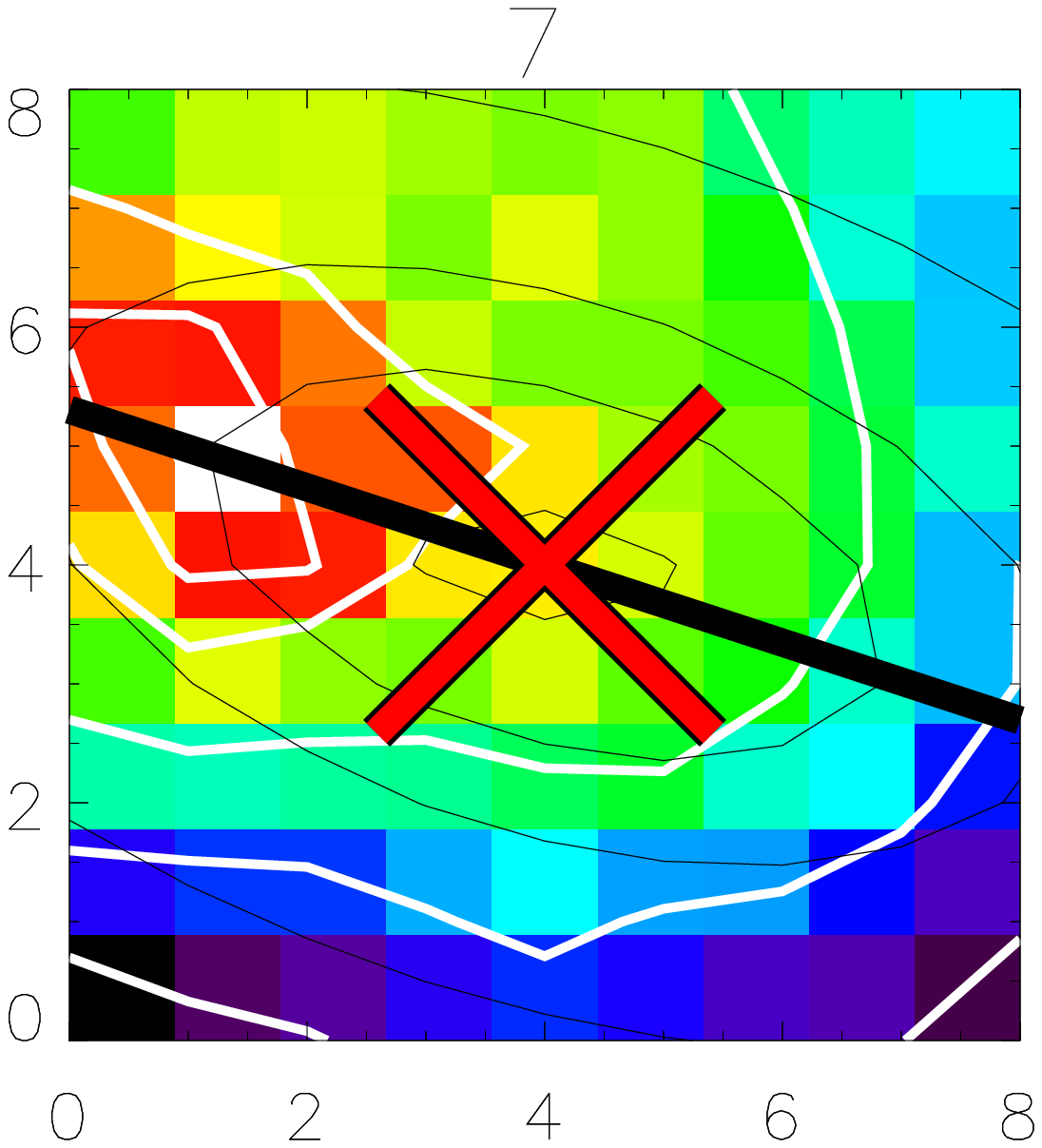}
\includegraphics[width=4cm]{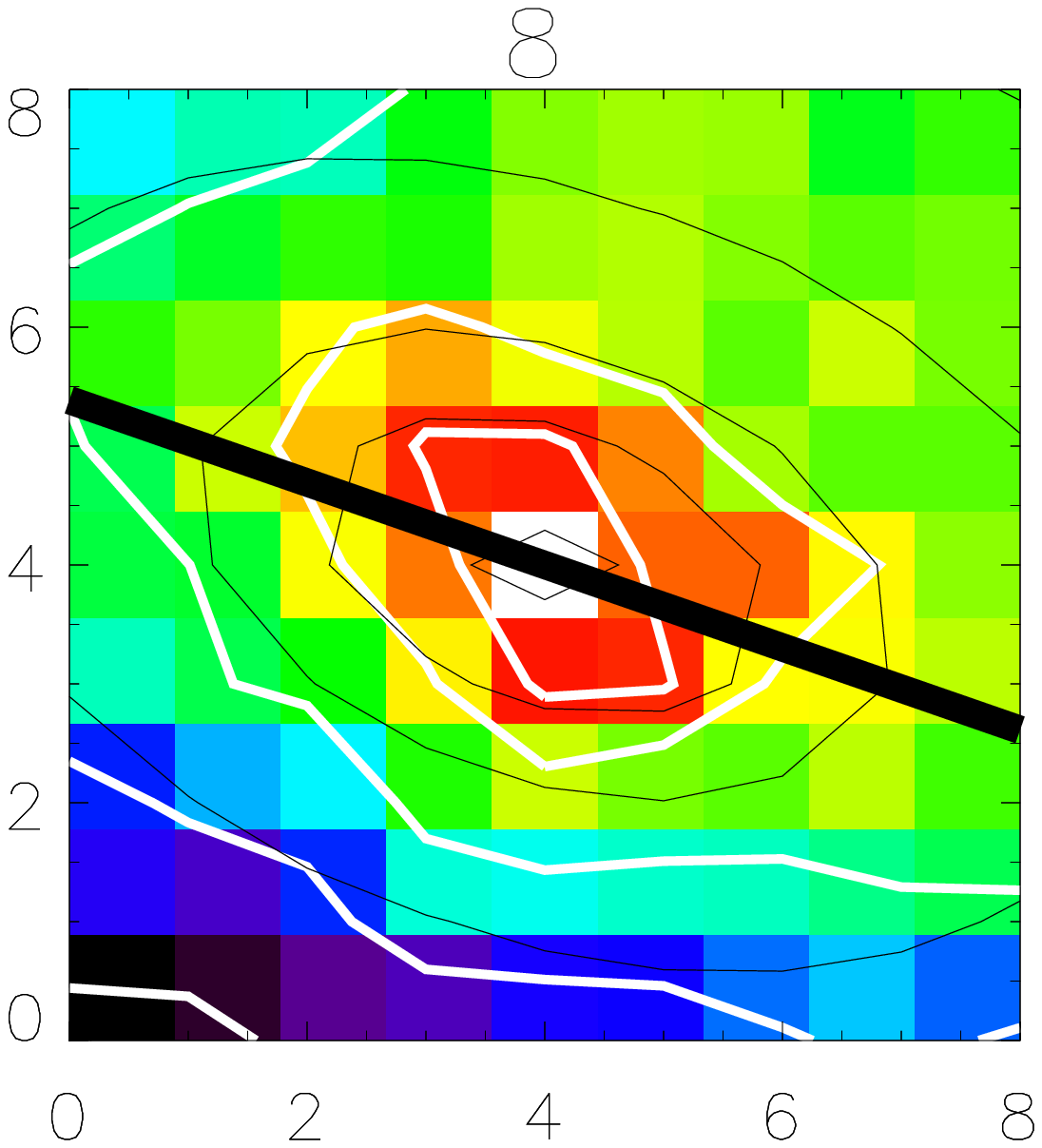}
\includegraphics[width=4cm]{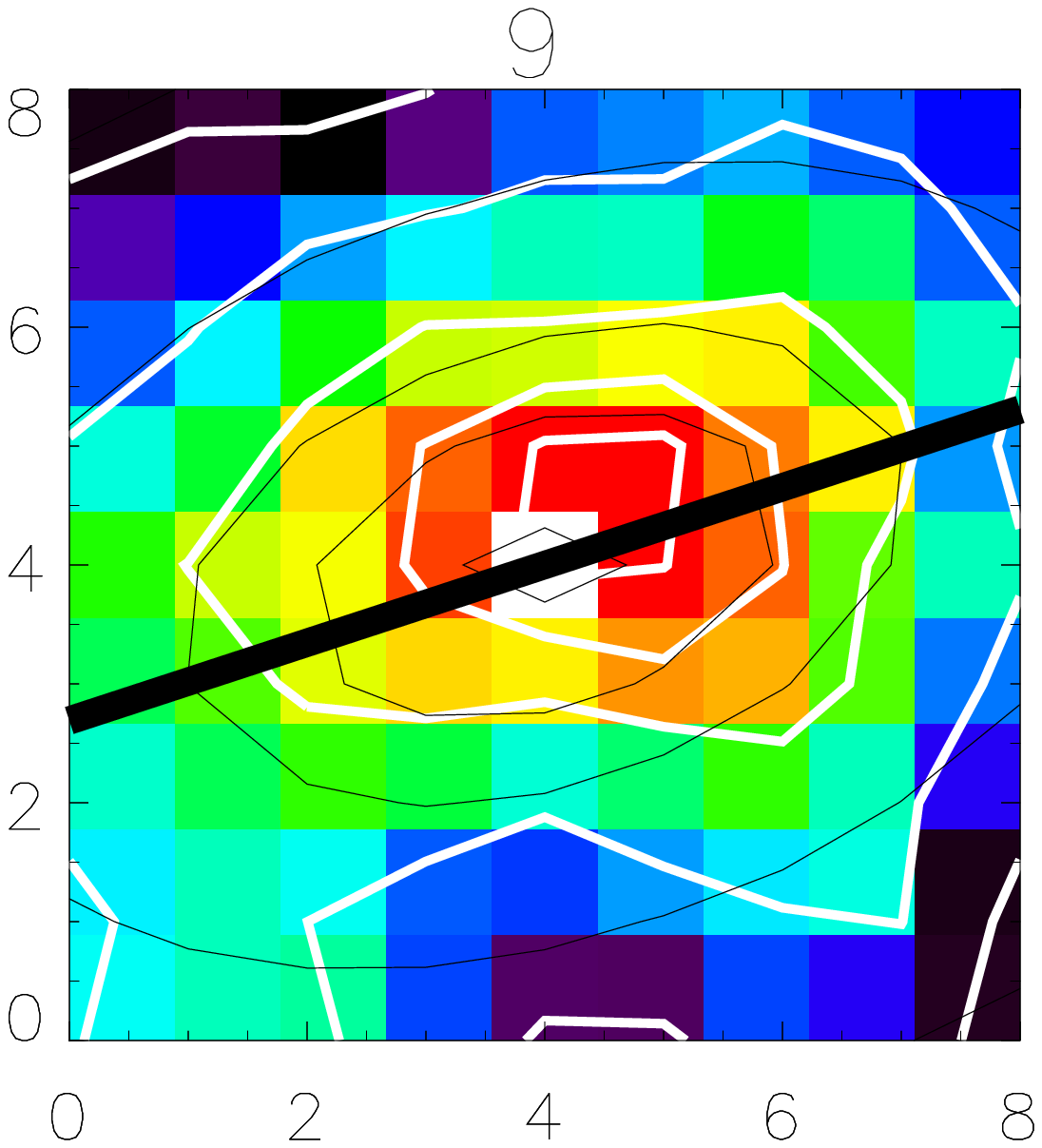}
\includegraphics[width=4cm]{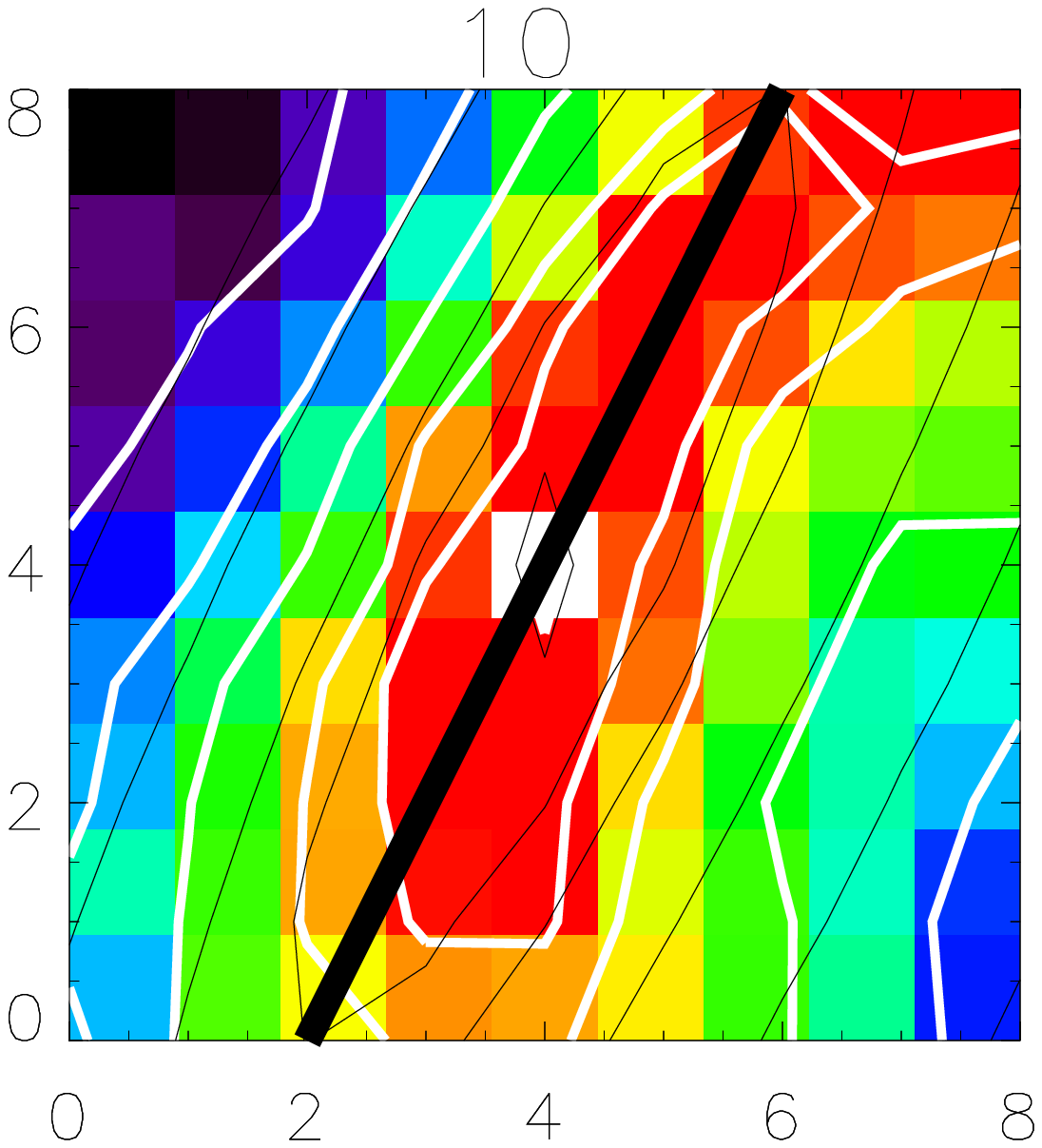}
\includegraphics[width=4cm]{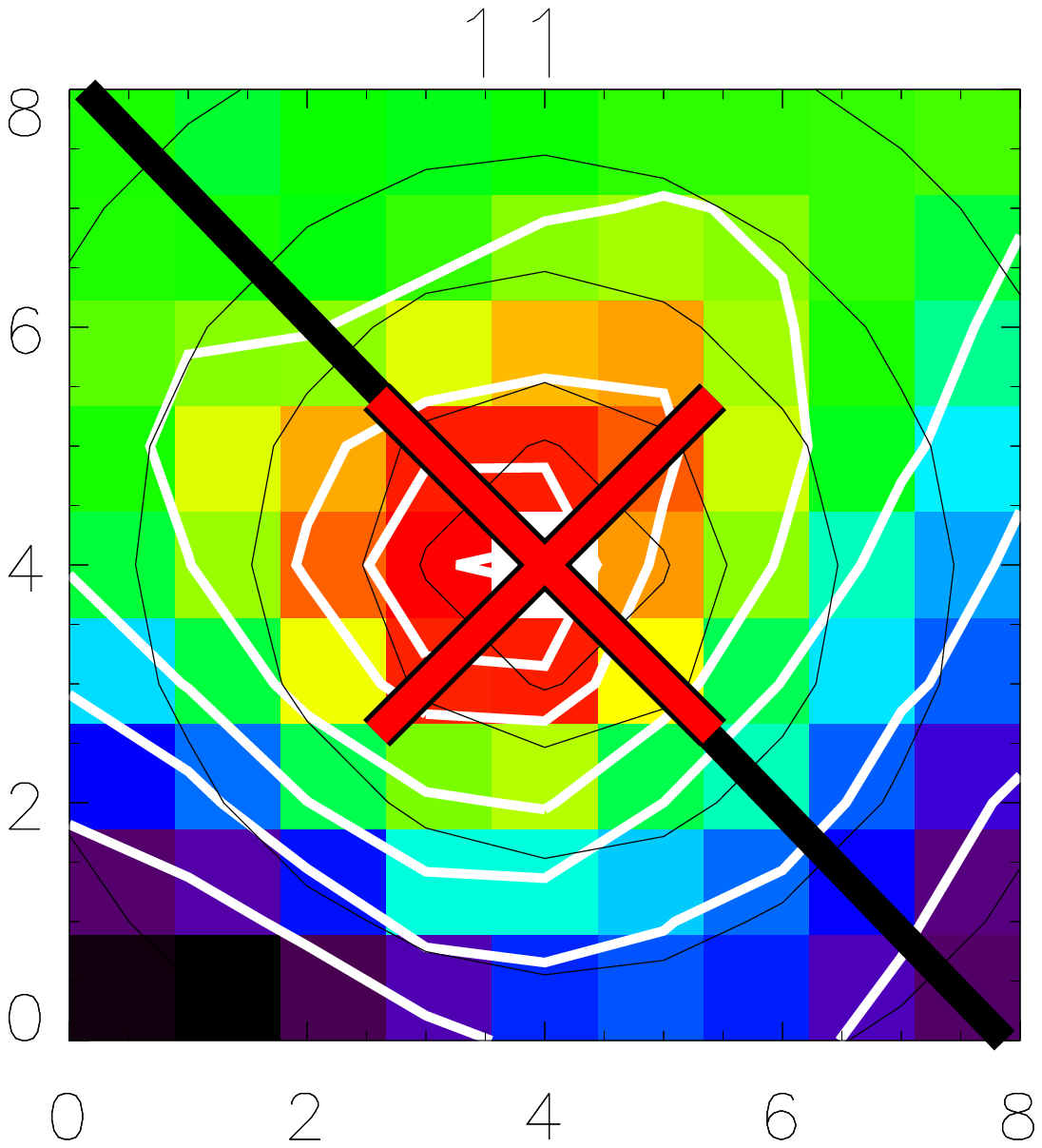}
\includegraphics[width=4cm]{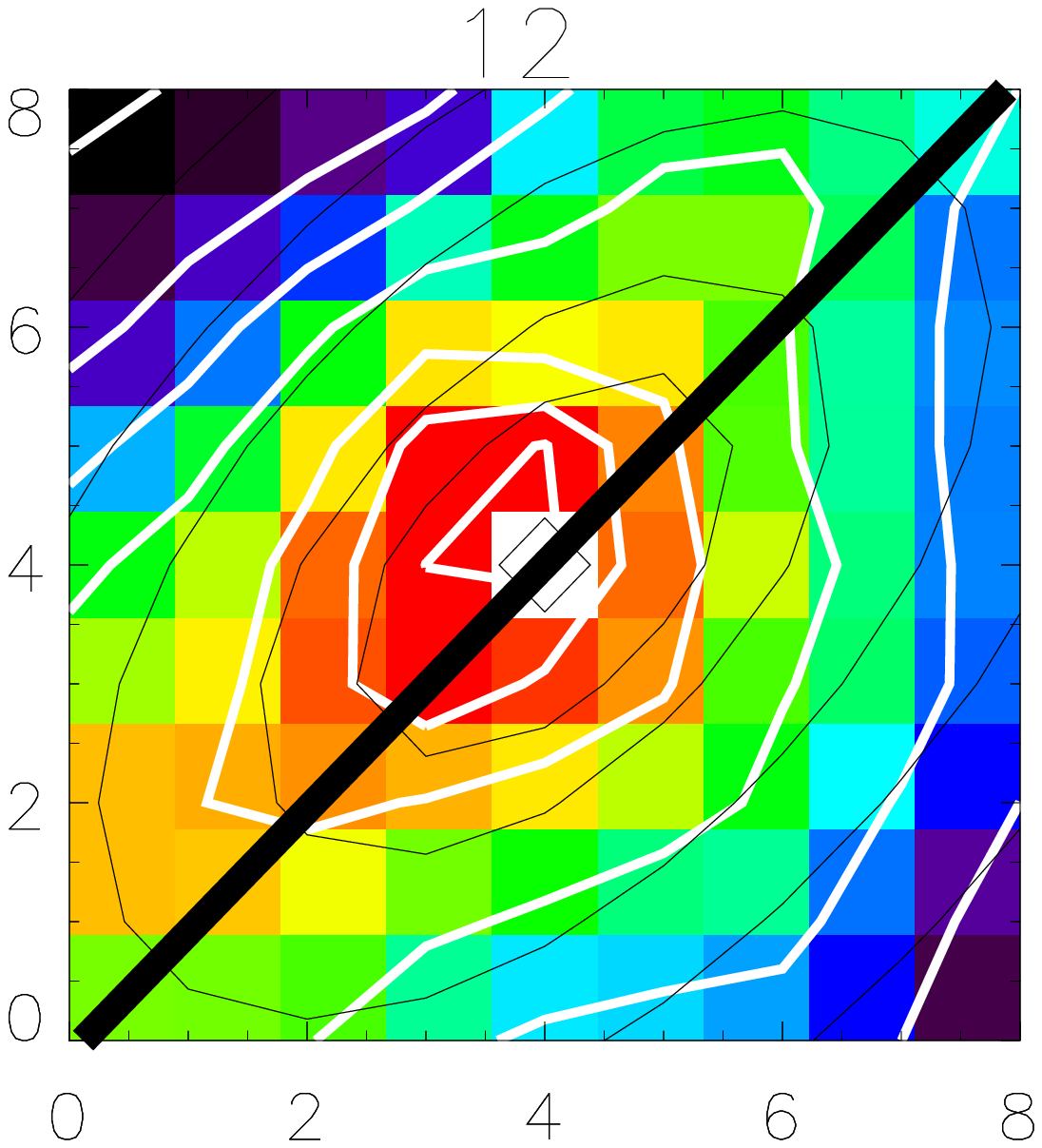}
\includegraphics[width=4cm]{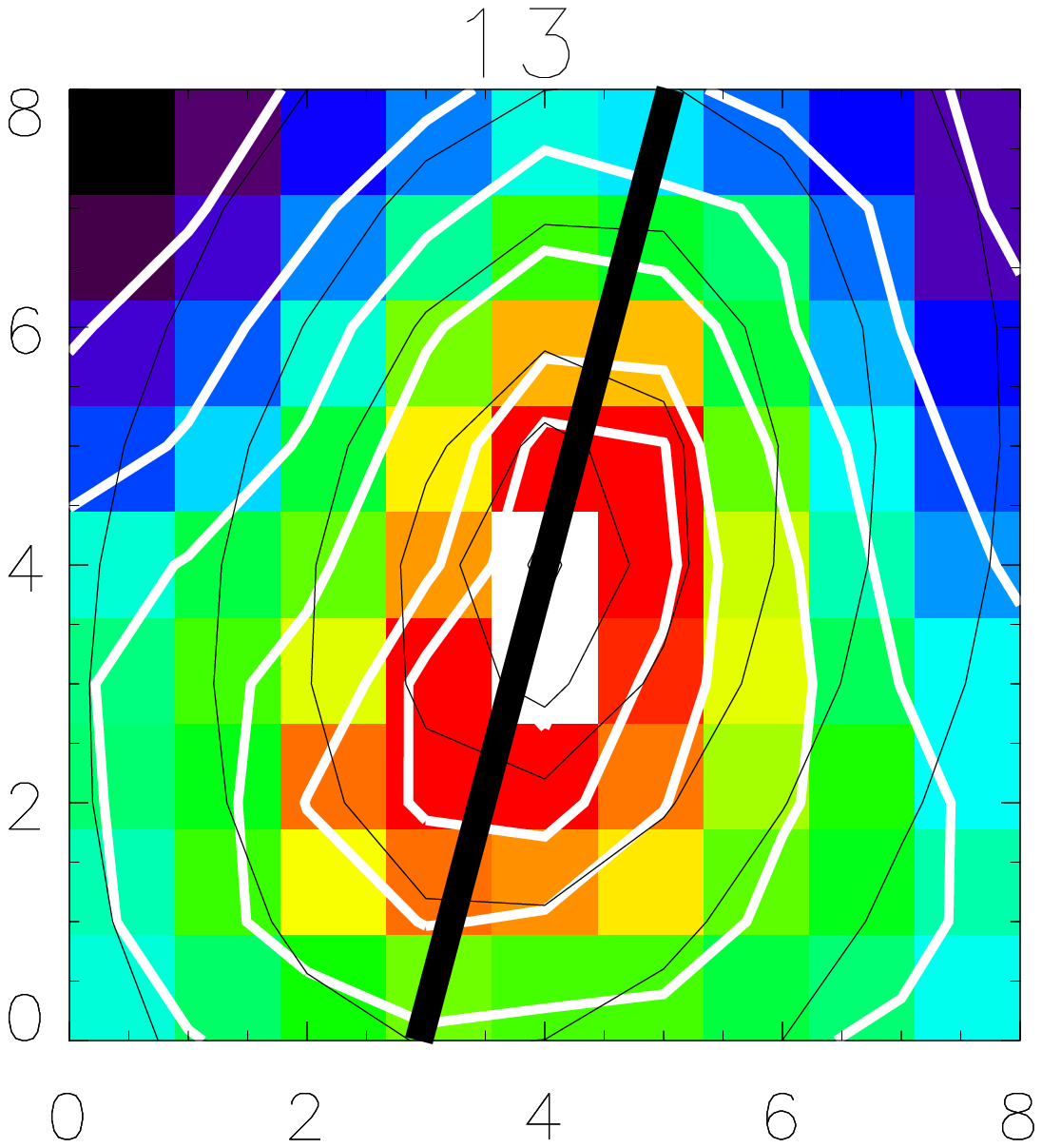}
\includegraphics[width=4cm]{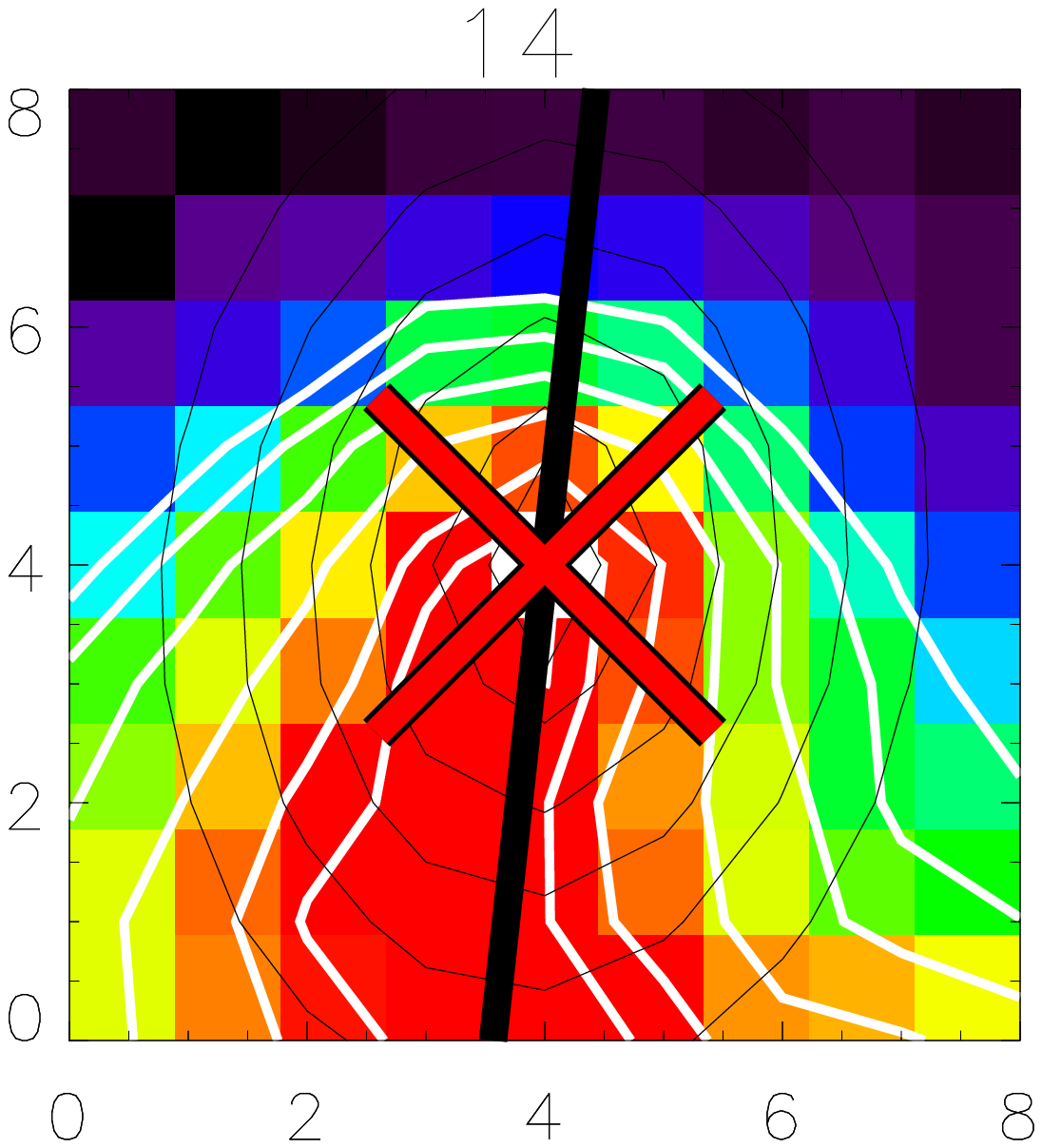}
\includegraphics[width=4cm]{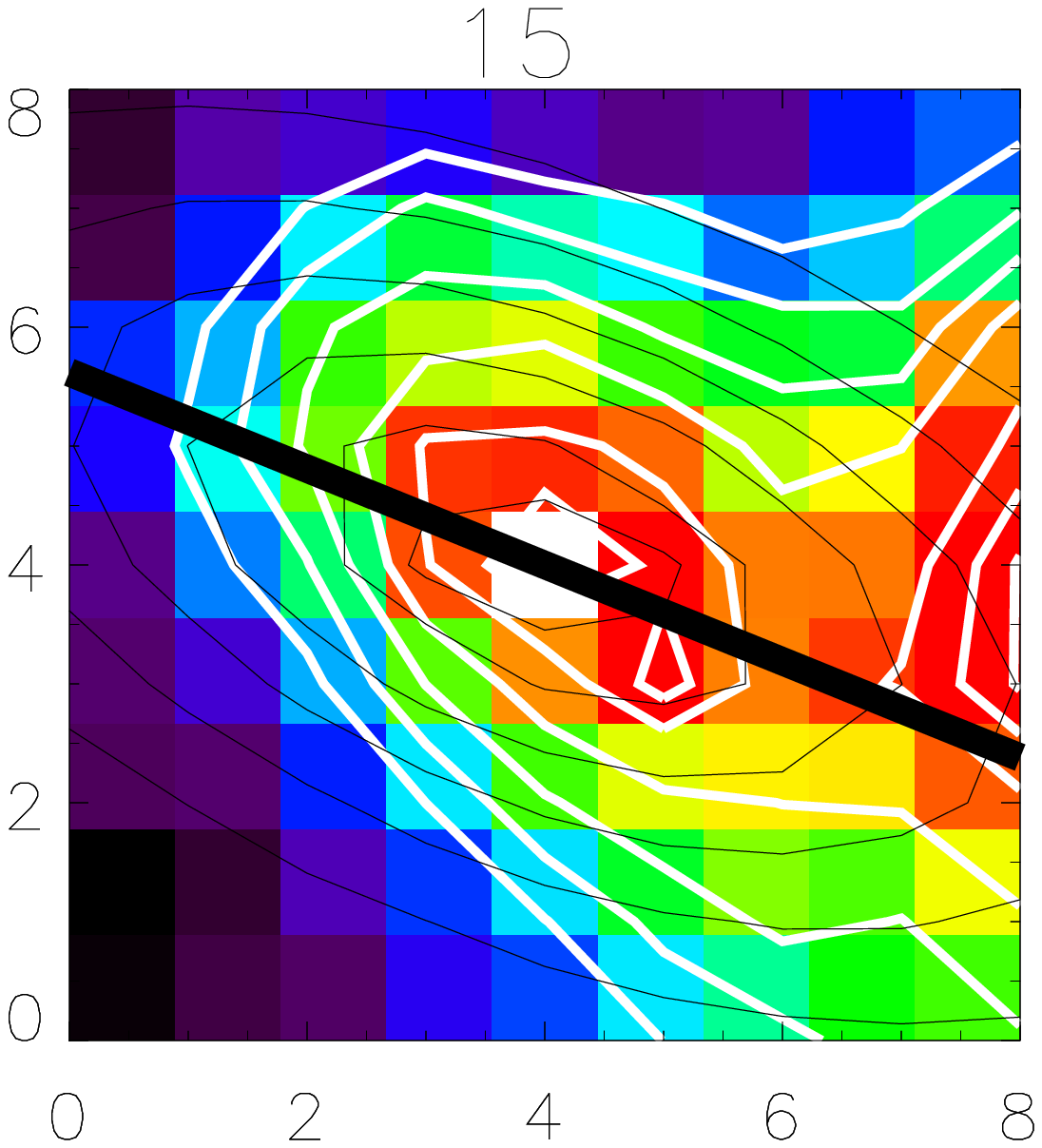}
\includegraphics[width=4cm]{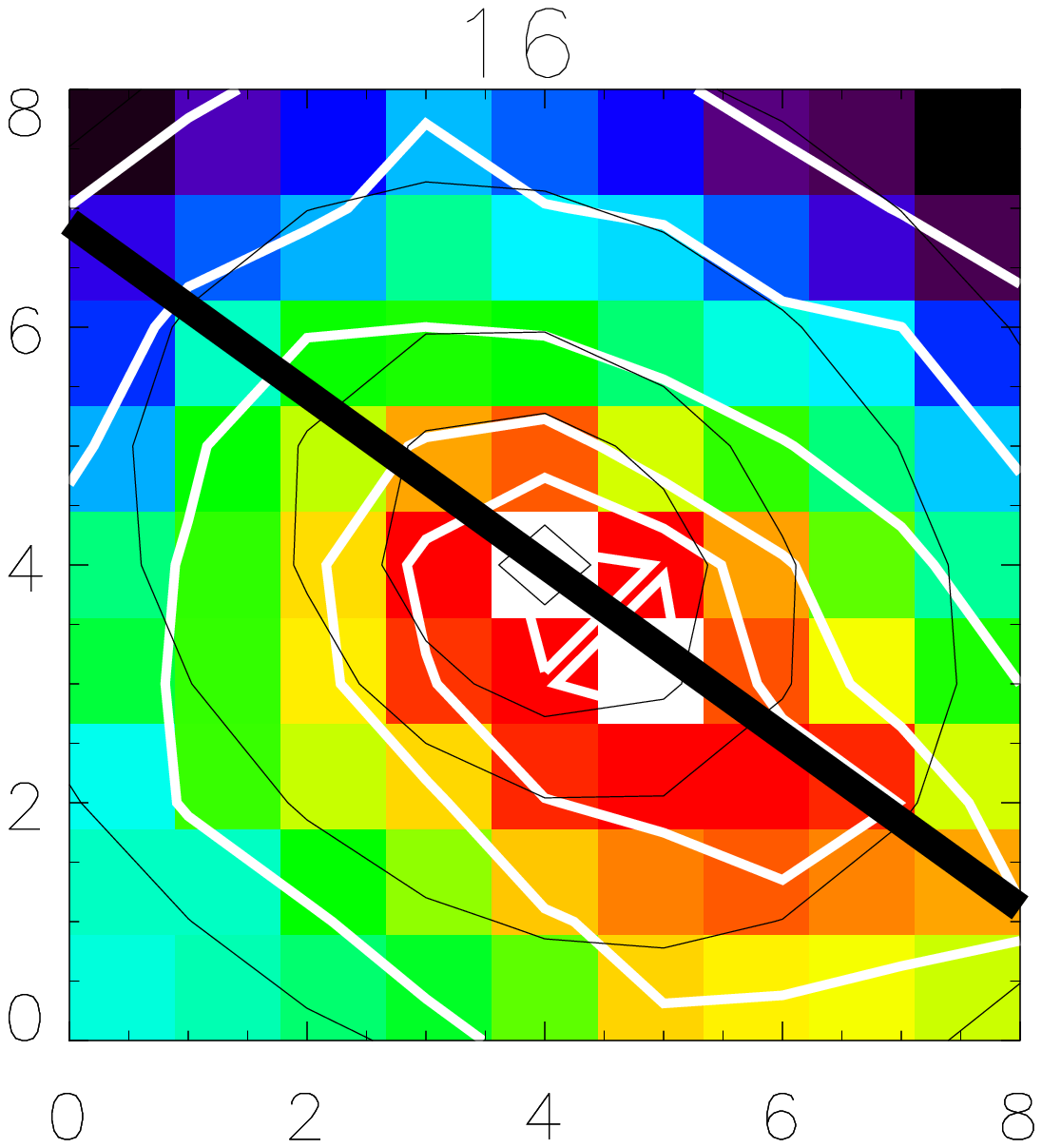}
\includegraphics[width=4cm]{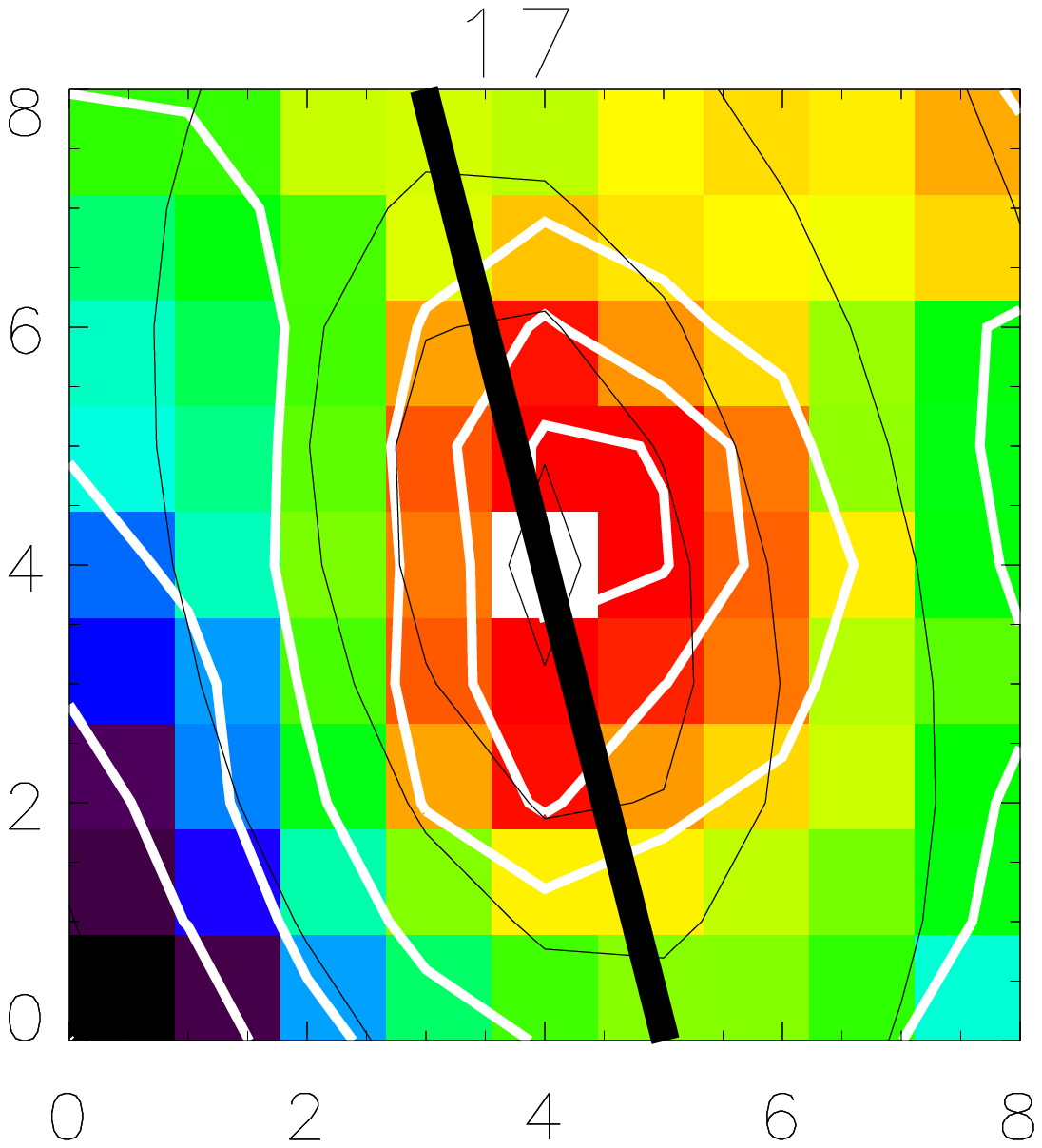}
\includegraphics[width=4cm]{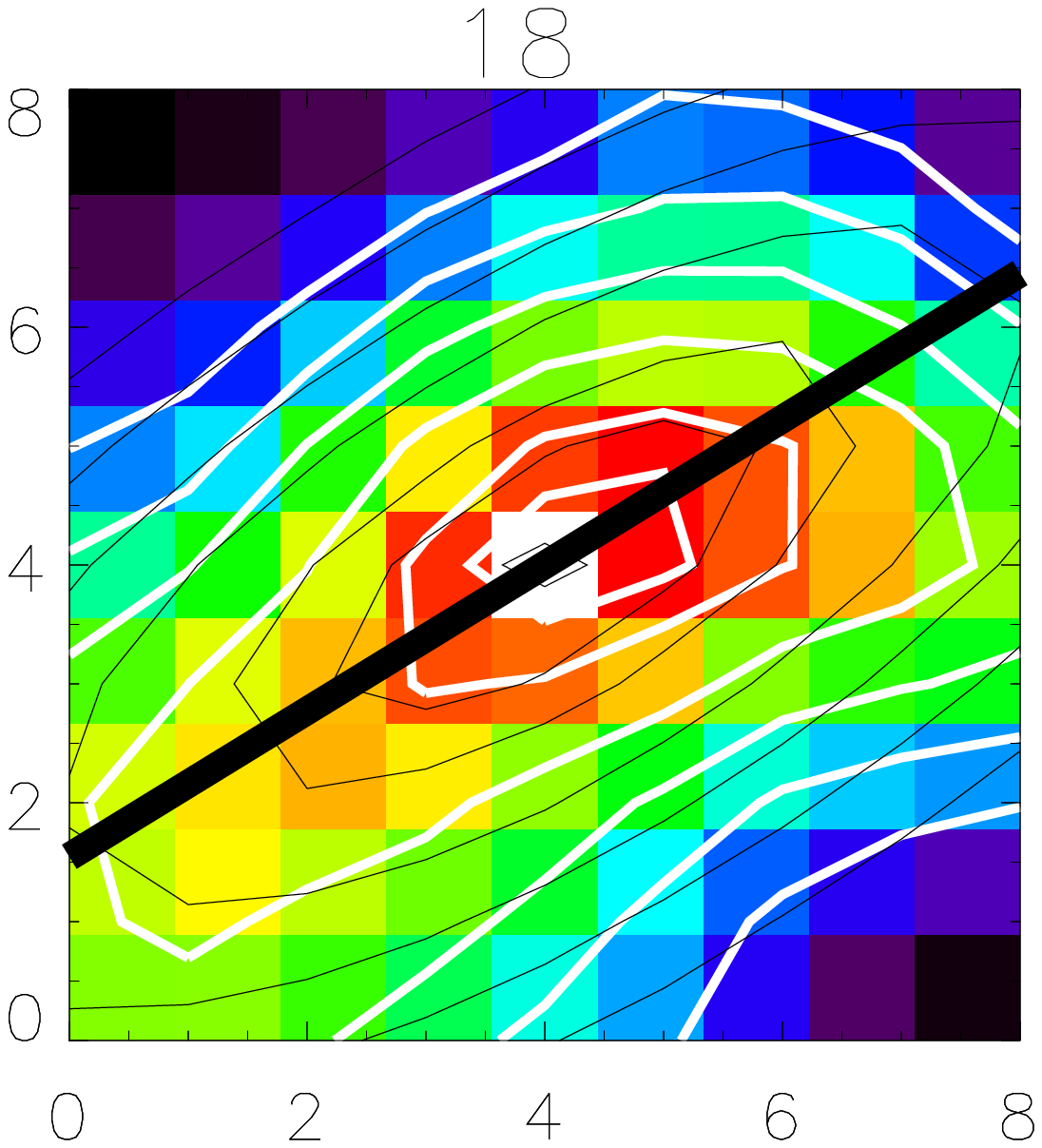}
\includegraphics[width=4cm]{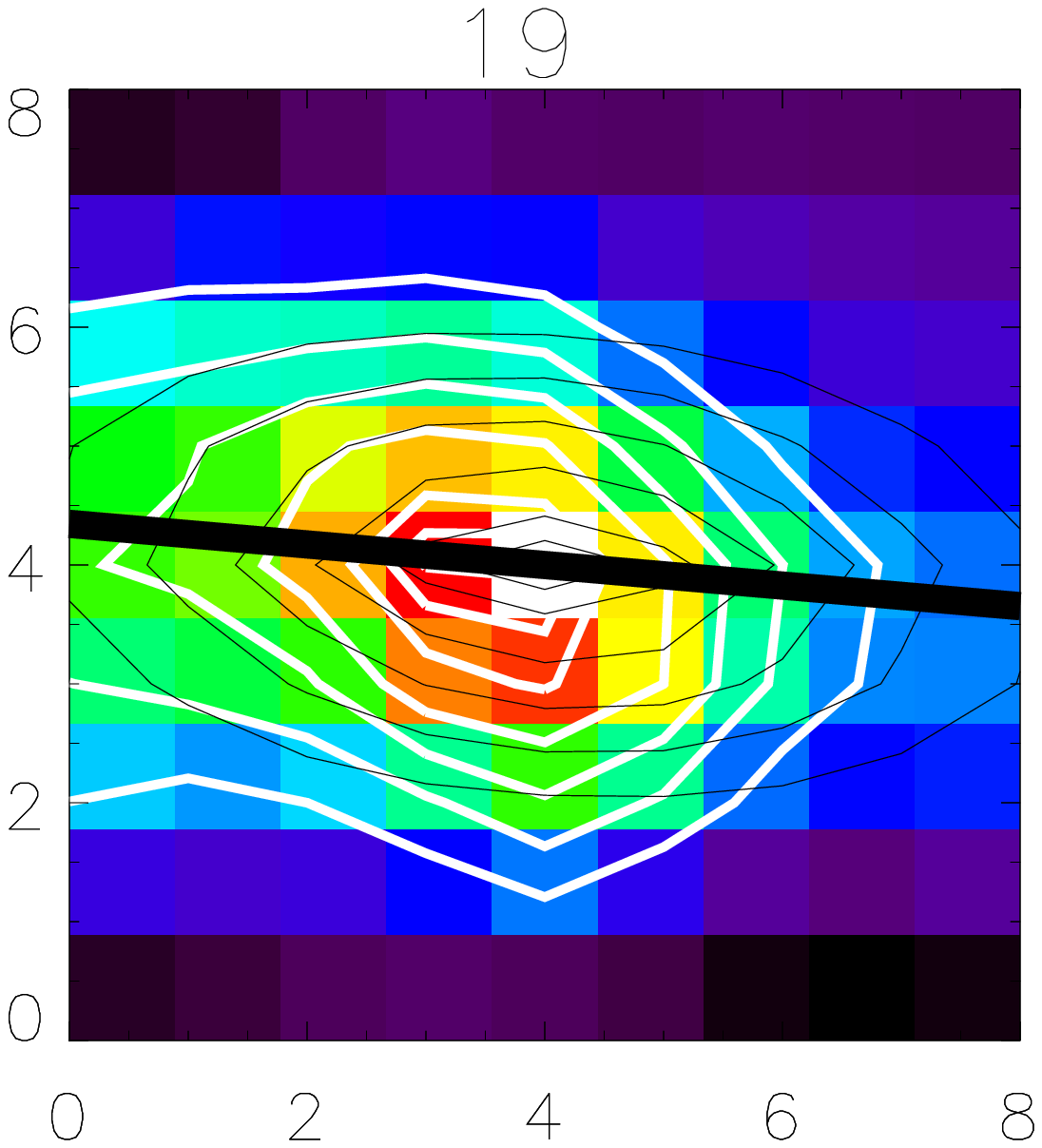}
\includegraphics[width=4cm]{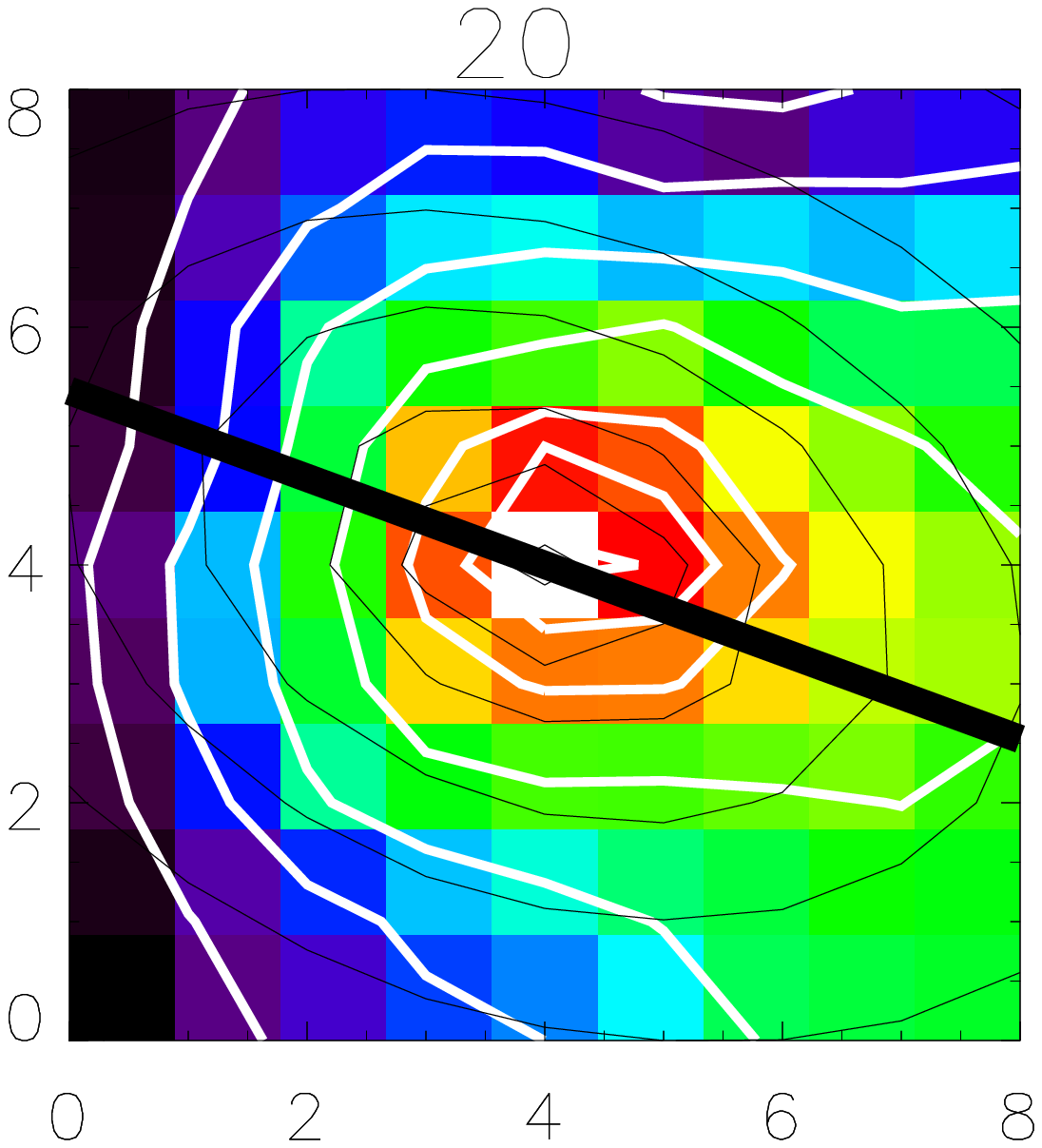}
\includegraphics[width=4cm]{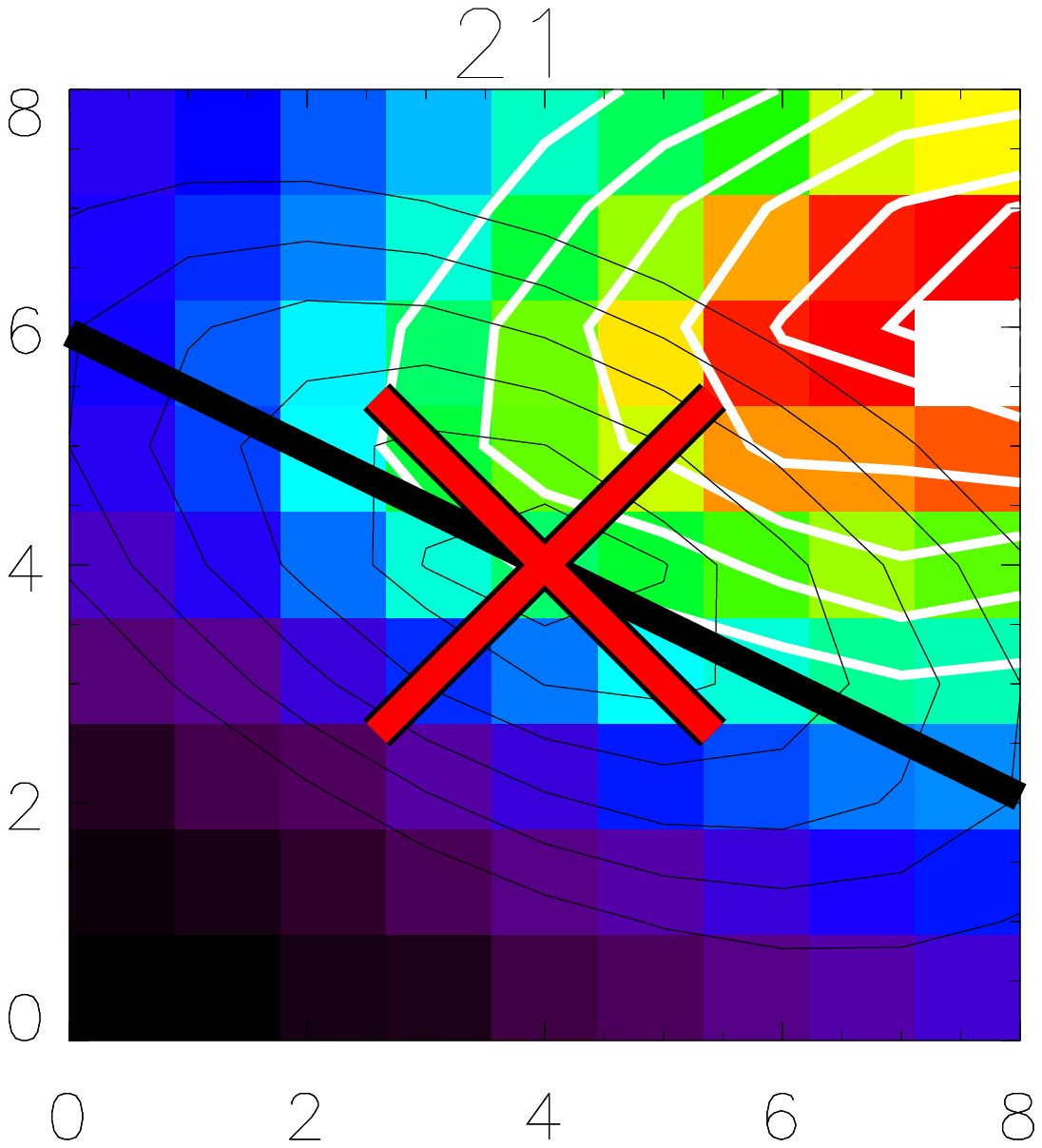}
\includegraphics[width=4cm]{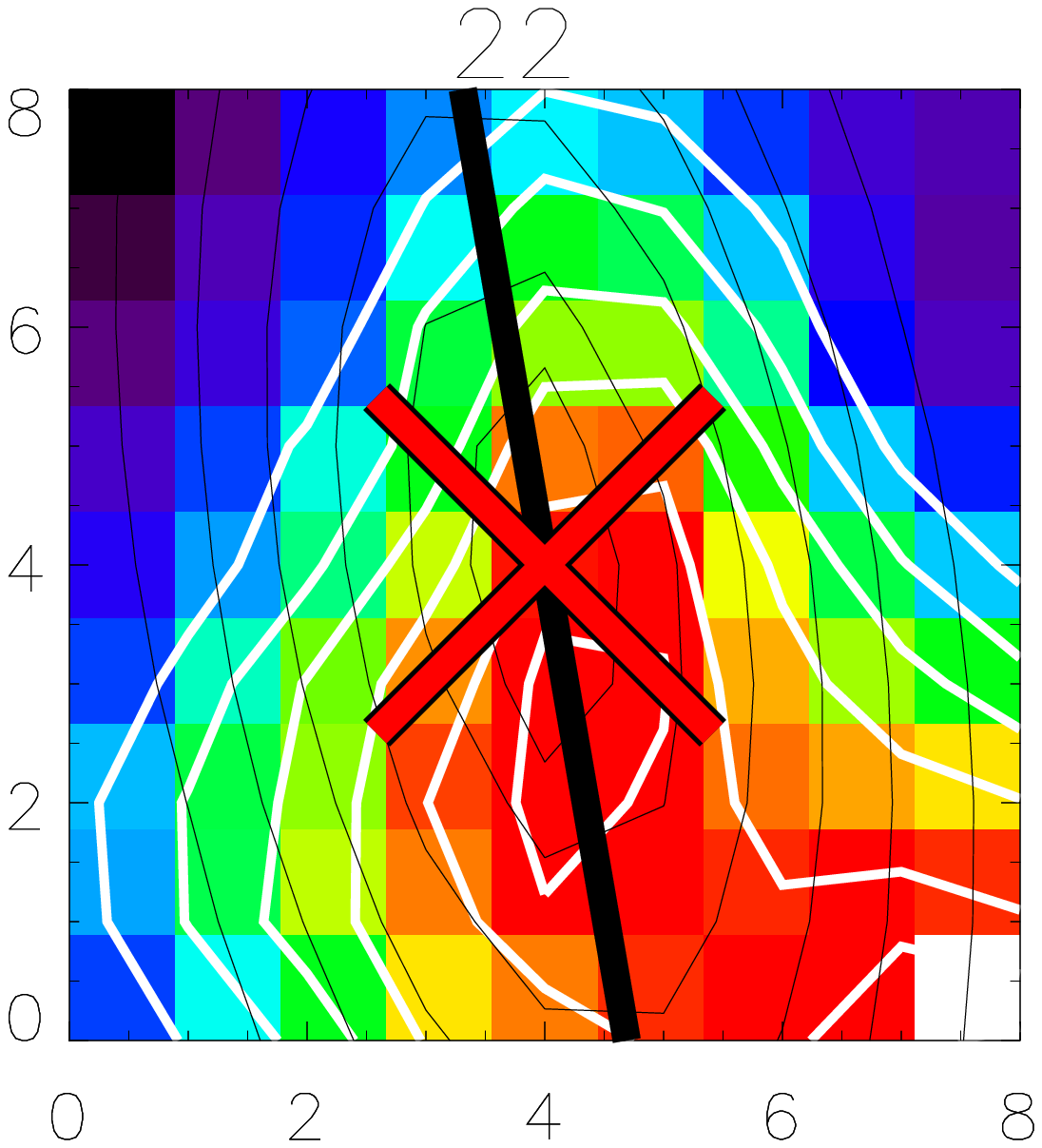}
\includegraphics[width=4cm]{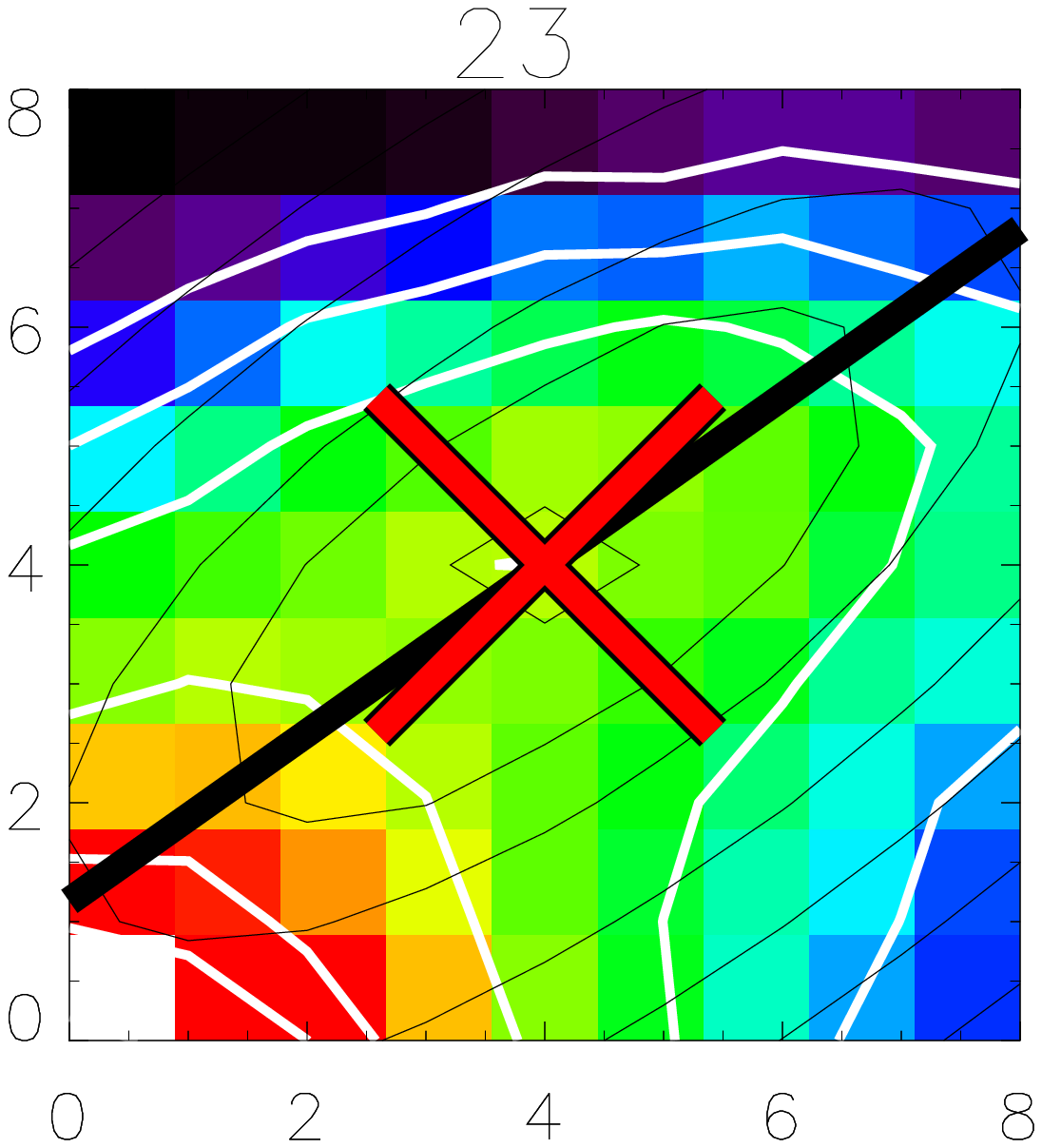}
\includegraphics[width=4cm]{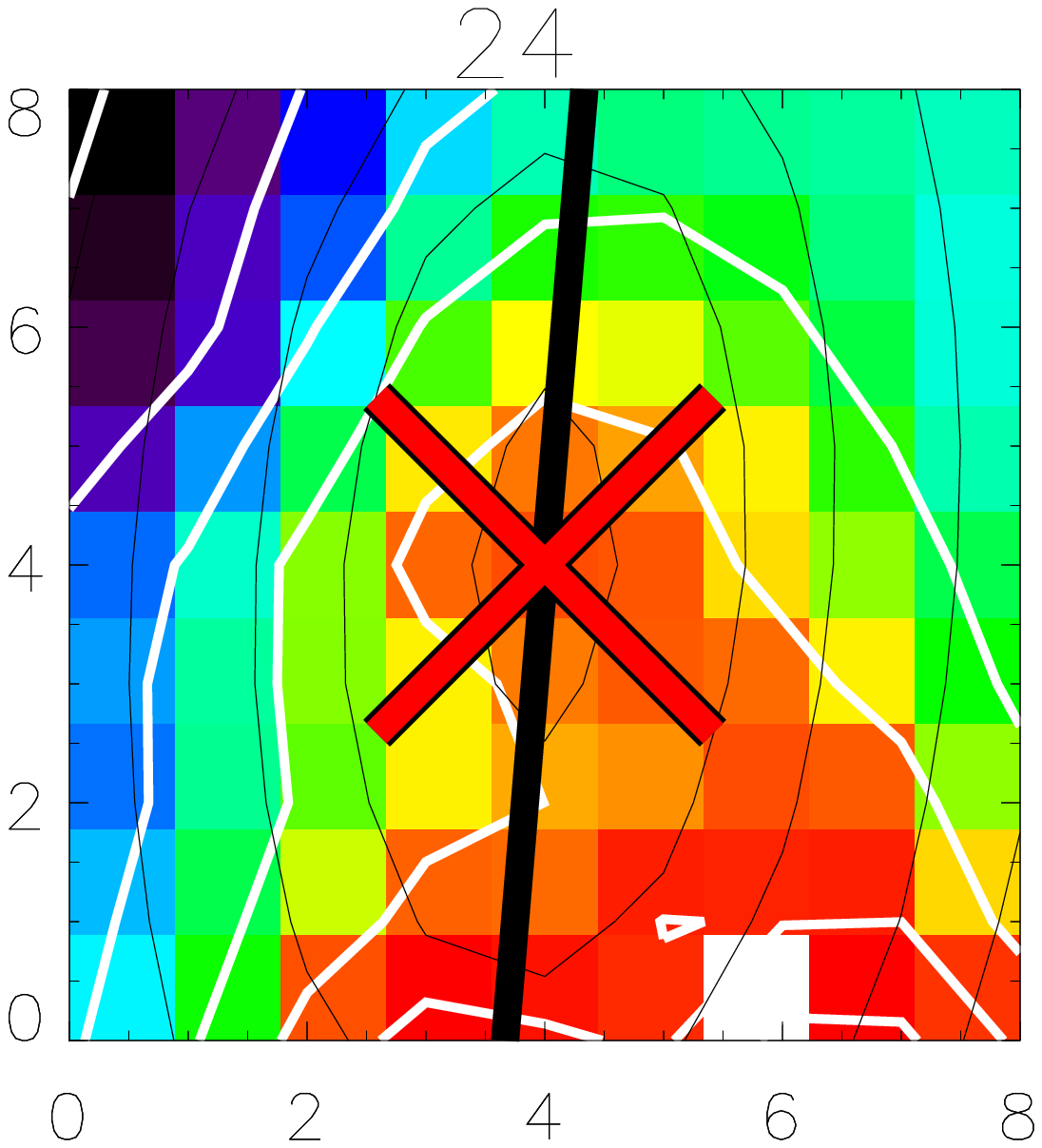}
\includegraphics[width=4cm]{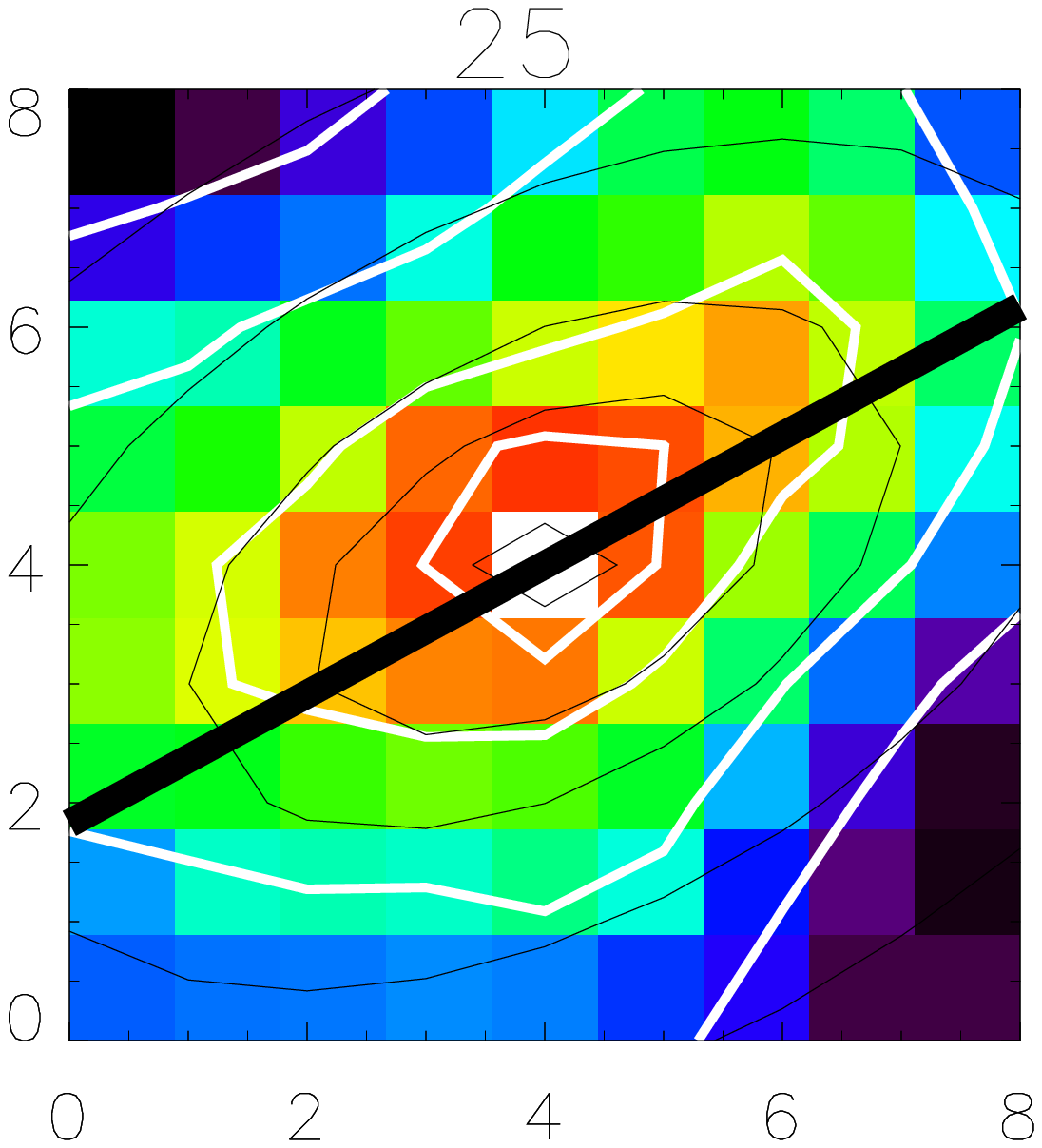}
\includegraphics[width=4cm]{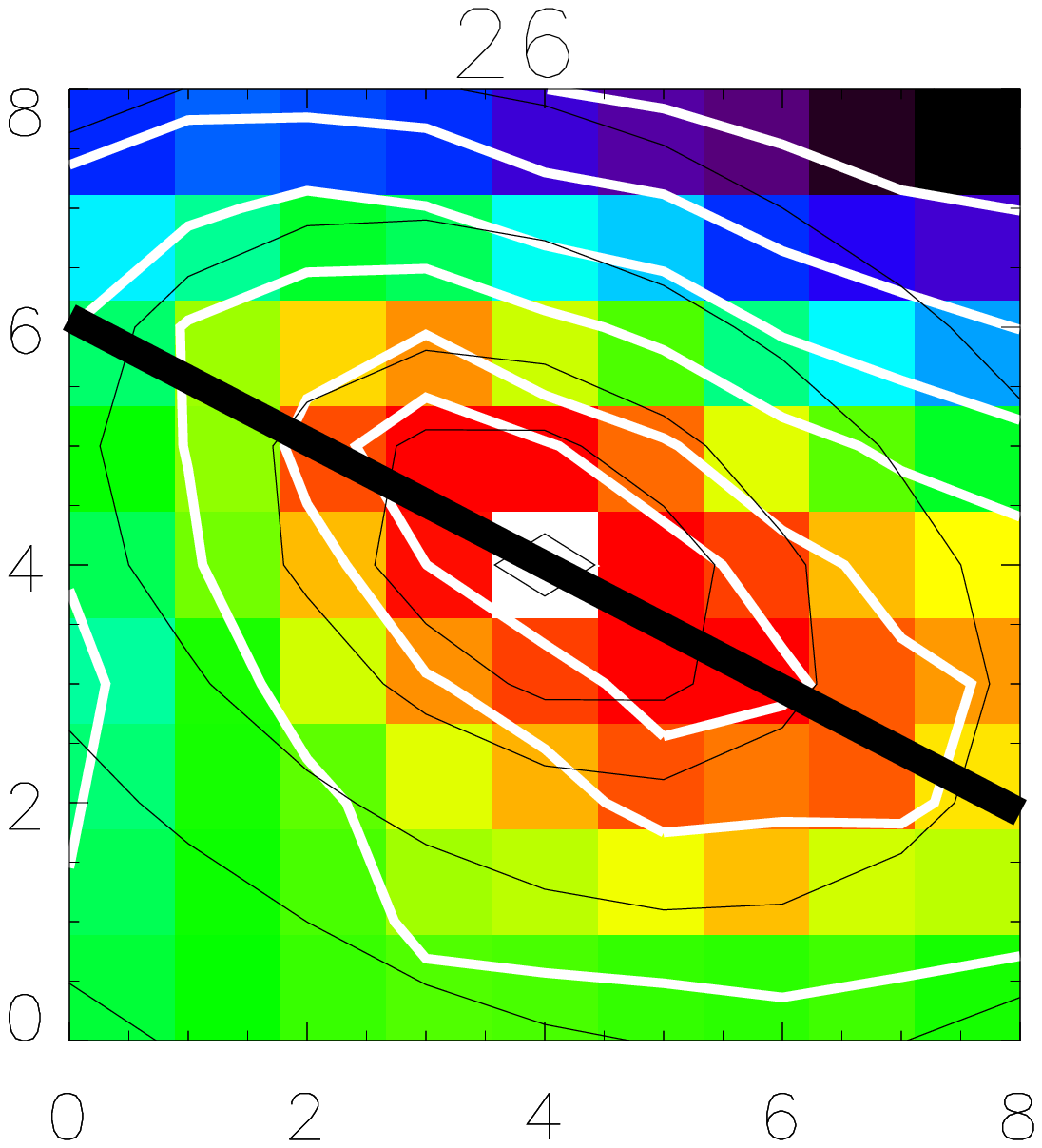}
\includegraphics[width=4cm]{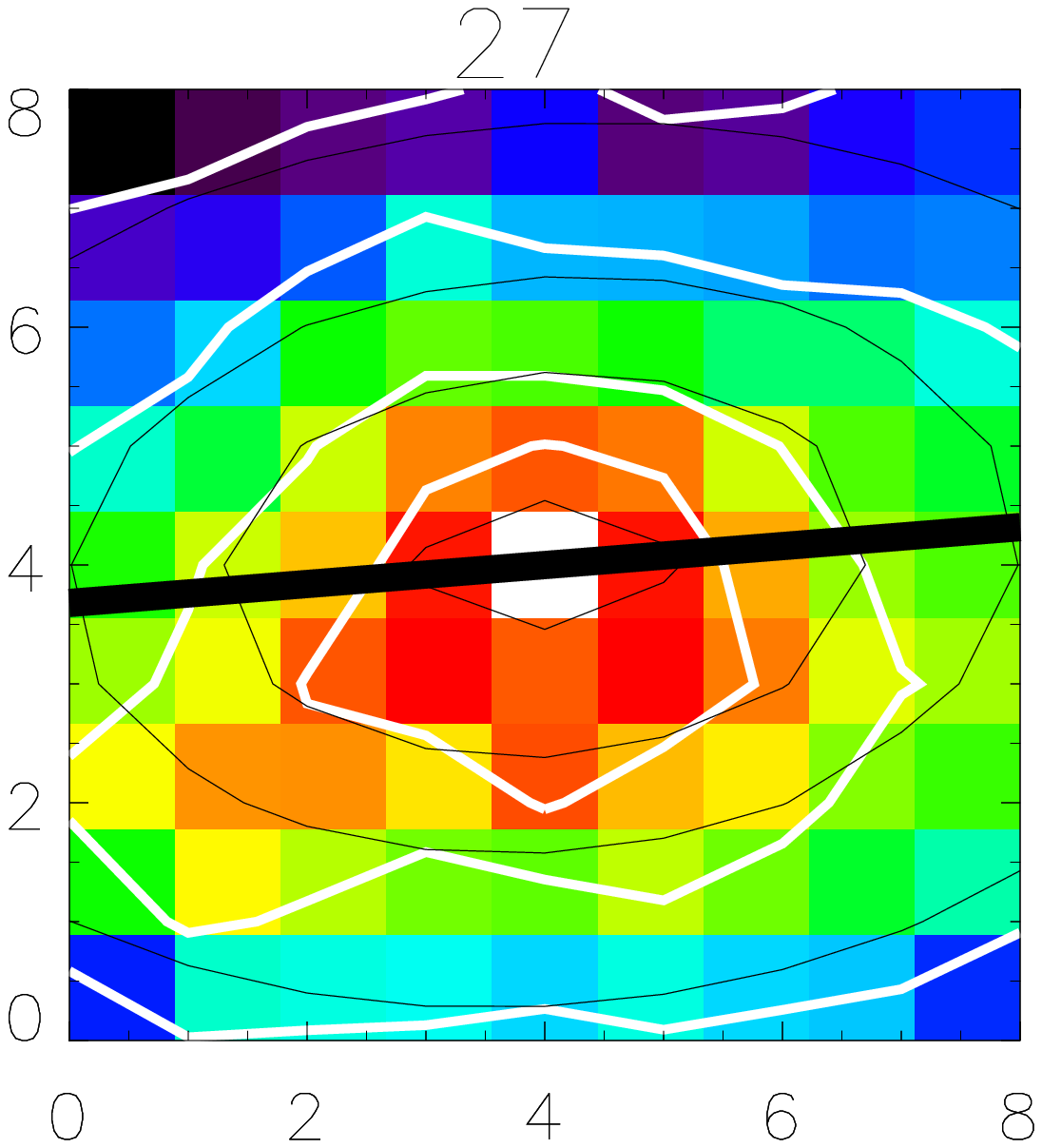}
\caption{Snapshot of the intensity structures centered on position of
  the cores listed in Tables \ref{tabcore1} and \ref{tabcore2}. The
  black line segment
  shows the elongation of the intensity structure obtained with the
  Lorentzian fitting method. White lines show intensity contours 
obtained at levels of 0.5, 0.6, 0.7, 0.8, 0.9, 0.95 and 0.99 of the peak.
Dark lines show similar contour lines obtained from the Lorentzian
fit models. The red crosses show the prestellar cores rejected from the
later analysis, as discussed in Section \ref{histoc}.
}
\label{stamps}
\end{figure*}

\begin{figure*}
\centering
\includegraphics[width=4.4cm]{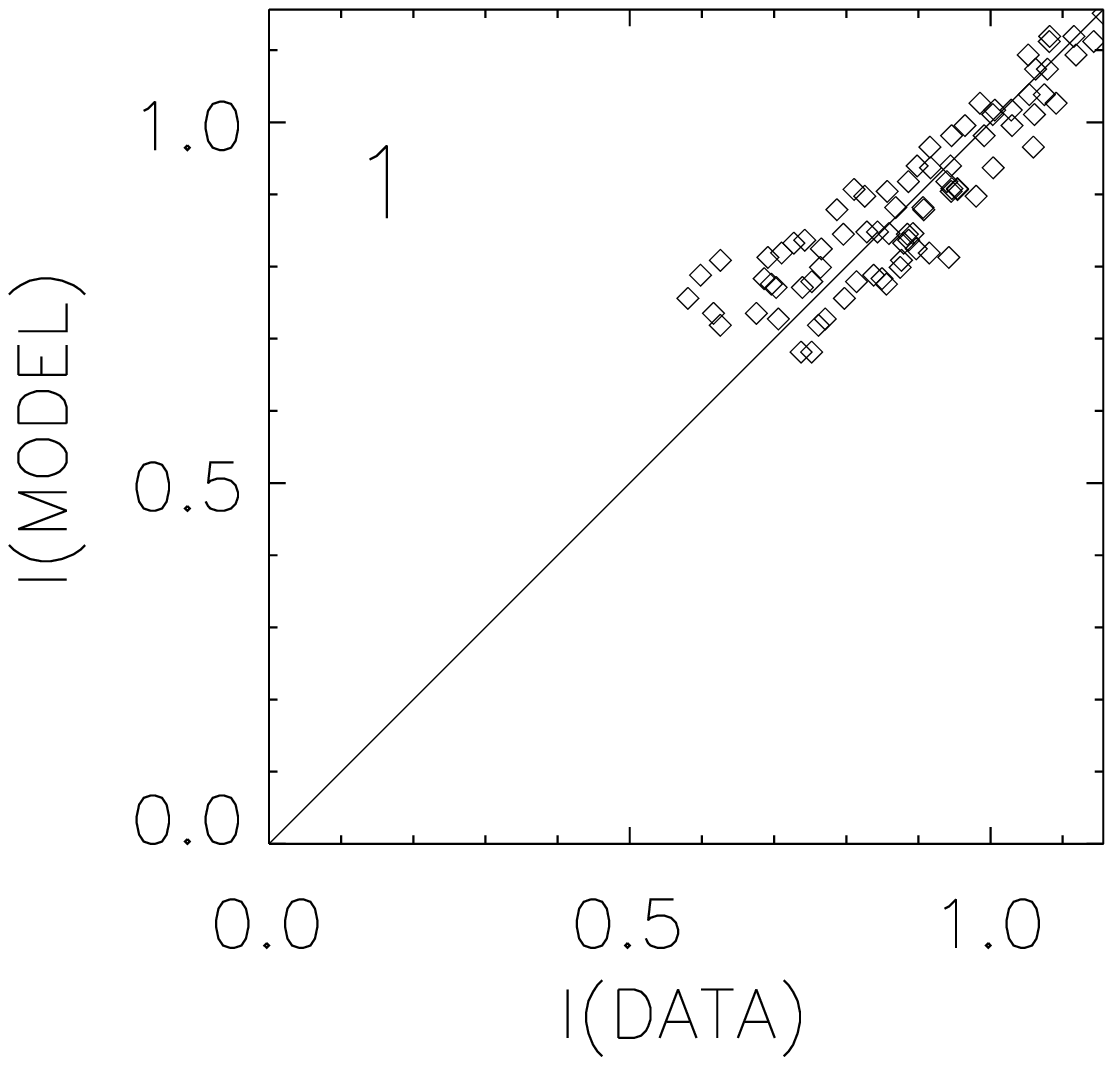}
\includegraphics[width=4.4cm]{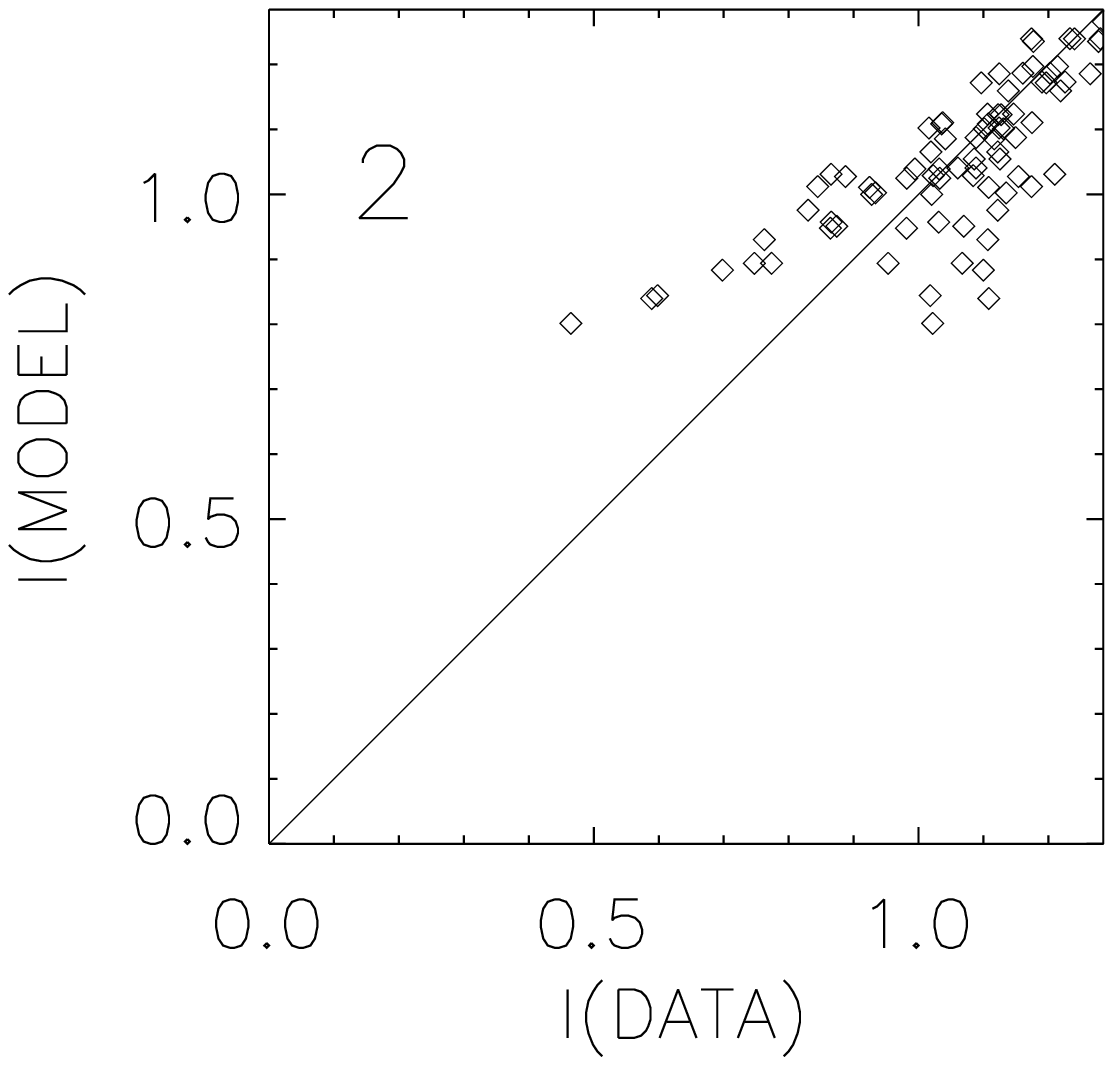}
\includegraphics[width=4.4cm]{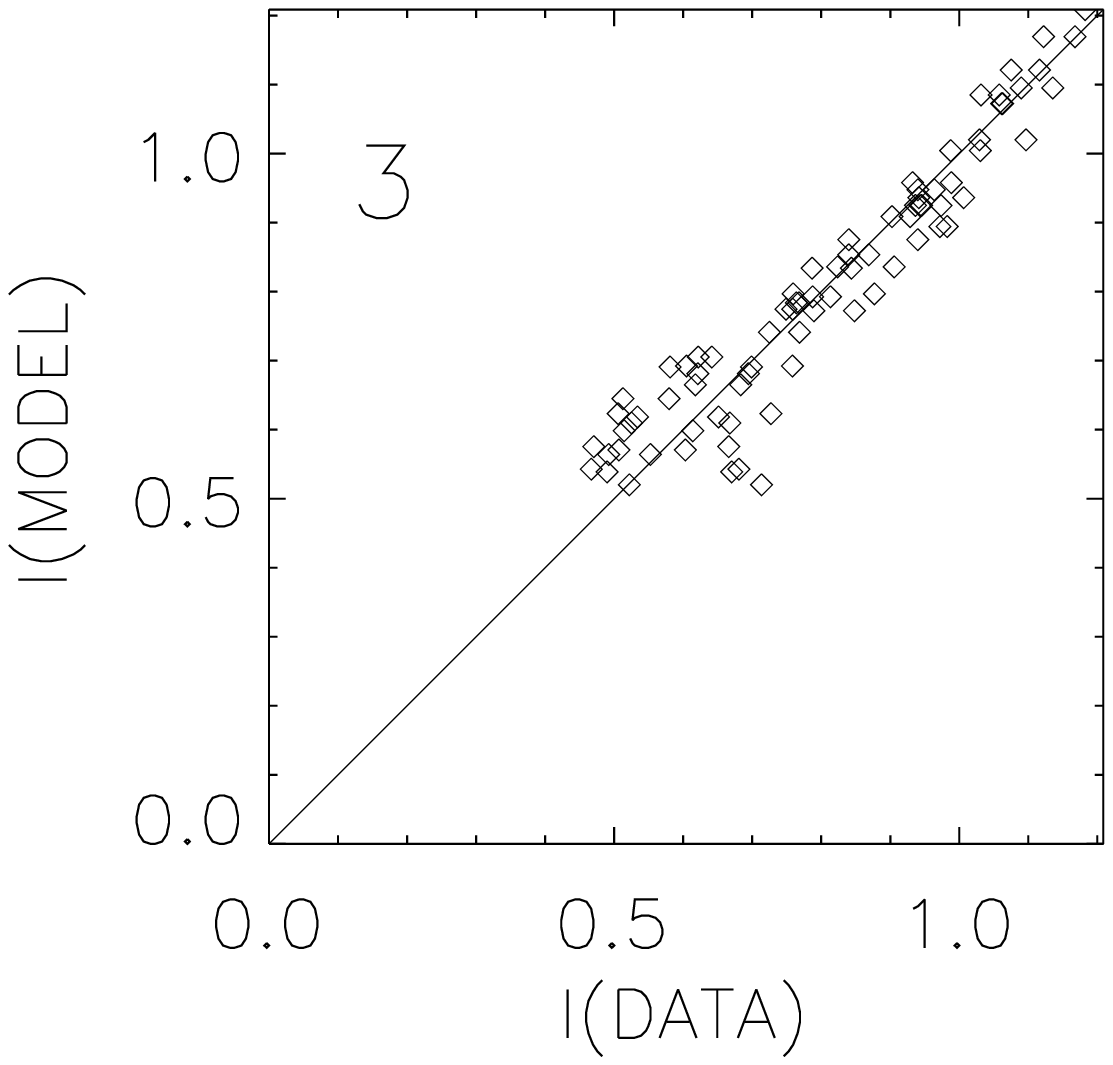}
\includegraphics[width=4.4cm]{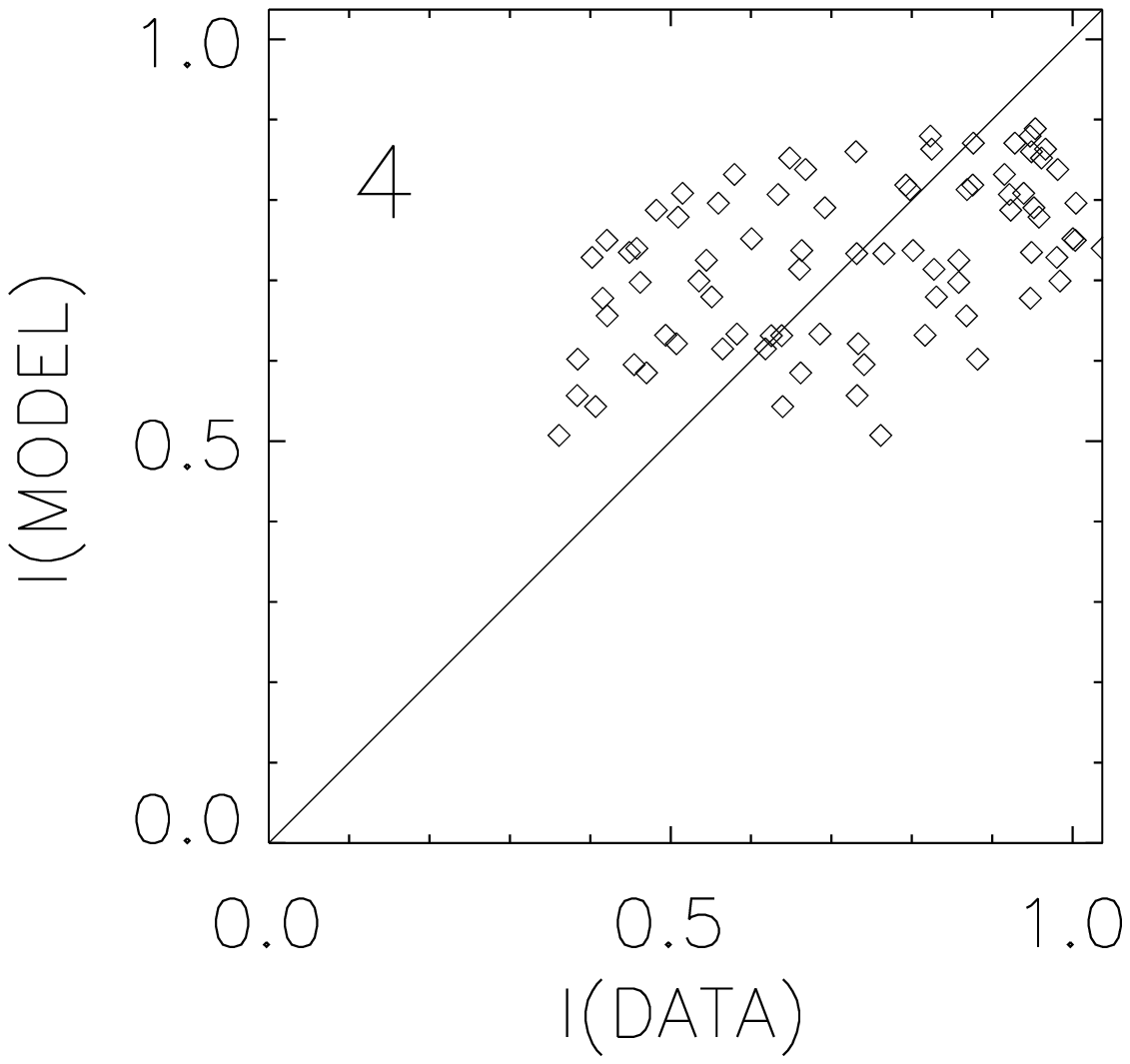}
\includegraphics[width=4.4cm]{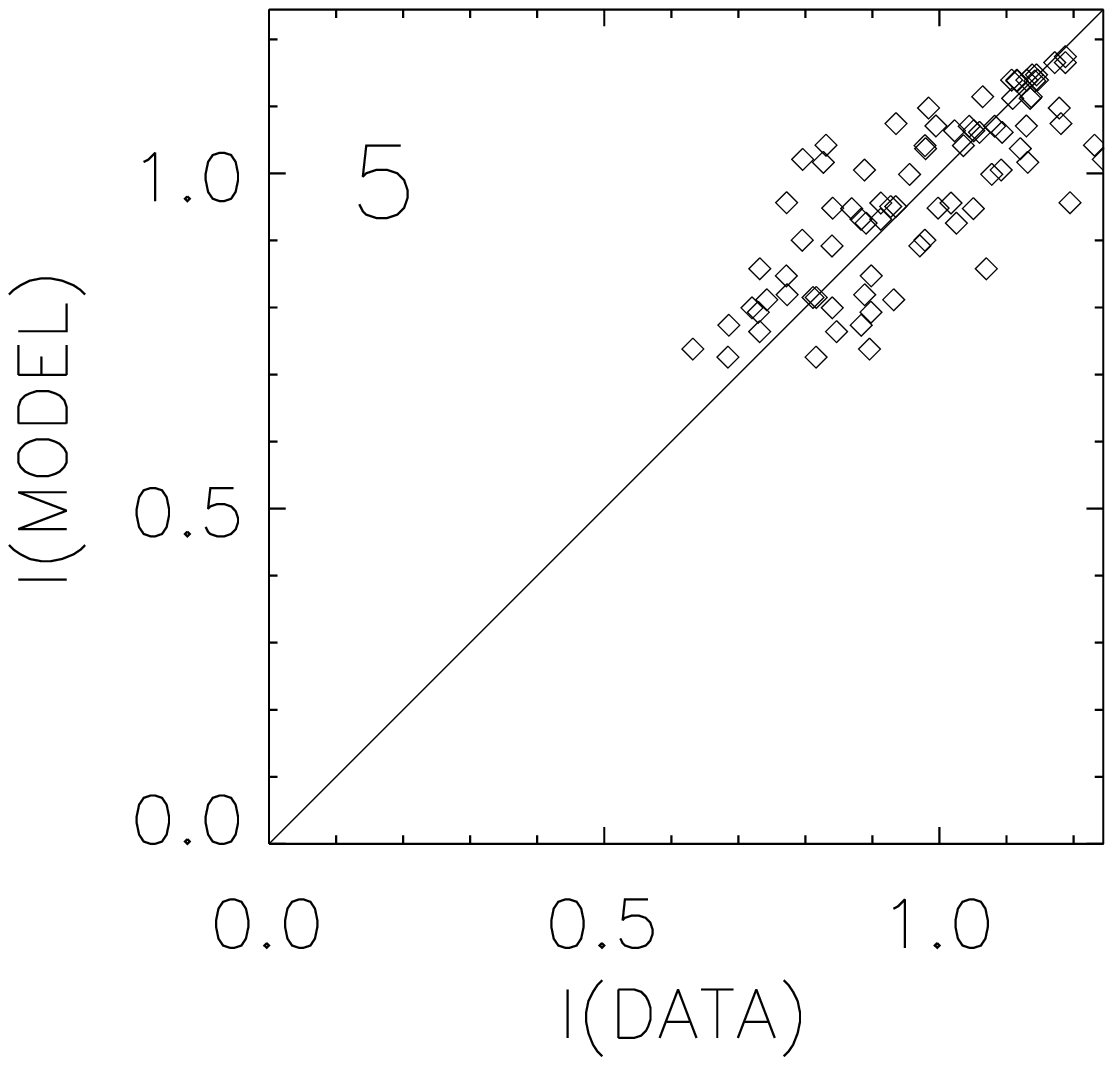}
\includegraphics[width=4.4cm]{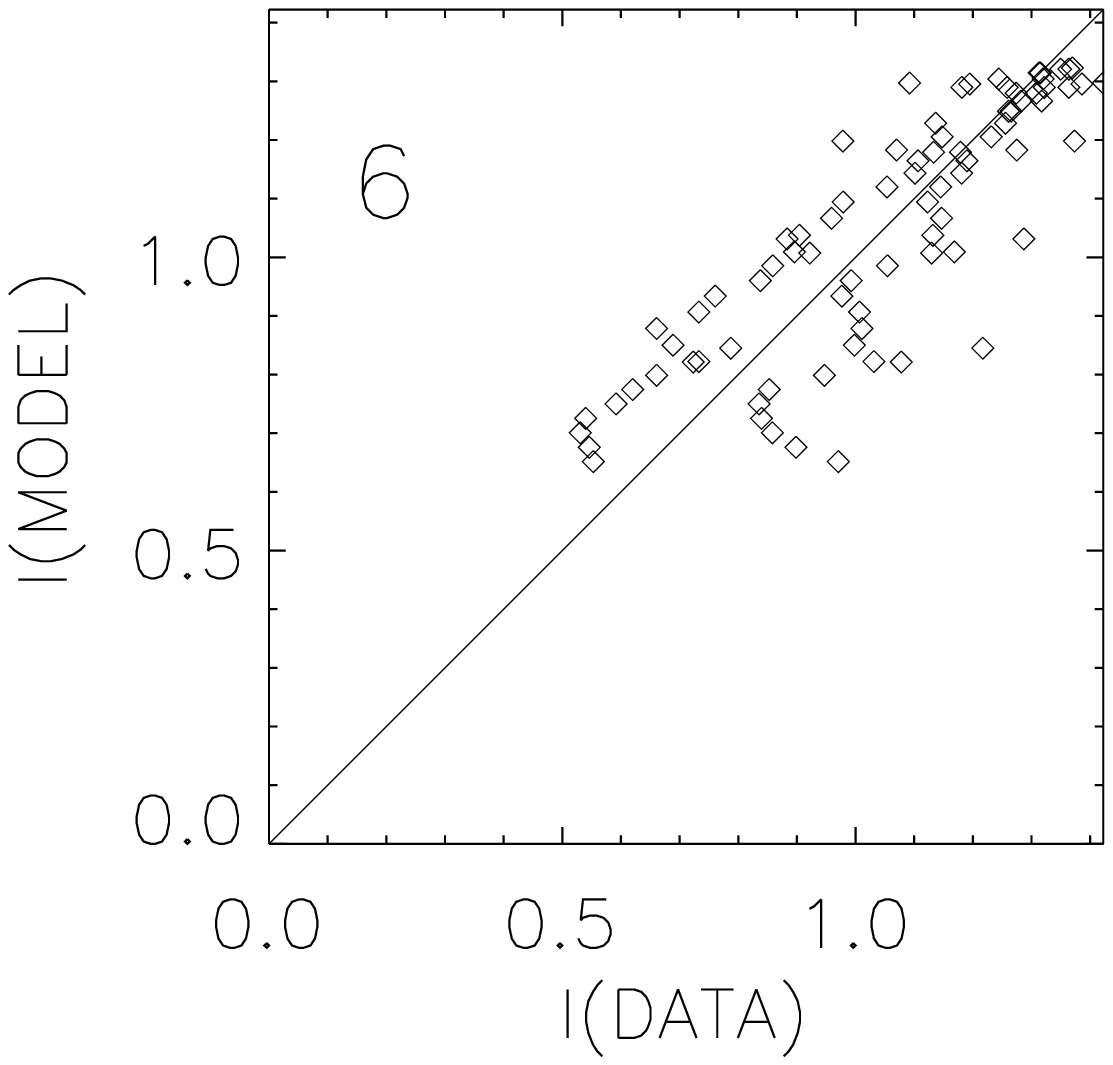}
\includegraphics[width=4.4cm]{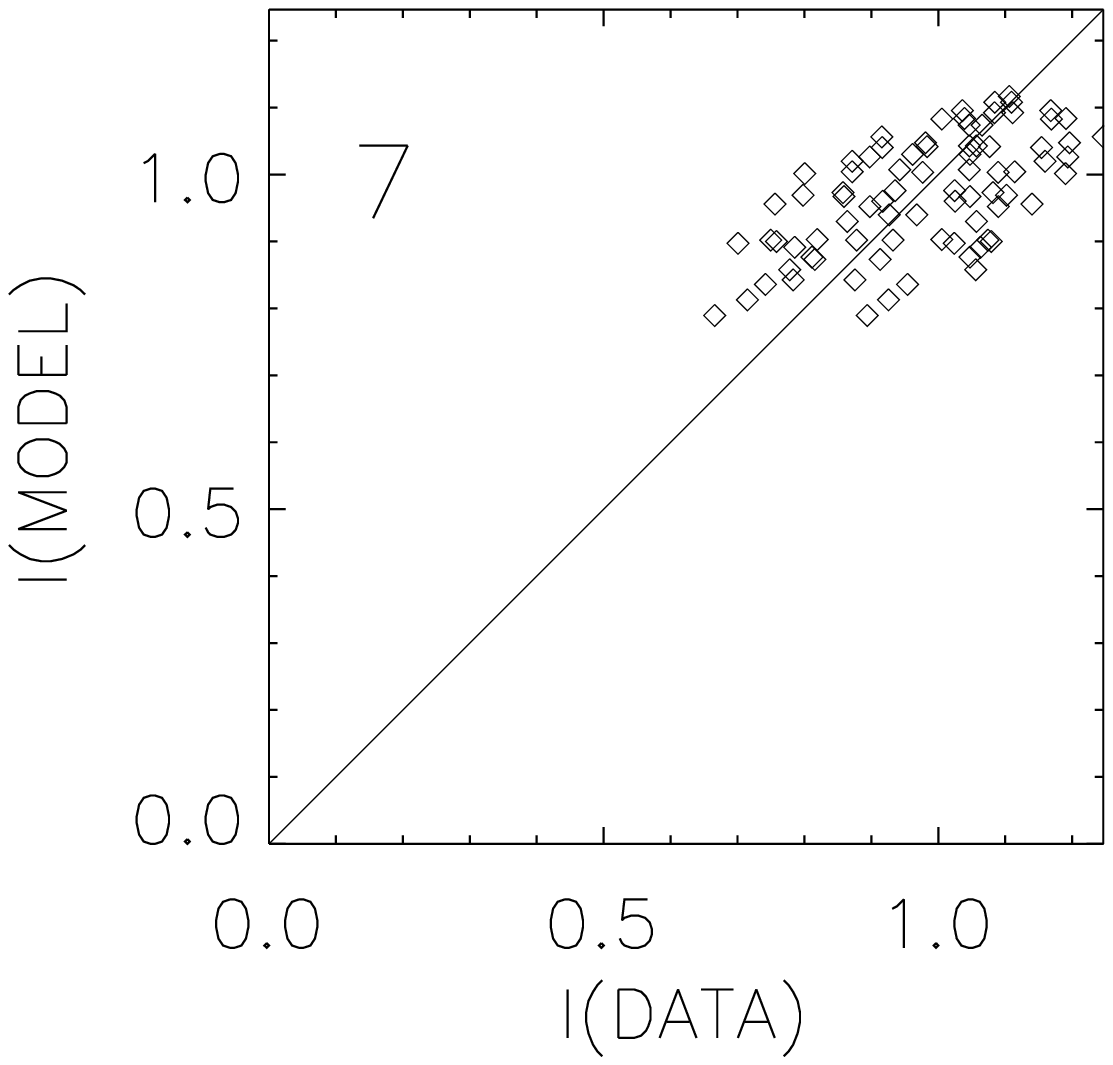}
\includegraphics[width=4.4cm]{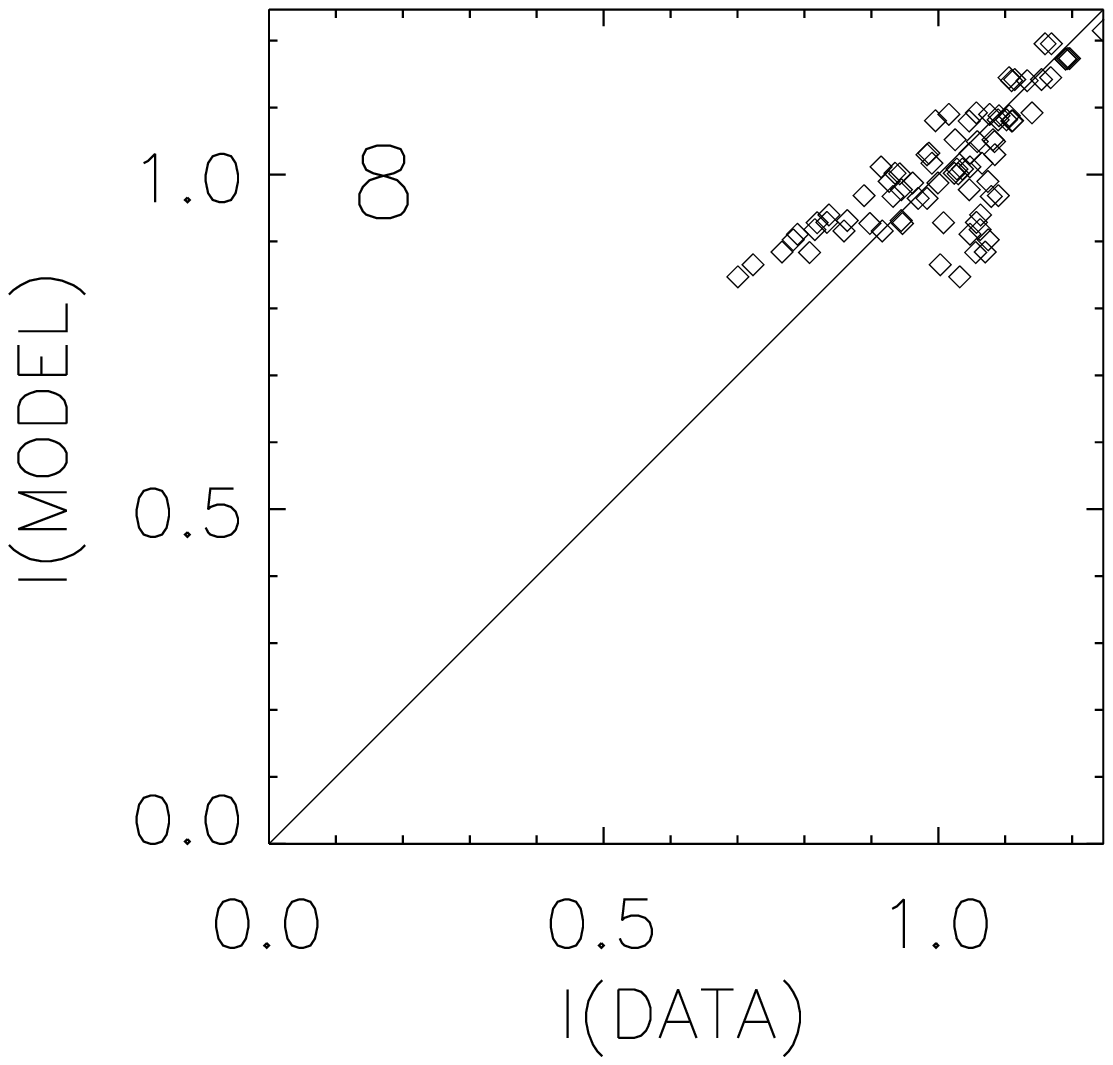}
\includegraphics[width=4.4cm]{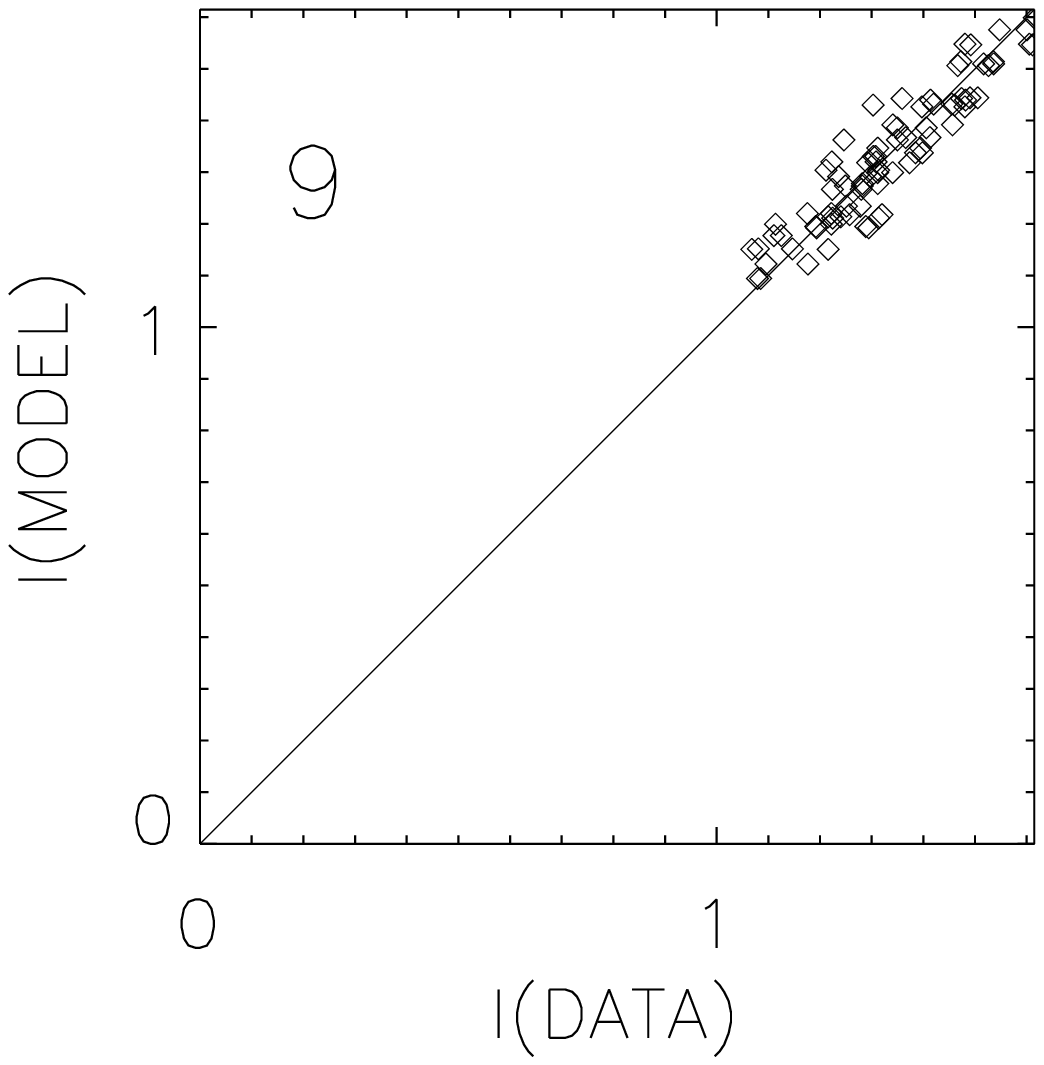}
\includegraphics[width=4.4cm]{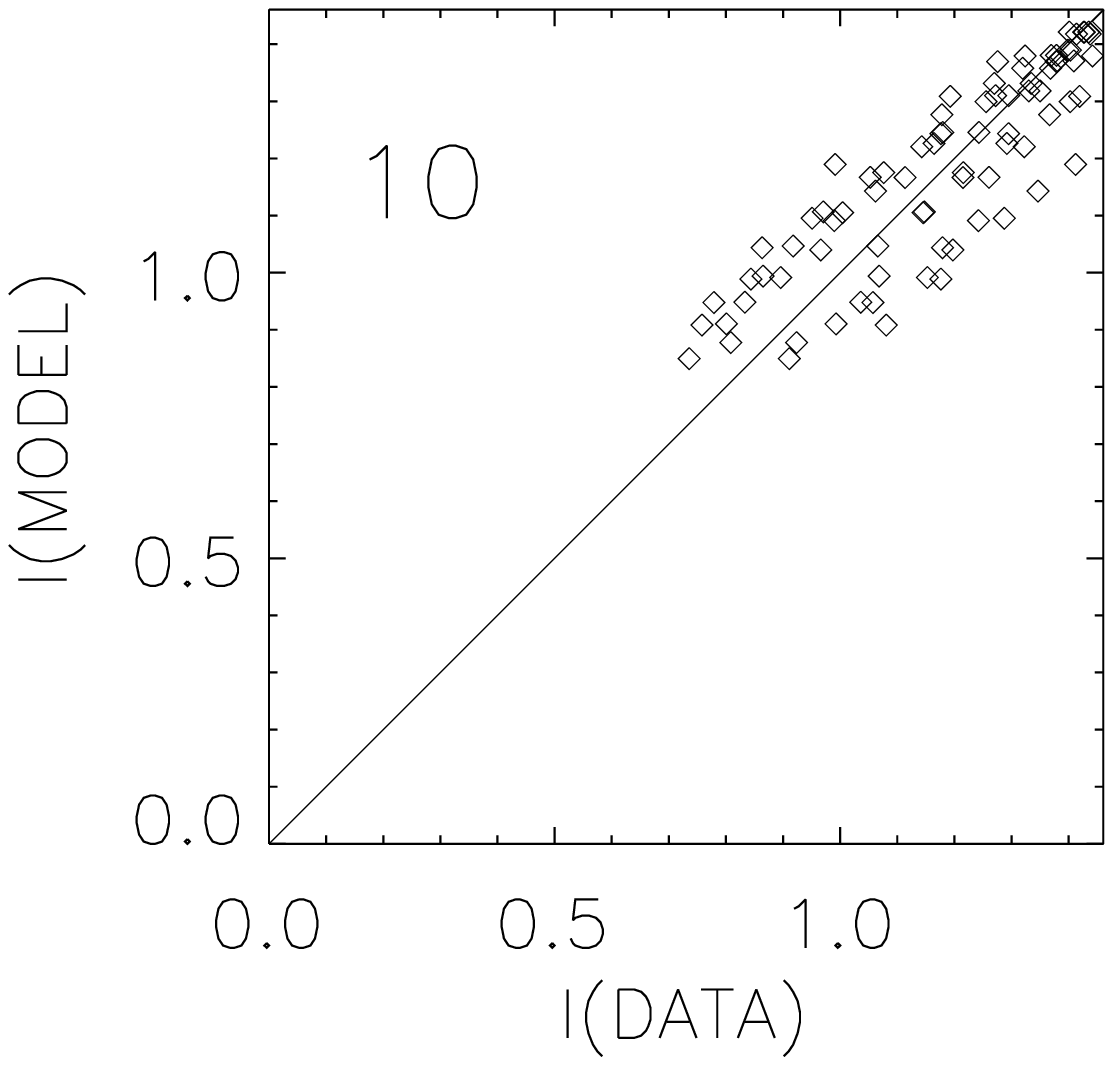}
\includegraphics[width=4.4cm]{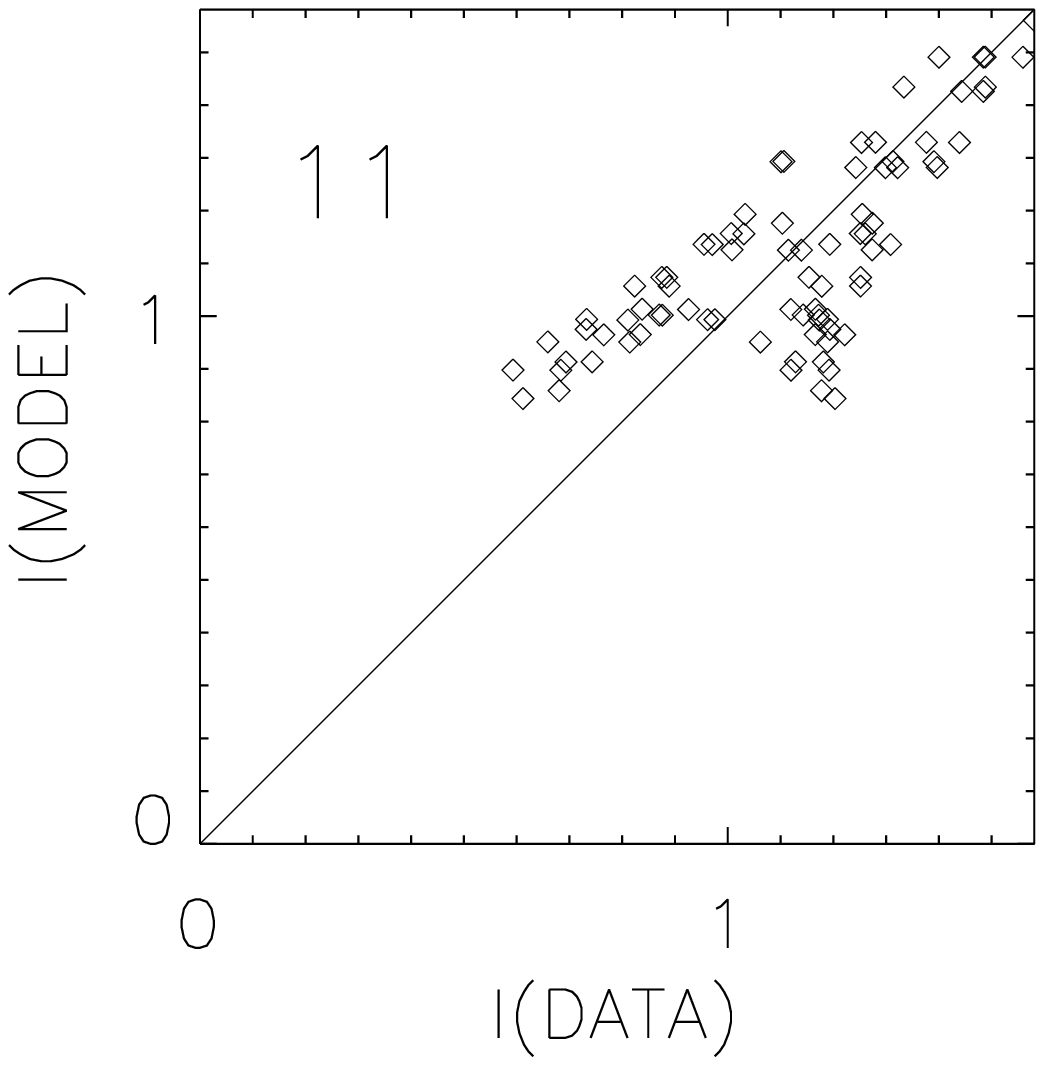}
\includegraphics[width=4.4cm]{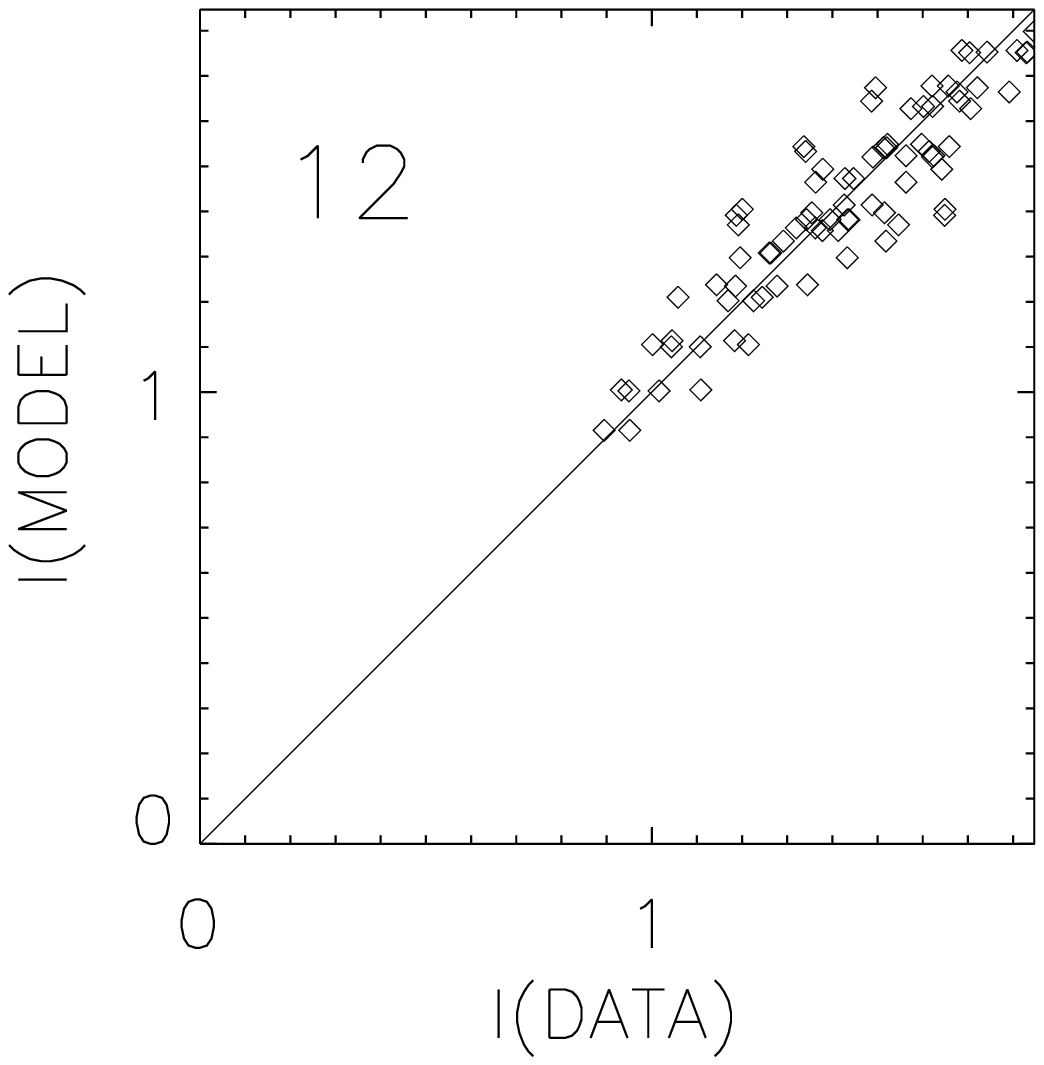}
\includegraphics[width=4.4cm]{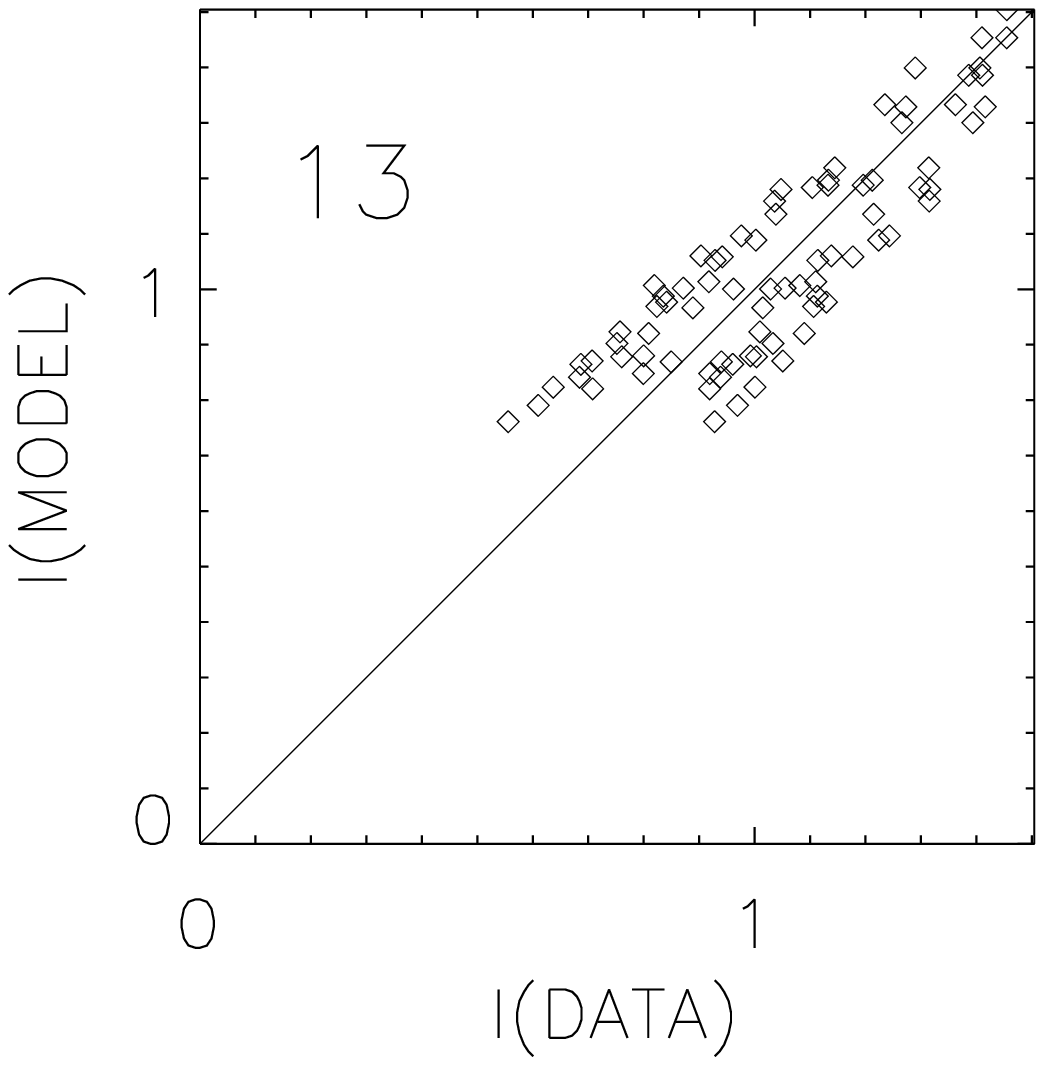}
\includegraphics[width=4.4cm]{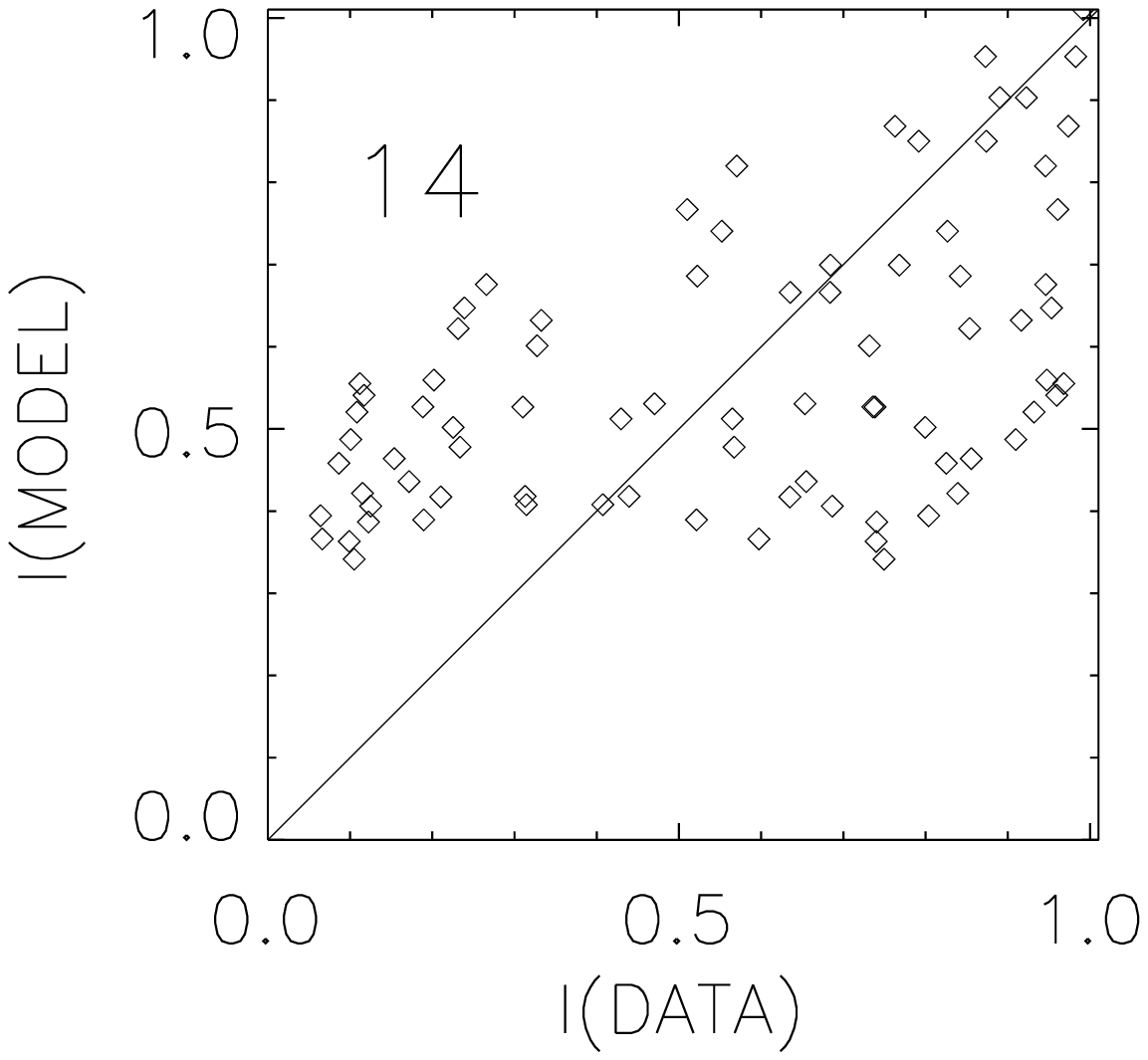}
\includegraphics[width=4.4cm]{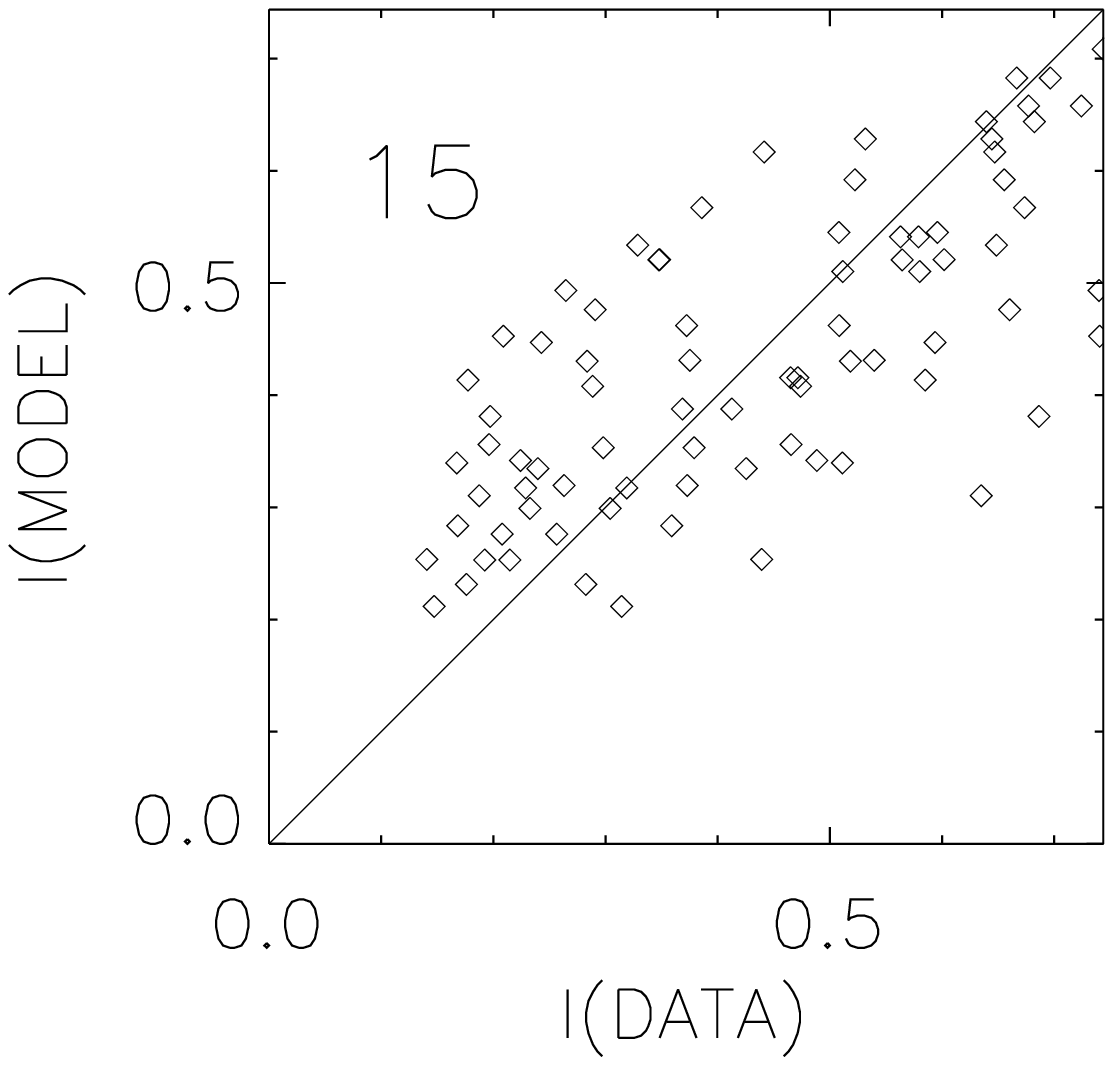}
\includegraphics[width=4.4cm]{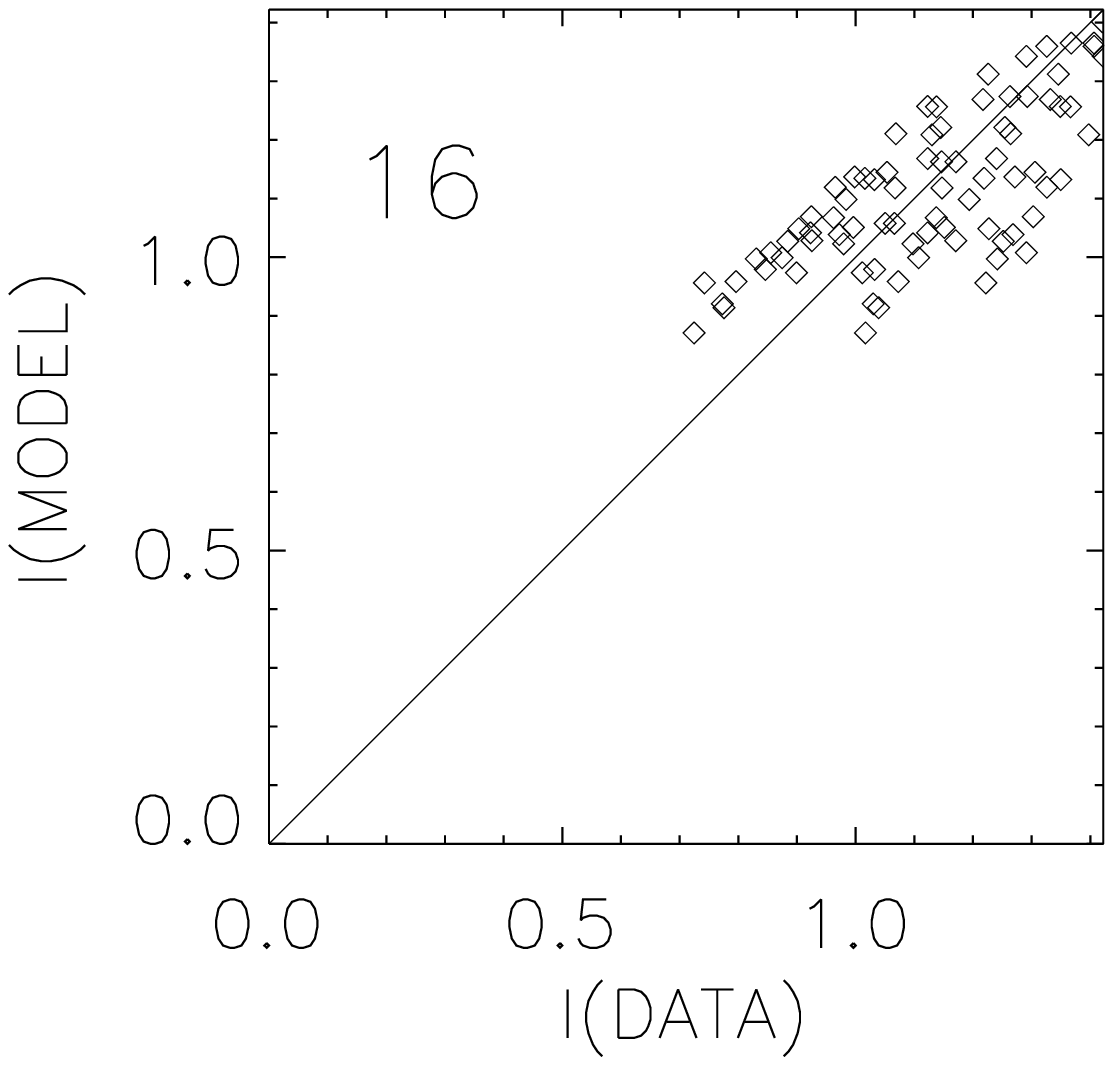}
\includegraphics[width=4.4cm]{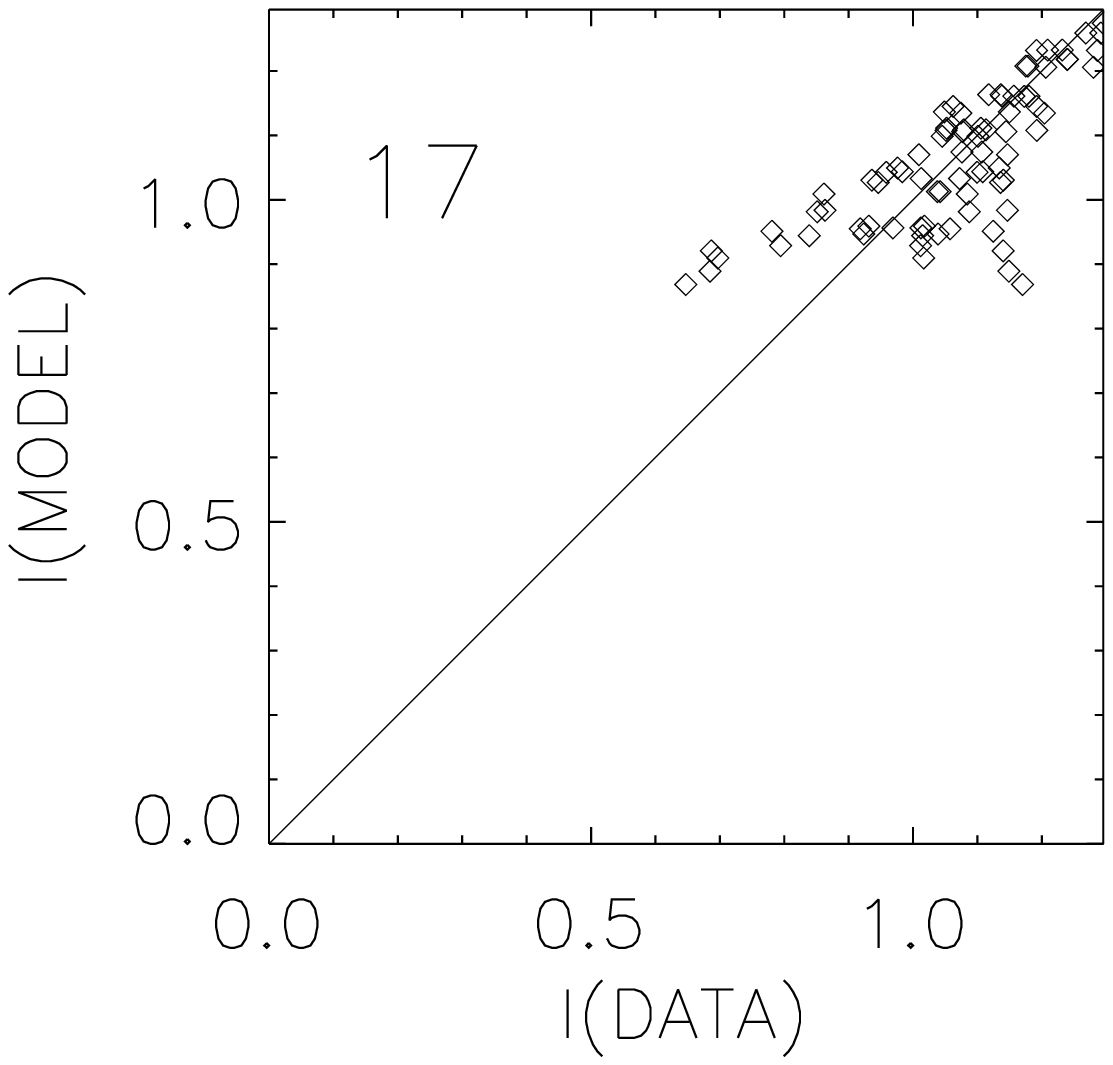}
\includegraphics[width=4.4cm]{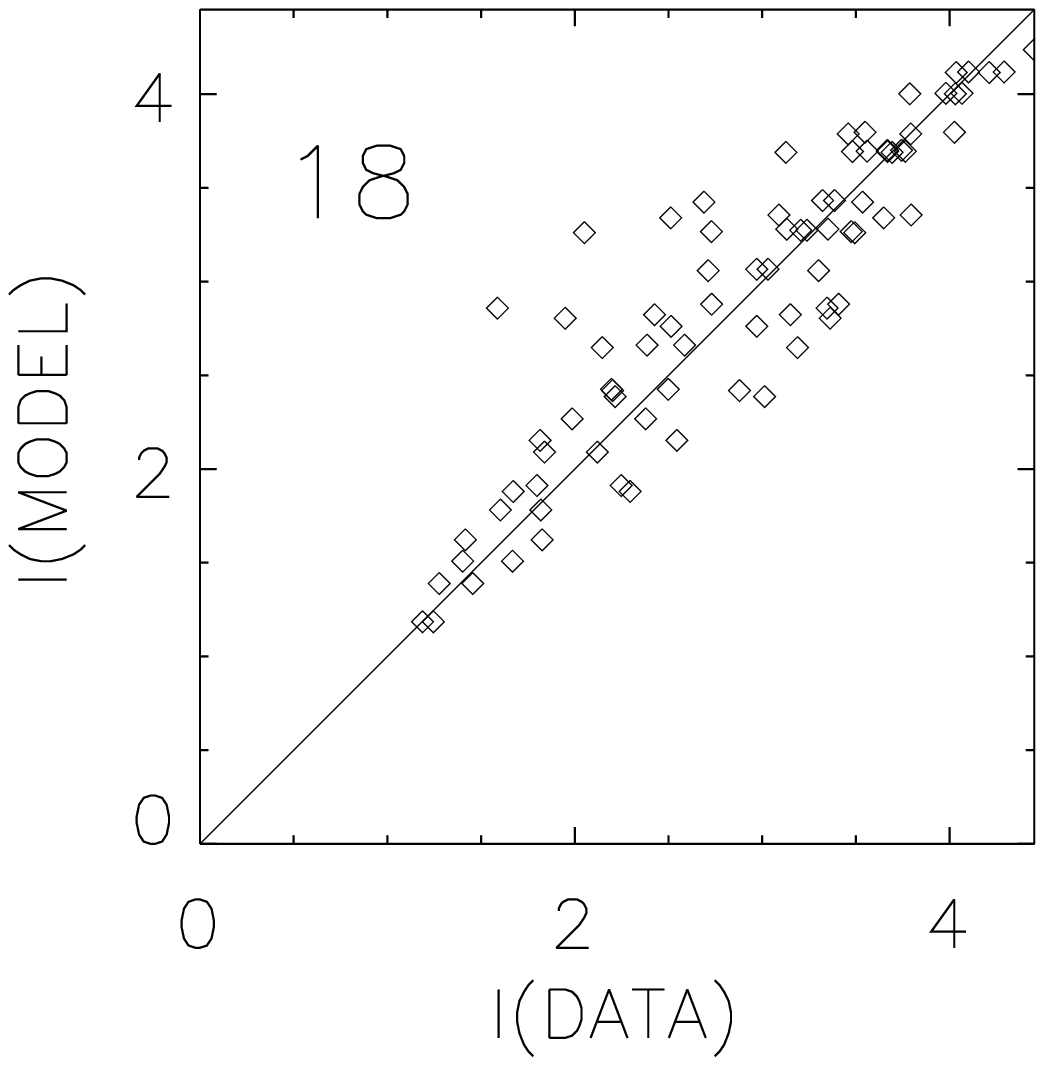}
\includegraphics[width=4.4cm]{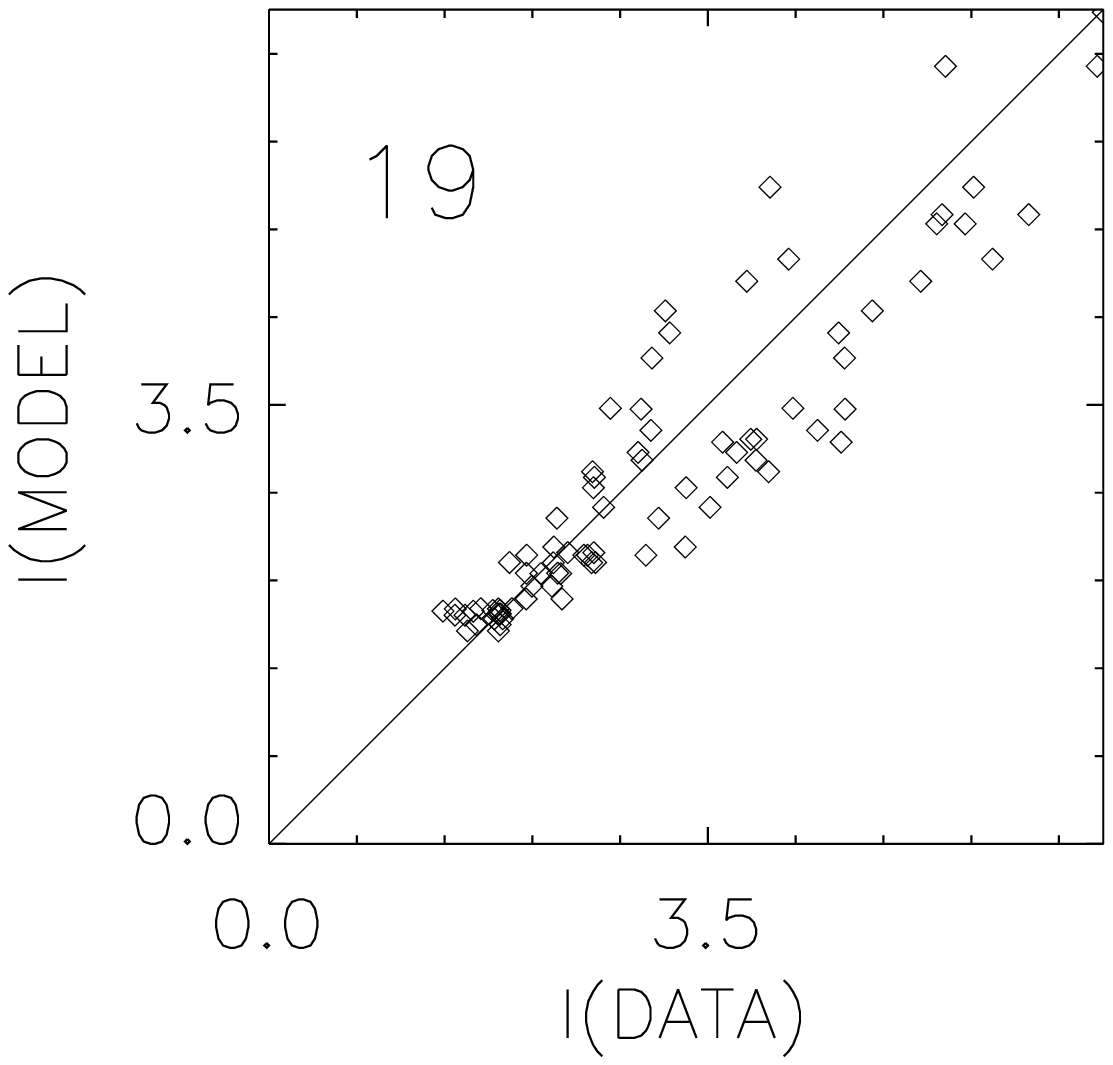}
\includegraphics[width=4.4cm]{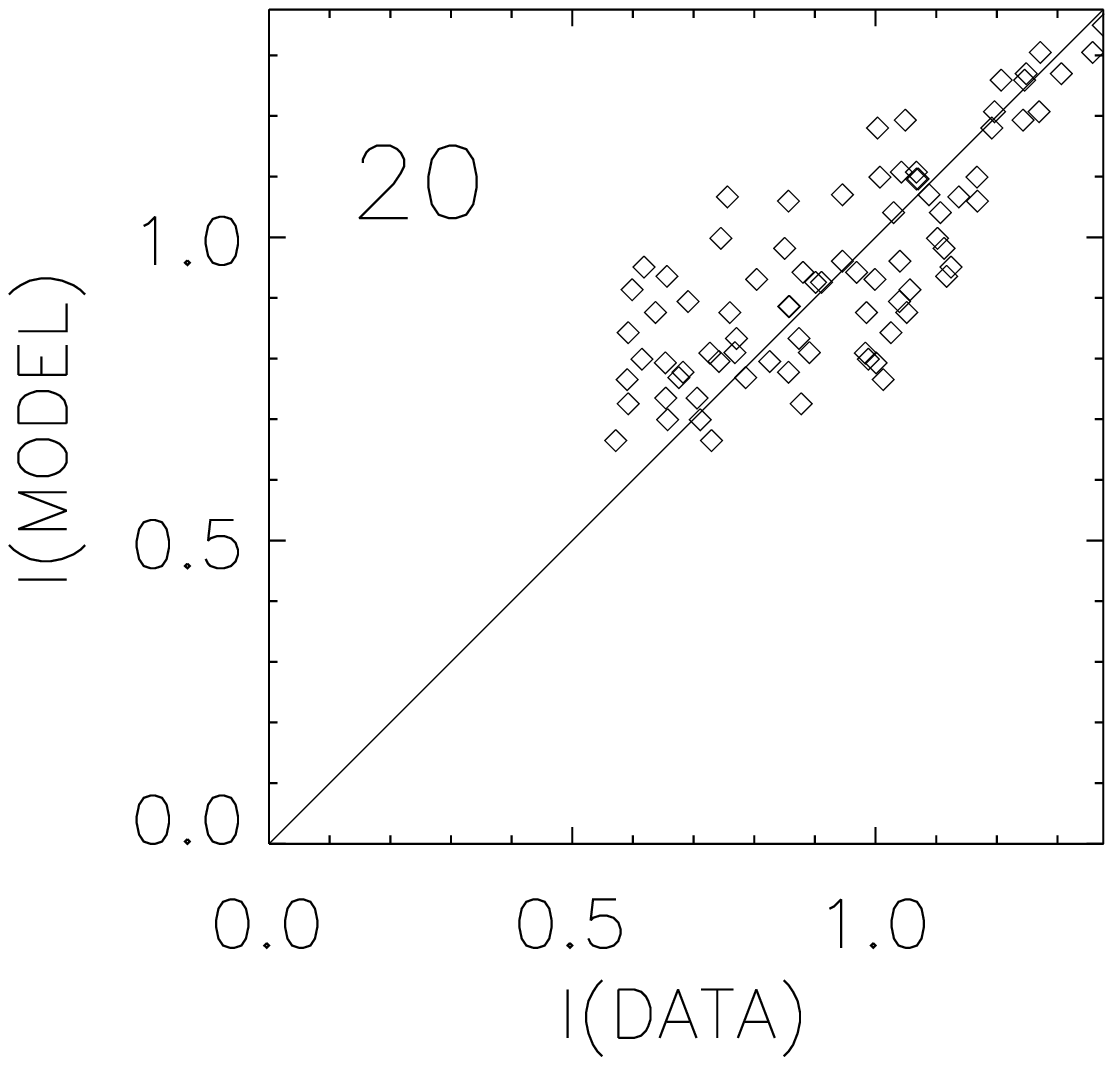}
\includegraphics[width=4.4cm]{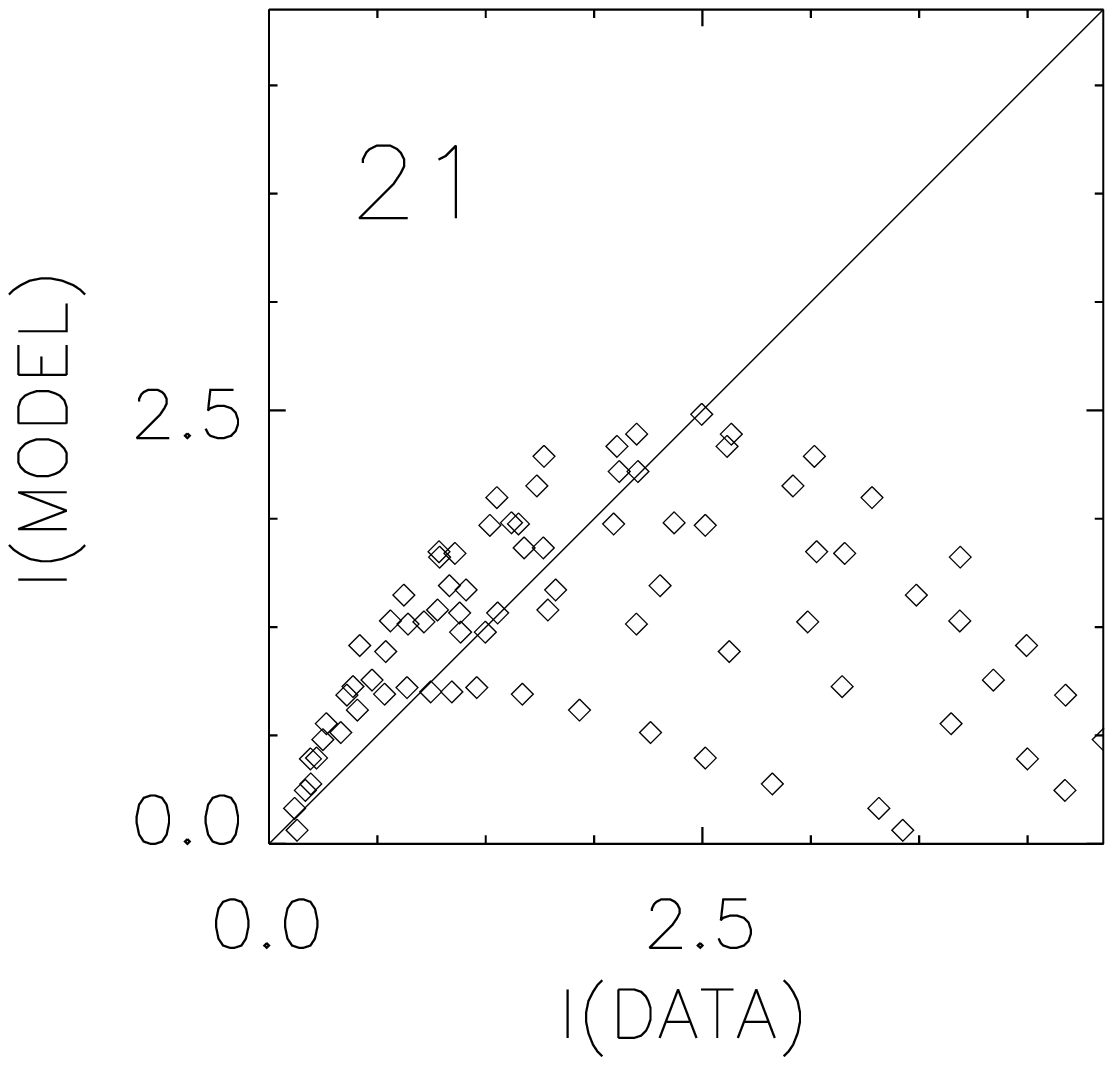}
\includegraphics[width=4.4cm]{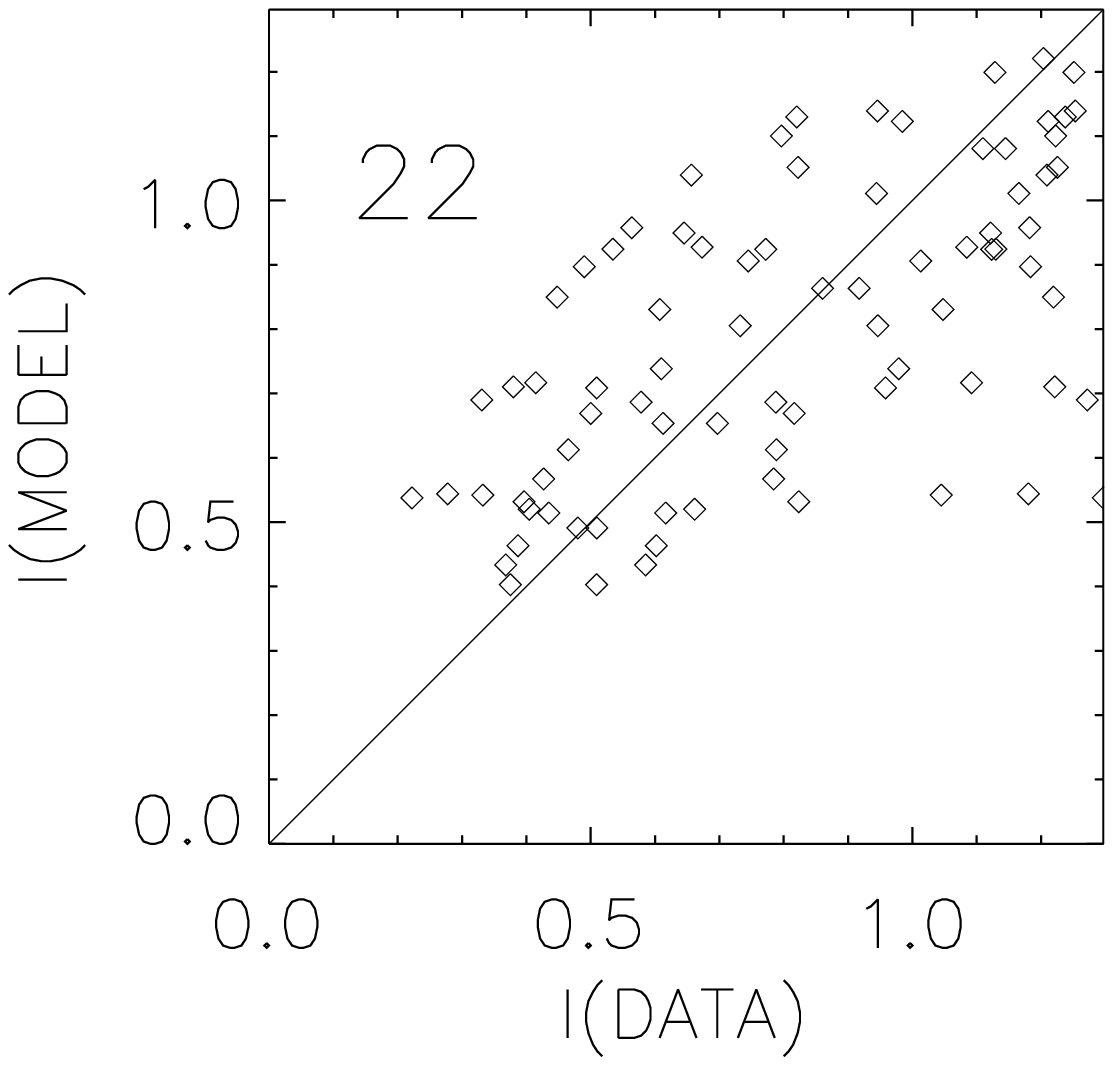}
\includegraphics[width=4.4cm]{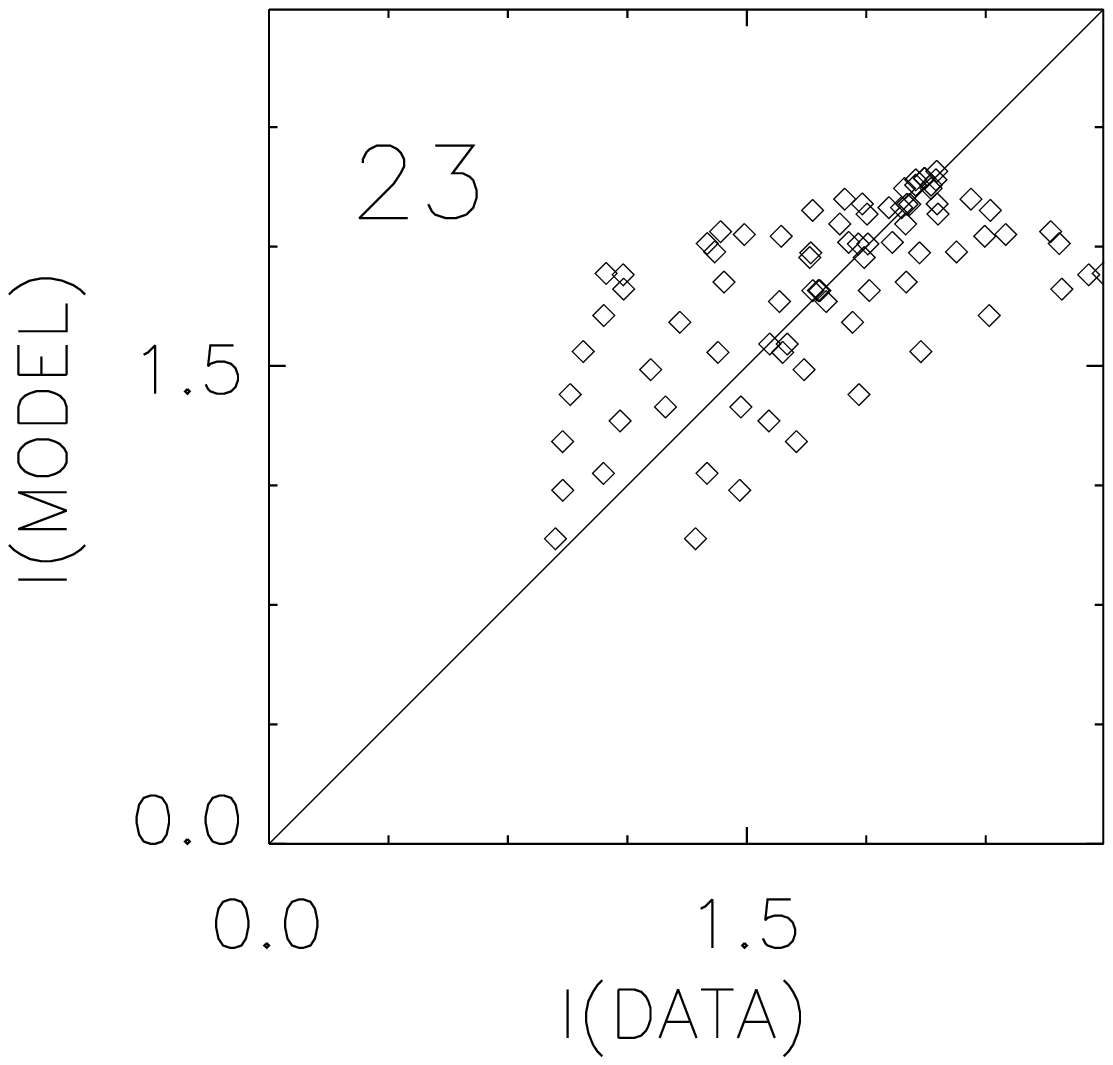}
\includegraphics[width=4.4cm]{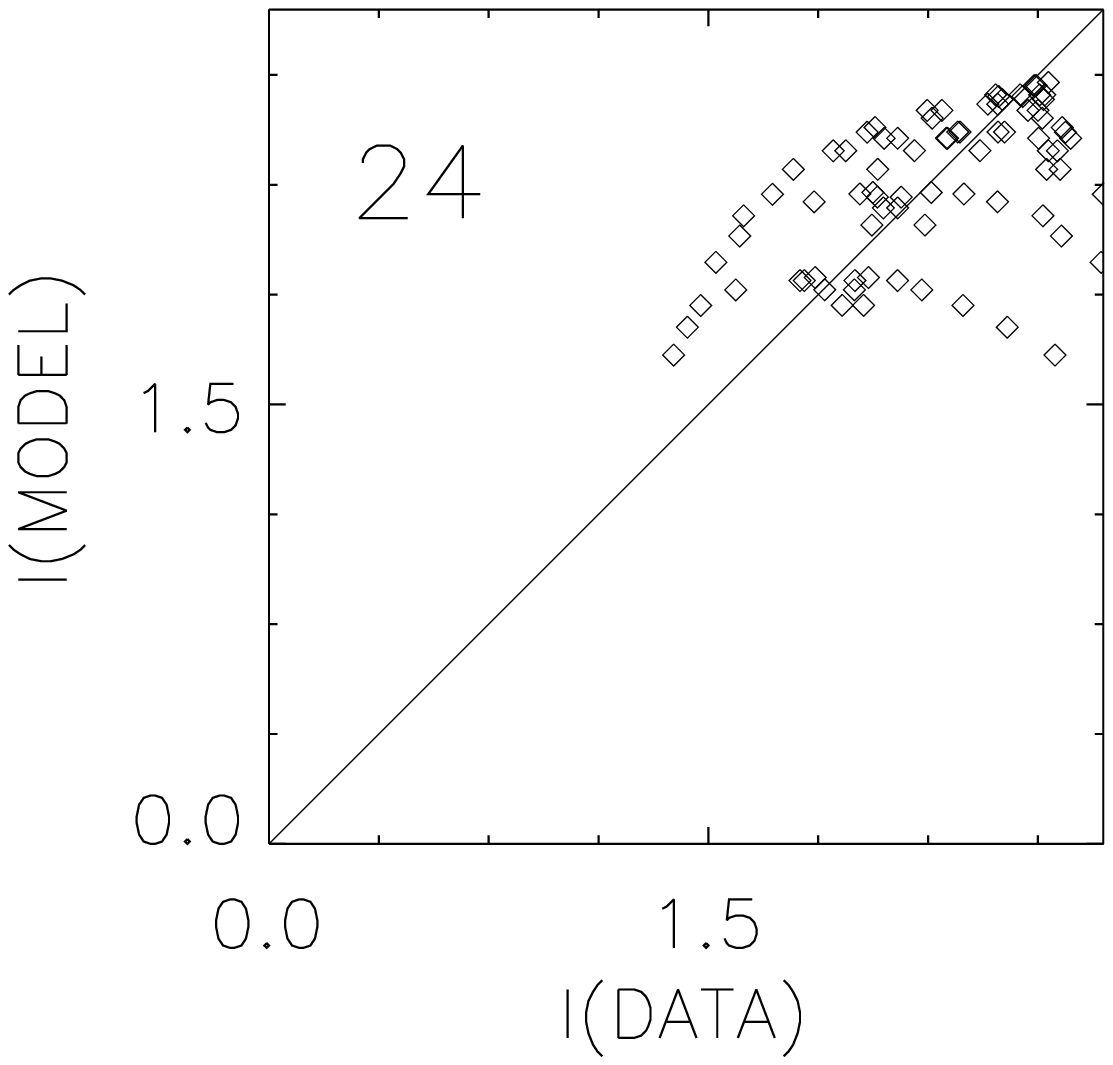}
\includegraphics[width=4.4cm]{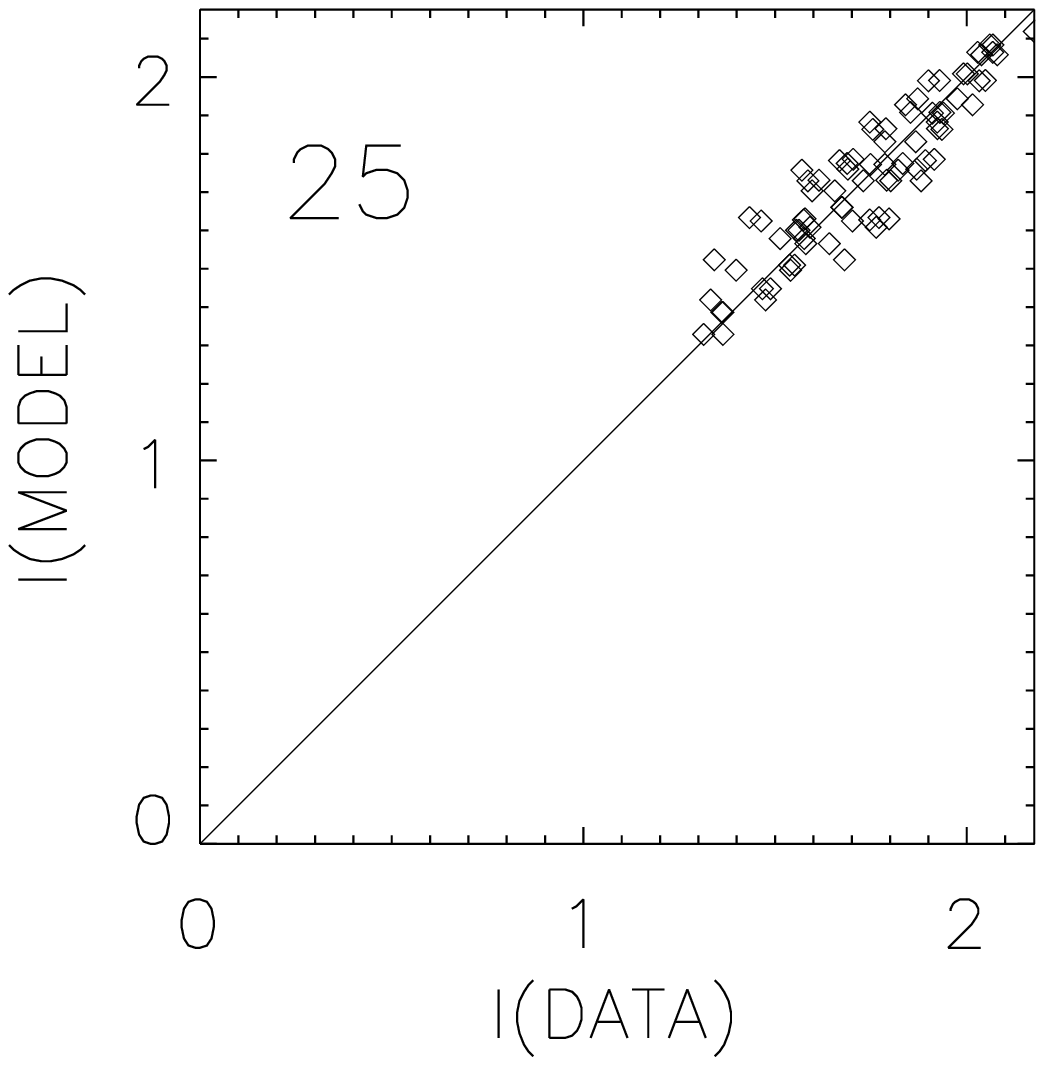}
\includegraphics[width=4.4cm]{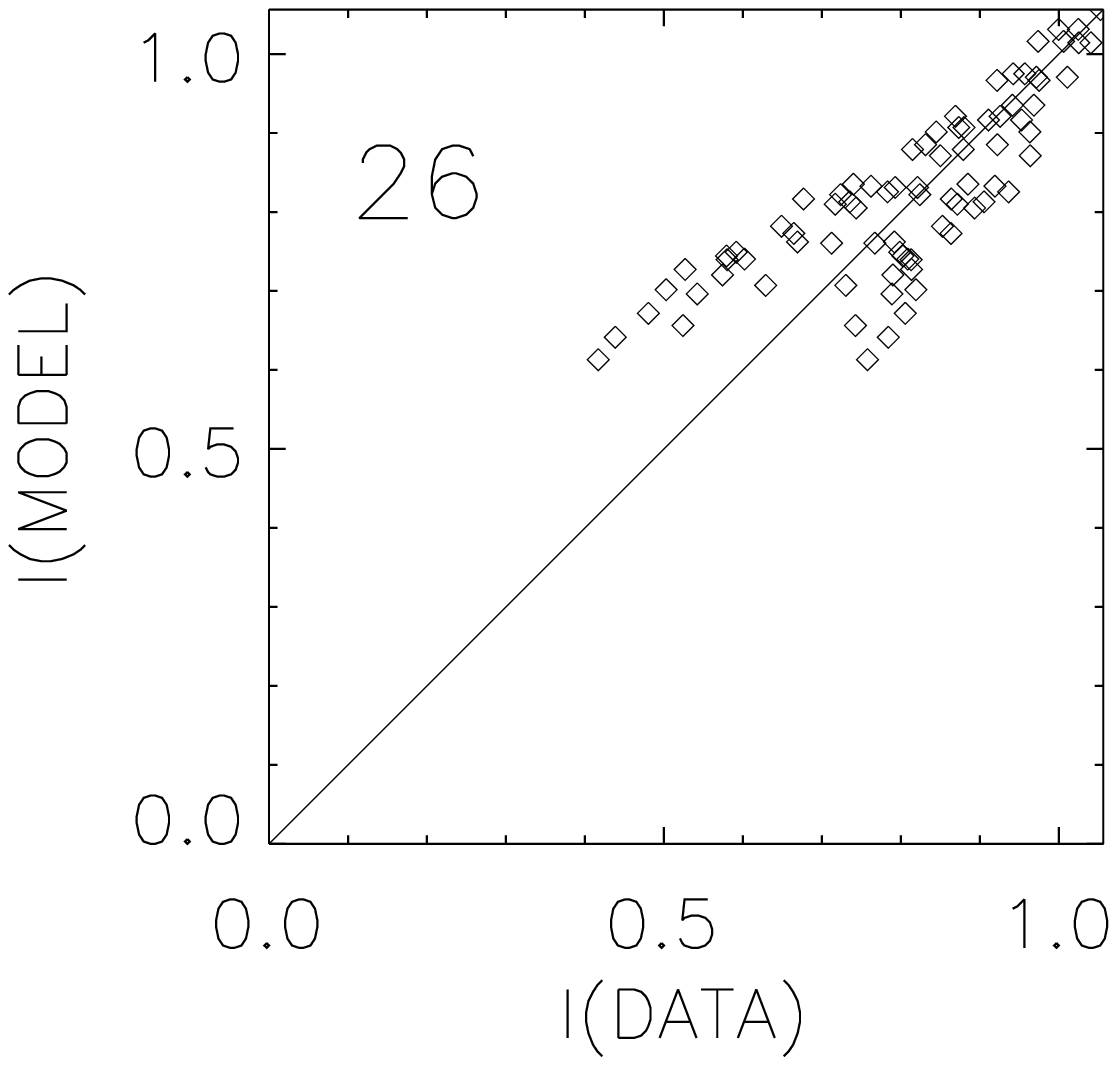}
\includegraphics[width=4.4cm]{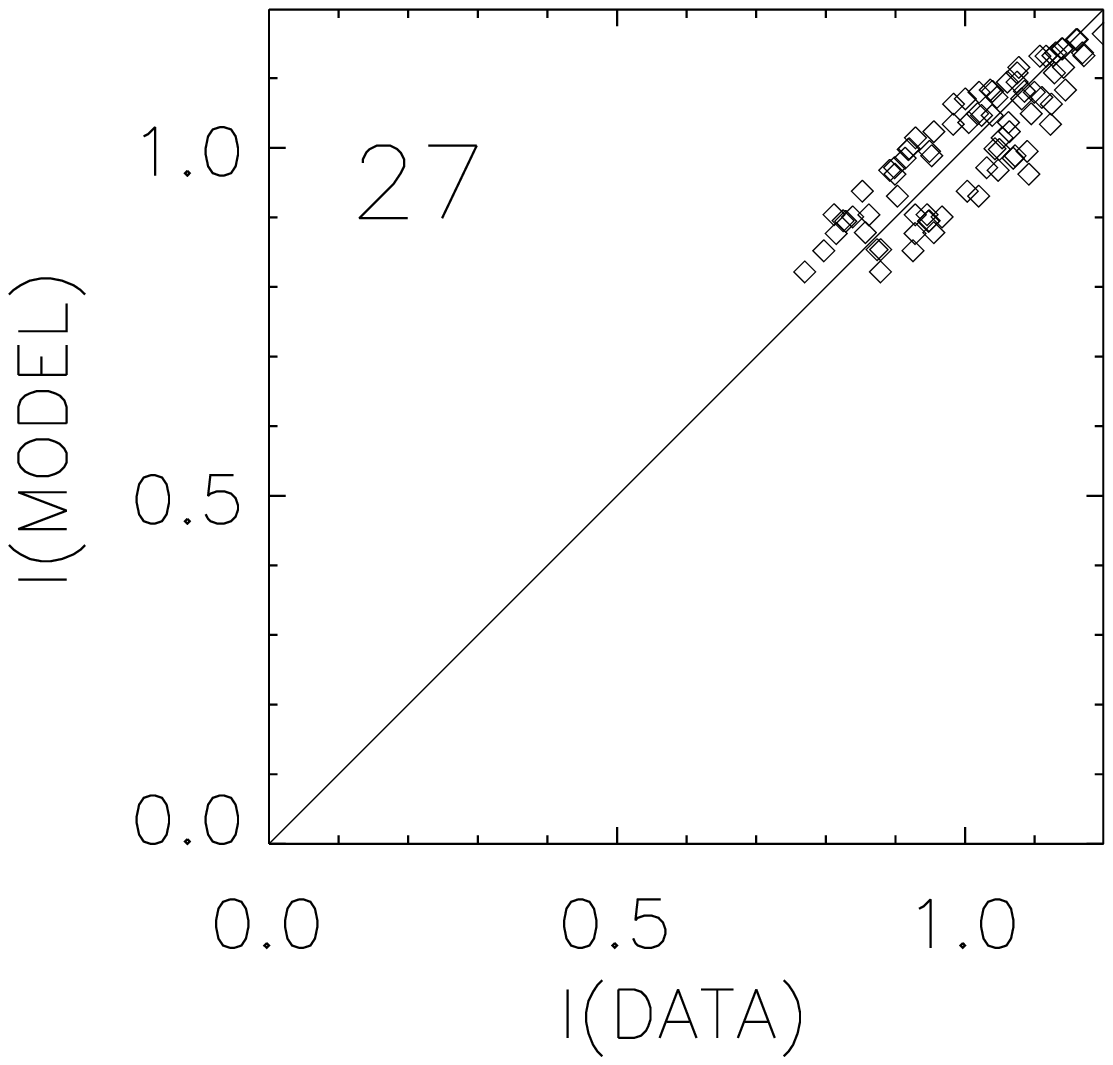}
\caption{Distribution of the modelled intensity obtained with Lorentzian
fits against the {\it Herschel} 350 $\mu$m observed intensity shown
for each region displayed in Fig. \ref{stamps}. There are 81 pixels 
shown in these panels. Strong correlations indicate that the
Lorentzian model provides a good fit.     
}
\label{correl_maps}
\end{figure*}

\subsection{Core Position Angle Estimates} \label{histoc}

To define the averaged orientation on the POS of the 
elongation of each prestellar core structure, 
we use the {\it Herschel} 350 $\mu$m intensity map. 
The spatial limit between the cores and the cloud envelopes
in which they are embedded is not always clearly defined, 
because of the limit of the resolution of the map. Also 
dust grain emission along the line-of-sight 
and the presence of other core structures can cause confusion. 
In their work, however, \citet{ryg13} consider a prestellar 
core to be defined as a gravitationally bound region of size $< 0.05$ pc.
We adopt the spatial scale, $l_{\rm min} = 0.05$ pc, 
as the smallest that should be
used for characterizing the prestellar core shapes.
With a distance to Lupus I of $\approx $ 155 pc \citep[][]{lom08}
this means that in the {\it Herschel} 350 $\mu$m intensity map, 
prestellar cores should be well sampled through 
kernels of $7 \times 7$ pixels with pixels of 
size $10{\arcsec} \times 10{\arcsec}$. To ensure 
that the structure of each prestellar core is fully included in our
analysis, we have decided to use sligthly larger kernels of $9 \times 9$ pixels
for estimating the elongations of the observed morphologies of the
cores. This ensures a large enough sample of pixels for the fitting procedure
that we detail below,
and also ensures direct comparison with the maps of the prestellar
cores displayed by \citet{ryg13} in their figure A.2.

Gravity is expected to be the dominant mechanism that shape core morphologies.
Therefore, we assume that the shape of a prestellar core can be modeled by a spheroid, and
that the size and orientation of the main axis of this spheroid (after
projection on the POS) can be 
approximated and described by the parameters associated with a 
two-dimensional (2D) Lorentzian distribution\footnote{In practice we
 also assumed and tested Gaussian distributions.
The results were similar to those obtained with Lorentzian distributions.}. 
Estimates of core elongations have been obtained with the IDL
mpfit2dpeak\footnote{http://www.exelisvis.com/docs/mpfit2dpeak.html.} 
routine assuming Lorentzian distributions. 
The fits have been obtained by only forcing the center of the
Lorentzian models to peak at
the position listed in Tables \ref{tabcore1} and
\ref{tabcore2}, and otherwise the remaining parameters (constant baseline level, peak value, 
half-width values along the short and long axis and position angle)
were left free in the fit.

Snapshots of the selected prestellar core regions and 
their long axis averaged  EPAs (as obtained with the Lorentzian fitting procedure)
are shown in Fig. \ref{stamps} with black line segments.
White lines show intensity contours 
obtained at levels of 0.5, 0.6, 0.7, 0.8, 0.9, 0.95 and 0.99 of the peak.
Dark lines show identical fractional intensity levels obtained from the Lorentzian
fit models. Estimates of the EPA and axis ratio 
obtained from the fits are displayed in columns 4 and 5 of Tables
\ref{tabcore1} and \ref{tabcore2}, respectively. 

In order to quantify the quality of the fits regarding the intensity 
structures imaged by {\it Herschel}, we calculated the linear 
Pearson correlation parameter (LPCP) of the Lorentzian model and
the observed intensity structure for each core.
This parameter is given in the last column in 
Tables \ref{tabcore1} and \ref{tabcore2}.
Figure \ref{correl_maps} show scatter plots of the Lorentzian modelled intensity 
(I(MODEL)) against the 350 $\mu$m observed emission (I(DATA)).
One can see strong correlations between the two intensities 
(e.g., for cores 1, 3 and 9), lack of correlations (e.g., for cores 4
and 21) and cases in between. The lack of correlation, i.e. a poor fit
of the Lorentzian model is mainly due to 
complex structure around the central position of the cores or to 
the presence of a secondary stronger peak nearby. In order to avoid 
a bias in the forthcoming analysis, we rejected the regions
for which the LPCP is lower than 
70 $\%$. This value provides a good compromise for rejecting ambiguous 
fits while ensuring that good quality fits are kept. 
We point out, however, that rejecting the regions for which the LPCP
is lower than 55 $\%$ would not have affected the general conclusions of this work.
We also rejected core 11, which has an axis ratio of unity from 
the Lorentzian fit, and so 
the EPA estimate cannot be trusted for this object. 
In Figure \ref{stamps}, all rejected cores are marked 
by a red cross.
   
\begin{figure}
\epsscale{1.}
\plotone{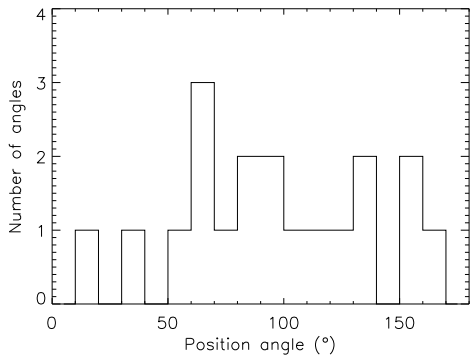}
\caption{Histogram of the distribution of the EPAs estimated for
  the list of sources displayed in Table \ref{tabcore1} and retained
  in our analysis, as discussed in Section \ref{histoc}.
\label{anghistoPAE}}
\end{figure}

The histogram of the distribution of the EPAs for the sample of
cores passing the tests discussed above are shown in Fig.
\ref{anghistoPAE}. The standard deviation is minimized for a
distribution centered around the median value of $95^{\circ}$, 
with a  standard deviation of $41 ^{\circ}$.
These results are discussed further in Section \ref{stat}.

\begin{deluxetable}{cccccc} 
\tablewidth{0pt}
\tabletypesize{\scriptsize}
\tablecaption{Location of cores and the estimates of their position
  angle elongation EPA 
and axis ratio as obtained from the Lorentzian fitting method
discussed in Section \ref{histoc}. Also
given in the last column is the linear Pearson coefficient parameter (LPCP) between
model and data distributions. 
\label{tabcore1}}
\tablehead{
\colhead{Index} & \colhead{RA (J2000)} &  \colhead{Dec (J2000)} &
\colhead{EPA} & \colhead{Axis} & \colhead{LPCP} \\
\colhead{} & \colhead{$(^{\circ})$} & 
\colhead{$(^{\circ})$} & \colhead{$(^{\circ})$} & \colhead{ratio} & \colhead{} 
}
\startdata
           1&      234.597&     -34.8721&      151&      1.2&     0.88\\
           2&      234.837&     -34.7280&      131&      1.8&     0.74\\
           3&      235.030&     -33.5602&      40&      3.3&     0.95\\
           5&      235.524&     -34.1556&      89&      1.8&     0.79\\
           6&      235.547&     -34.1519&      98&      8.1&     0.84\\
           8&      235.578&     -33.8461&      71&      1.7&     0.75\\
           9&      235.654&     -33.8625&      108&      1.7&     0.93\\
          10&      235.706&     -33.9900&      153&      4.5&     0.87\\
          12&      235.729&     -34.0750&      136&      1.8&     0.91\\
          13&      235.817&     -34.0778&      165&      1.6&     0.85\\
          15&      236.029&     -34.6499&      68&      2.1&     0.71\\
          16&      236.171&     -34.3421&      54&      1.3&     0.72\\
          17&      236.183&     -34.2955&      14&      1.9&     0.73\\
          18&      236.247&     -34.2861&      121&      2.6&     0.91\\
          19&      236.300&     -34.2855&      85&      2.2&     0.92\\
          20&      236.325&     -34.2093&      70&      1.5&     0.79\\
          25&      236.485&     -34.4937&      118&      1.8&     0.92\\
          26&      236.566&     -34.5111&      62&      1.6&     0.80\\
          27&      236.631&     -34.5514&      95&      1.6&     0.86\\
\enddata
\end{deluxetable}
 
\begin{deluxetable}{cccccc} 
\tablewidth{0pt}
\tabletypesize{\scriptsize}
\tablecaption{List of cores rejected from our analysis because we
  cannot reliably estimate the EPA, as discussed in
  section \ref{histoc}. Information displayed is as in Table \ref{tabcore1}. \label{tabcore2}}
\tablehead{
\colhead{Index} & \colhead{RA (J2000)} &  \colhead{Dec (J2000)} &
\colhead{EPA} & \colhead{Axis} & \colhead{LPCP} \\
\colhead{} & \colhead{$(^{\circ})$} & 
\colhead{$(^{\circ})$} & \colhead{$(^{\circ})$} & \colhead{ratio} & \colhead{} 
}
\startdata
           4&      235.042&     -34.9250&      ...&      1.4&     0.47\\
           7&      235.565&     -33.8499&      ...&      2.0&     0.60\\
          11&      235.704&     -34.2224&      ...&      1.0&     0.73\\
          14&      236.000&     -34.6417&      ...&      1.4&     0.50\\
          21&      236.325&     -34.2889&     ...&      2.0&     0.10\\
          22&      236.342&     -34.2541&      ...&      2.4&     0.61\\
          23&      236.353&     -34.3681&      ...&      2.5&     0.63\\
          24&      236.367&     -34.3806&      ...&      2.0&     0.55\\
\enddata
\end{deluxetable}

\begin{figure}
\epsscale{1.}
\plotone{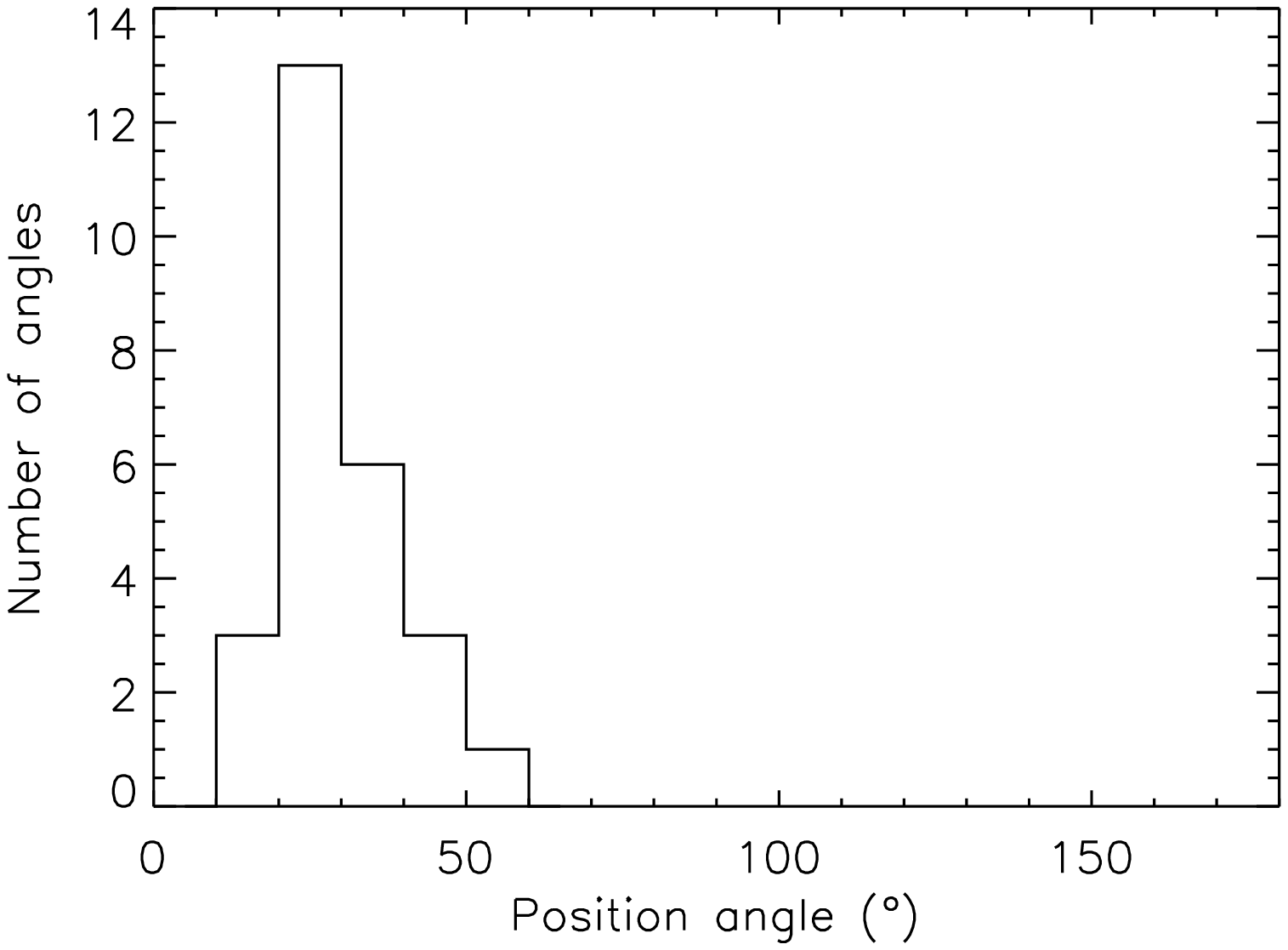}
\caption{Histogram of the inferred magnetic field orientations,
as derived from the BLASTPol 2010 350 $\mu$m polarization data.
The inferred field direction is obtained by adding $90^{\circ}$
to the measured polarization angle.
\label{anghisto350}}
\end{figure}

\subsection{Histogram of Inferred Magnetic Field Position Angles} \label{histob}

The histogram of inferred magnetic field orientations 
is obtained by shifting all the measured 350 $\mu$m polarization
angles by an angle of $90^{\circ}$ \citep[][]{mat13}. This histogram is shown in Fig.
\ref{anghisto350}. The distribution is strongly
peaked, with an average value of $\theta_{\rm B} \approx
29^{\circ}$ and a standard deviation of $\approx 10^{\circ} $. 

\section{STATISTICAL RESULTS} \label{stat}

\begin{figure}
\epsscale{1.5}
\plotone{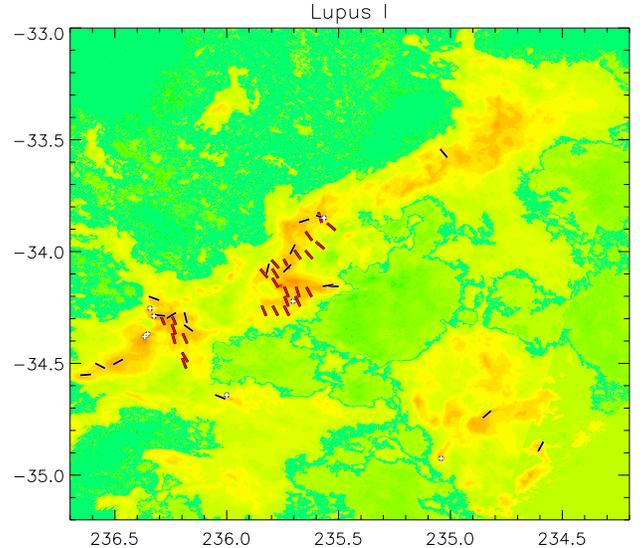}
\caption{
Map showing magnetic field orientations red lines inferred
from the BLASTPol 350 $\mu$m data. The locations 
of the prestellar cores discussed by
\citet{ryg13} are shown with black lines (where orientations could be
robustly derived) and white crosses (where no clear orientation could
be fit). The map in the background shows the {\it Herschel} 350 $\mu$m 
dust emission intensity map.
\label{map350}}
\end{figure}

A visual summary of our results is shown on the Lupus I map displayed in Fig. \ref{map350}. 
The orientations of the average elongations of the prestellar core
structures are shown with black lines. 
The locations of the cores displayed in Table 
\ref{tabcore2}, which are rejected by our analysis, are indicated with white crosses. 
The inferred projected magnetic field orientations are shown with
red lines.

\subsection{Review The Morphology Discussed in Matthews et al.} \label{review}

The morphology of the main filament
running in the Lupus I region from bottom left to top right in
Fig. \ref{map350} has been discussed by \citet{mat13}.
The relation between the large-scale magnetic field, as probed with starlight
polarimetry by \citet{riz98} in the diffuse ISM surrounding Lupus I, 
the magnetic field structure probed in denser regions of the filament
with the BLASTPol 2010 data, and the Lupus I main filamentary shape 
observed on large-scale, has also been discussed by these authors. 
Consistency is found between the mean magnetic field orientation in 
dense and diffuse regions of the ISM, and the elongation of the main
filament is found on average to be nearly perpendicular to the large-scale magnetic
field structure as seen in projection on the POS. 

\citet{mat13} investigated the relation between the
large-scale magnetic field and 
an arced filament model that they introduce for
more accurately describing the shape of the main filament in Lupus I. This bent
filament is parameterized by an arc of a circle centered at RA (J2000)
$=231.77^{\circ}$, Dec (J2000)$=-37.67^{\circ}$, with a radius 
of $=4.92^{\circ}$. With this model, the authors found that the 
magnetic field orientations probed along high column density regions with submm data,
and in the diffuse ISM with optical data, bracket the filament normal, differing 
from it by 9.8$^{\circ}$ and 8.6$^{\circ}$, respectively.

\subsection{Core Elongation Distribution vs. Large-Scale Molecular Cloud Structures} \label{cores_filaments}

The histogram of the EPAs for the 19
cores passing the analysis test discussed in Section \ref{histoc}
 are shown in Fig. \ref{anghistoPAE}. 
Data for these 19 cores are listed in Table \ref{tabcore1} . 

The standard deviation is minimized for a
distribution centered around the median value of $95^{\circ}$, but 
whether or not we try to minimize this parameter, high standard
deviation values of the order of $41^{\circ}$ are obtained,
i.e. values quite close to the value of $\approx 52^{\circ}$
expected for a strictly random distribution \citep[][]{ser62}. 
Therefore, the EPAs appear to have
no special global alignment with respect to the 
large scale main filament 
model discussed by \citet{mat13}, nor do they appear to have a special
alignment with respect to the pattern of secondary filaments
that often run orthgonally to this large scale main filament.  

\begin{figure}
\epsscale{1.}
\plotone{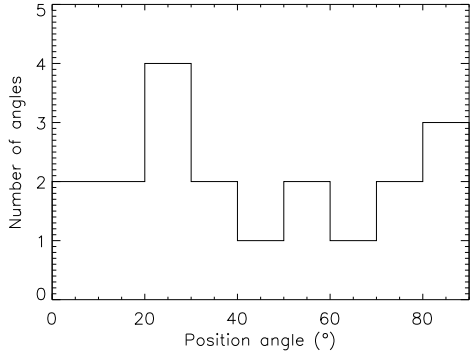}
\caption{Histogram of the distribution of the offset angles between
  the core EPAs and the normal to the arced filament model
 discussed by \citet{mat13} .
\label{offsetshisto}}
\end{figure}

By using the description of  the arced filament model proposed by \citet{mat13} 
we calculated the offset PA between each core's EPA and the local 
normal to the bent filament. Figure \ref{offsetshisto} shows the
histogram of these PA offsets, obtained for the 19
cores passing the selection of Section \ref{histoc}
and listed in Table \ref{tabcore1}. 
Here again we find a distribution
consistent with random, which means that no specific
orientation can be seen between the EPAs and the large-scale
bent filament model. 

\subsection{Core Elongation Distribution vs. Magnetic Field Structures} \label{cores_filaments}

Due to problems with a damaged blocking IR filter during flight and resultant systematics, limiting the 2010 data set
\citep[as discussed by][]{mat13}, the information provided by the BLASTPol 350  $\mu$m 
polarimetry data is pixelized in pixels of size $2.5{\arcmin} \times 2.5{\arcmin}$. 
This corresponds approximately to kernels of $16 \times 16$ pixels in the
{\it Herschel} 350  $\mu$m intensity map, i.e., about 3.2 times larger 
than the regions used for characterizing the core morphology structures.
Because of this we decided not to compare the 
orientation of each individual core to its local magnetic field structure.
As a consequence, in the following we only compare the distribution of the 
elongation of the cores to the mean magnetic field orientation.

The elongation of the main
filament is found on average to be nearly perpendicular to the large-scale magnetic
field structure as seen in projection on the POS \citep[][]{mat13}.
Our result therefore also implies that there is no specific
orientation of the average elongation of the cores with respect to
the large-scale structure of the magnetic field that might shape the main filament. 

However, since the structure of the magnetic fields in the high
density regions (which have not been probed with submm polarimetry) is
not clear, we have searched for possible correlations between the 
elongations of subsets of cores having POS displacements from submm pseudo-vectors
smaller than various threshold values. Given that the mean 
width of the Lupus I main filament is of order $7.5{\arcmin}$,
we first calculated mean and median EPA estimates
for cores having displacements of less than $7.5{\arcmin}$ from
the nearest pseudo-vector. The standard deviation of the distribution
of EPAs for this subset is higher than 
that obtained for the complete set of selected cores,
i.e., it too is consistent with a random distribution. When the
same calculations are performed for a displacement lower than the size of the
BLASTPol 2010 beam ($2.5{\arcmin}$), 
the size of the sample ($N=6$) starts to be quite small,
but the same conclusion can be drawn. As an ultimate test, we
calculate a mean EPA of $103^{\circ}$ for the subset of cores
matching within a BLASTPol beam. This means there is an average
offset angle of about $16^{\circ}$ between the mean short axis
direction of the sample of cores and 
the mean orientation of the magnetic field of $\approx 29^{\circ}$. 
Although this result is obtained for a
sample of two cores only, it is consistent
within the uncertainties with the results obtained
by \citet{war09} and \citet{tas09}, but obviously no strong
conclusions can be drawn from this.

A summary of all these results is displayed 
in Table \ref{tabsummary}.

\begin{deluxetable}{llccc} 
\tablewidth{0pt}
\tabletypesize{\scriptsize}
\tablecaption{Statistics on PA Distributions. \label{tabsummary}}
\tablehead{
\colhead{Field} & \colhead{Sample} &  
\colhead{Mean} & \colhead{Median} &
\colhead{$\sigma$} 
\\
\colhead{} & \colhead{Size} & 
\colhead{$(^{\circ})$} & \colhead{$(^{\circ})$}
&\colhead{$(^{\circ})$} 
}
\startdata
EPA(Cores)$^{(\rm a)}$&          19&       96&       95&       41\\
PA(B-Field) $^{(\rm b)}$&          26&       29&       28&       9\\
EPA(Cores)$^{(\rm c)}$&          12&       114&       108&       46\\
EPA(Cores)$^{(\rm d)}$&           6&       112&       108&       43\\
EPA(Cores)$^{(\rm e)}$&           2&       103&       108&       7\\
\enddata
\tablenotetext{(a)}{Average position angle estimate for the sample of 
cores listed in Table \ref{tabcore1}.}
\tablenotetext{(b)}{Average inferred magnetic field position angle
  estimate.} 
\tablenotetext{(c)}{Average EPA for the sample
 of cores with distance less than 7.5${\arcmin}$
from a polarization pseudo-vector as seen on the POS.}   
\tablenotetext{(d)}{Same as (c) but for a distance lower than 
2.5${\arcmin}$, similar to the effective beam size of the BLASTPol 2010 data.}
\tablenotetext{(e)}{Same as (c) but for a distance less than 1.25${\arcmin}$.}
\end{deluxetable}

\section{DISCUSSION} \label{discussion}

Our core orientation characterization method is based 
on 2D Lorentzian fits, which means that
no assumptions on whether the 3D structure of the cores 
is oblate or prolate have been made.
We chose not to investigate these aspects because, 
due to the integration of the signal along the LOS, 
it is not possible to spatially separate the contribution 
of the dust emission provided by the filament from 
that originating in the cores. This is particularly the case in crowded regions
at different evolutionary stages \citep[see figure A1 of][]{ryg13}
where overlapping cores add confusion. 
It has been possible, however, to define EPAs
for 19 sources from the 27 objects in the sample, as listed in Table
\ref{tabcore1}.

For the remaining targets of the sample listed in Table \ref{tabcore2} it was not possible to define 
EPA values with high confidence. In the case of core 11, we believe
this is a result of 
projection effects due to the complex three-dimensional structures of this core. 
The shape of this object might be that of an oblate disk seen
face-on or of a prolate ellipsoid pointing end up, as indicated by
the axis ratio of unity.
The probability of this is expected to be low \citep[][]{gam03}
but cannot be totally rejected \citep[][]{tas09}, in particular if
the amount of dust emission between the observer and a 
given core is negligible. For prestellar cores 4, 7, 14, 21, 22, 23 and 24, the
problems with modelling their 2D projected shapes is more likely 
coming from the complex dust emission intensity distribution 
along their LOSs, in particular in regions containing high
numbers of cores with embedded sources \citep[][]{ryg13}.

With all the limitations mentioned above and the results discussed previously, 
the general picture emerging from our analysis is that 
the sample of selected prestellar cores 
surviving our analysis looks quite randomly 
oriented on the POS, and therefore randomly oriented
with respect to the 
main filament of the Lupus I molecular 
cloud, as well as with respect to 
the mean magnetic field structure probed 
in various density regimes in this region.
This last finding is in agreement with the analysis of simulated cores
(or clumps) provided by \citet{gam03}. These authors study the
formation of three-dimensional analogs of cores using 
self-consistent, time-dependent numerical models of molecular clouds.
Their models include decay of initially supersonic turbulence in an 
isothermal, self-gravitating, magnetized fluid. All simulated cores are not
expected to be self-gravitating and their axes are    
not strongly aligned with the large-scale magnetic field.

\begin{figure}
\epsscale{1.4}
\plotone{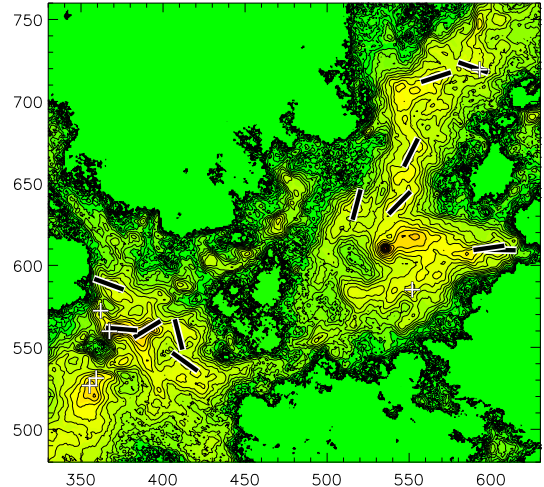}
\caption{Zoom in on the region centered around RA (J2000)$=236.0^{\circ}$,
 Dec (J2000)$=-34.2^{\circ}$ where the magnetic field has been probed
 with BLASTPol in Lupus I (see Fig. \ref{map350}). Contour lines showing spatial
 variations of the column density have been overlaid for comparison
 with the core elongations (shown with black lines) with 
respect to the 2D morphology of the cloud structures 
in their neighborhood.
\label{zoomc}}
\end{figure}

\citet{mat13} suggest a correlation between the main shape of the filament 
and the mean magnetic field on large scales, but secondary filaments
are also observed, which make the picture of Lupus I a complex one once 
smaller scales are considered. To illustrate this aspect, we plotted
in Fig. \ref{zoomc} intensity contours showing spatial
variations of the column density overlaid on
the core elongations (shown with black lines). 
In this part of the Lupus I cloud, a
secondary filament crosses the main cloud filament, as discussed 
by \citet{mat13}, near where the magnetic field has been probed 
by BLASTPol (see Fig. \ref{map350}). 
Is is clear from this Figure that the cloud sub-structure is complex, so 
that the morphology of the cores may be determined 
by their local environment, i.e., the local physics of the filament,
rather than by the large-scale morphology of the filament. 

Regarding the effects of the magnetic fields on
sub-parsec scales, 
various studies \citep[e.g.,][]{hil99} have shown that a ''polarization hole'',
or in other words, a decrease of the polarization fraction as a
function of the intensity, is
measured toward the majority of molecular cloud cores observed 
with submm polarimetry. Therefore, it is not yet clear whether or not 
magnetic fields are probing deep into the cores, in particular above visual extinction of about
10 magnitudes, where dust grain alignment might
be inefficient \citep[][]{laz07,pel09}. Apparent depolarization may also
occur because of a lack of angular resolution, the effect of which is
to smooth complex
small scale magnetic fields structures thereby producing a net low degree
of polarization. 
A mean offset of $\approx 30^{\circ} $ between the short axis of cores
embedded in six distinct Bok Globules and the magnetic field
orientation in their local diffuse environment
was discovered by \citet{war00,war09}.
Similar results have  been subsequently
obtained by \citet{tas09} for a sample of 24 molecular clouds.
The latter study concentrates on high mass molecular cloud cores
that are larger and denser
regions than Bok Globules, but these
authors also find that the magnetic field orientation is close to the
shortest cloud axis
by showing on average a deviation of $24^{\circ} $.

A similar qualitative conclusion is inferred by \citet{hul13} 
from their study of a sample of low mass protostellar cores whose cores embedded
disks are not expected to be aligned with the magnetic fields in the
cores but this picture is not so clear when compared to results 
obtained from the analysis by \citet{cha13} showing
correlation for low mass cores. In addition anti-correlation
cases can be expected for high mass cores 
\citep[see][]{poi06},
which makes this subject still open to more investigation.

We expect that the BLASTPol 2012 data collected during December 2012
-- January 2013, currently under analysis, may be able to address these issues. 

\section{SUMMARY} \label{conclusion}

In this work we first calculated the average 
elongation position angles, EPAs, for a sample of
prestellar cores identified by \citet{ryg13}. We then compared the
distribution of the core EPA values to the mean shape of the 
large-scale filaments in Lupus I.

The average orientation of the cores, as seen on the POS is obtained by 
fitting 2D Lorentzian models to the 350 $\mu$m 
{\it Herschel} dust emission intensity map centered at the position 
of the prestellar sources.  

We find the EPAs to be consistent with a random 
distribution, which means no specific orientation of the morphology 
of the cores is observed with respect to a large-scale filament shape model 
for Lupus I. Similar results are found when the average elongation 
of each core is compared to the closest normal of a large-scale 
bent filament model discussed by \citet{mat13}.

As a second step we compared this distribution with the mean orientation of
the magnetic fields probed with 350 $\mu$m polarimetry 
in the high density regions of Lupus I with the BLASTPol experiment.
Here again we do not find any correlation with respect to the large-scale magnetic
field structure.

Our main conclusion is that the local filament dynamics -- including
secondary filaments that often run orthogonally to the primary
filament -- and possibly small-scale variations in the local magnetic 
field direction, could be the dominant factors to explain the final orientation of each core.

The BLASTPol collaboration acknowledges support from NASA 
(through grant numbers NAG5-12785, NAG5-13301, 
NNGO-6GI11G, NNX0-9AB98G, and the Illinois Space Grant 
Consortium), the Canadian Space Agency (CSA), 
the Leverhulme Trust through the Research Project Grant 
F/00 407/BN, Canada’s Natural Sciences and Engineering 
Research Council (NSERC), the Canada Foundation for 
Innovation, the Ontario Innovation Trust, the Puerto Rico 
Space Grant Consortium, the Fondo Institucional para la 
Investigaci\'on of the University of Puerto Rico, and the 
National Science Foundation Office of Polar Programs. C.B. 
Netterfield also acknowledges support from the Canadian 
Institute for Advanced Research. Finally, we thank the 
Columbia Scientific Balloon Facility (CSBF) staff for their outstanding work.




\begin{thebibliography}{}
 
\bibitem[Angil\`e et~al.(2014)]{ang14}Angil\`e, E. and the BLASTPol team. In prep.

\bibitem[Ballesteros-Paredes et~al.(2011)]{bal11}Ballesteros-Paredes, J., Hatmann, L.W., V\'asquez-Semadeni, E., Heitsch, F., Zamora-Avil\'es, M.A. 2011, \mnras, 411, 65.

\bibitem[Bonnell et~al.(2013)]{bon13}Bonnell, I.A., Dobbs, C.L., Smith R.J.2013, \mnras, 430, 1790. 

\bibitem[Chapman et~al.(2013)]{cha13}Chapman, N.L., Davidson, J.A., Goldsmith, P.F., Houde, M., Kwon, W. et al., 2013 \apj, in press.

\bibitem[Falceta-Gon\c{c}alves et~al.(2008)]{fal08}Falceta-Gon\c{c}alves, D., Lazarian, A., Kowal, G. 2008, \apj, 679, 537-551. 

\bibitem[Gammie et~al.(2003)]{gam03}Gammie, C.F., Lin, Y.-T., Ostriker, E. C., Stone, J.M. 2003, \apj, 592, 203. 

\bibitem[Girart et~al.(2013)]{gir13}Girart, J.M., Frau, P., Zhang, Q., Koch, P.M., Qiu, K., Tang, Y.-W., Lai, S.-P., Ho, P.T.P. 2013, \apj, 772, 69.

\bibitem[Hildebrand et~al.(1999)]{hil99}Hildebrand, R.H., Dotson,  J.L., Dowell, C.D., Schleuning, D.A., Vaillancourt, J.E. 1999, \apj, 516, 834.

\bibitem[Heitsch et~al.(2009)]{hei09}Heitsch, F., Stone, J.M., Hartmann, L.W. 2009, \apj, 695, 248.

\bibitem[Hara et~al.(1999)]{har99}Hara, A., Tachihara, K., Mizuno, A., Onishi, T., Kawamura, A., Obayashi, A., Fukui, Y. 1999, PASJ, 51, 895.

\bibitem[Hull et~al.(2013)]{hul13}Hull, C.L.H., Plambeck, R.L., Bolatto, A.D., Bower, G.C., Carpenter, J.M. et al., 2013 \apj, 768, 159.

\bibitem[Lazarian(2007)]{laz07}Lazarian, A. 2007, JQSRT, 106, 225.

\bibitem[Le{\~a}o et al.(2013)]{lea13}Le{\~a}o, M.~R.~M., de Gouveia Dal Pino, E.~M., Santos-Lima, R., Lazarian, A.\ 2013, ApJ, in press. 

\bibitem[Lombardi et~al.(2008)]{lom08}Lombardi, L., Lada, C.J., Alves, J. 2008, \aa, 480, 785.

\bibitem[Matthews et~al.(2014)]{mat13}Matthews, T. G., Ade. P. A. R., Angil\`e, F. E., Benton, S. J., Chapin, E. L. et. al 2014, \apj, 784, 116.

\bibitem[Molinari et~al.(2014)]{mol14}Molinari, S., Bally, J., Glover, S., Moore, T., Noriega-Crespo, A. et al., arXiv:1402.6196.

\bibitem[Moncelsi et~al.(2014)]{mon14}Moncelsi, L., Ade, P., Angil\`e, F. E., Benton, S.J., Devlin, M. et al., 2014 MNRAS, 437, 2772.

\bibitem[McKee \& Ostriker (2007)]{mck07}McKee, C.F., Ostriker, E.C. 2007, ARAA, 45, 565.

\bibitem[Nakamura \& Li (2011)]{nak11}Nakamura, F., Li, Z.-Y. 2011, \apj, 740, 36.

\bibitem[Pascale et~al.(2012)]{pas12}Pascale, E., Ade, P.A.R., Angil\`e, F.E. et. al 2012, SPIE Vol. 8444, 15.

\bibitem[Pelkonen et~al.(2009)] {pel09}Pelkonen, V.-M., Juvela, M., Padoan, P. 2009, \aap, 502, 833. 

\bibitem[Poidevin \& Bastien(2006)]{poi06}Poidevin, F. \& Bastien, P. 2006, \apj, 650, 945.



\bibitem[Poidevin et~al.(2013)]{poi13}Poidevin, F., Falceta-Gon\c{c}alves, D., Kowal, G., De Gouveia Del Pino, E., Magalh\~{a}es, A.M., 2013, \apj, 777, 112. 

\bibitem[Rizzo et~al.(1998)]{riz98}Rizzo, J., Morras, R., Arnal, E. 1998, MNRAS, 300, 497.

\bibitem[Rygl et~al.(2013)]{ryg13}Rygl, K. L. J., Benedettini, M., Schisano, E. et al. 2013, A\&A, 549, L1.

\bibitem[Ostriker et~al.(2001)]{ost01}Ostriker, E. C., Stone, J.M., Gammie, C.F. 2001, \apj, 546, 980.

\bibitem[Serkowski(1962)]{ser62}Serkowski, K. 1962, Advances in Astronomy and Astrophysics, ed. Z. Kopal (New York: Academic), 290.

\bibitem[Schneider et~al.(2013)]{sch13}Schneider, N., Andr \'e, Ph., K\"{o}nyves, V., Bontemps, S., Motte, F. et al. 2013, \apj, 766, L17.


\bibitem[Tassis et~al.(2009)]{tas09}Tassis, K., Dowell, C.D., Hildebrand, R.H., Kirby, L., Vaillancourt, J.E. 2009, MNRAS, 399, 1681.

\bibitem[Tassis et~al.(2012c)]{tas12c}Tassis, K., Talayeh, H., Willacy, K. 2012, \apj 760, 57.

\bibitem[Tassis et~al.(2012b)]{tas12b}Tassis, K., Willacy, K., Yorke, H. W., Turner, N. J. 2012, \apj 754, 6.

\bibitem[Tassis et~al.(2012a)]{tas12a}Tassis, K., Willacy, K., Yorke, H. W., Turner, N., J. 2012, \apj 745, 68.

\bibitem[Ward et~al.(2014)]{war14}Ward, R. L., Wadsley, J., Sills, A. MNRAS 2014, 439, 651.
	
\bibitem[Ward-Thompson et~al.(2000)]{war00}Ward-Thompson, D., Kirk,J. M., Crutcher, R. M.; Greaves, J. S., Holland, W. S., Andr \'e, P. 2000, MNRAS 537, 135.

\bibitem[Ward-Thompson et~al.(2009)]{war09}Ward-Thomspon, D., Sen , A.K., Kirk, J.M., Nutter, D. 2009, MNRAS, 398, 394.

\end{thebibliography}
\end{document}